\documentclass{article}

\usepackage{silence}
\usepackage{amsmath,amsfonts}
\usepackage{amsthm}
\usepackage{etoolbox}
\usepackage{hyperref}
\usepackage{algorithm}
\usepackage{algpseudocode}
\usepackage{xcolor}
\usepackage{grffile}
\usepackage{graphicx}
\usepackage{multirow}
\usepackage[inline,shortlabels]{enumitem}
\usepackage{xspace}
\usepackage{xifthen}
\usepackage{datetime}
\usepackage[skip=0pt,labelformat=simple]{subcaption}
\usepackage{thmtools,thm-restate}
\usepackage[normalem]{ulem}
\usepackage{mathtools}
\usepackage{mdframed}
\usepackage{relsize}
\usepackage{bold-extra}
\usepackage{cleveref}
\usepackage{caption}
\usepackage{quoting}

\newtoggle{paper}

\togglefalse{paper}

\iftoggle{paper}{}{%
  \usepackage[a4paper]{geometry}
  \usepackage[T1]{fontenc}
  \usepackage[tt=false, type1=true]{libertine}
  \usepackage[varqu]{zi4}
  \usepackage[libertine]{newtxmath}
  \usepackage[numbers]{natbib} 
  \usepackage{array}
  \usepackage[labelfont=bf]{caption}
}

\iftoggle{paper}{%
\setlist{nosep,leftmargin=*}
}{}

\newcounter{savedEquation}
\newenvironment{steps}{%
  \setcounter{savedEquation}{\value{equation}}
  \setcounter{equation}{0}
  
  \align
}{%
  \endalign
  \setcounter{equation}{\value{savedEquation}}
}
\newcommand{\stepref}[1]{Step~(\ref{#1})}

\DeclareEmphSequence{\itshape,\bfseries\itshape}

\declaretheoremstyle[headindent=0pt,headfont=\normalfont\bfseries,bodyfont=\itshape]{bolded}

\declaretheorem[style=bolded]{problem}

\declaretheorem[style=bolded]{corollary}

\declaretheorem[style=bolded]{example}

\declaretheorem[style=bolded,name=Cost Model]{costmodel}
\declaretheorem[style=bolded]{query}

\newcommand{\Bigggl}{\ensuremath{\left(\rule{0pt}{2em}\right.}}
\newcommand{\Bigggr}{\ensuremath{\left.\rule{0pt}{2em}\right)}}

\algtext*{EndFunction}
\algtext*{EndFor}
\algtext*{EndWhile}
\algtext*{EndIf}

\algnewcommand\InlineFor[2]{%
  \algorithmicfor\ {#1}\ \algorithmicdo\ {#2}%
}
\algnewcommand\IfThenElse[3]{%
  \algorithmicif\ {#1}\ \algorithmicthen\ {#2}\ \algorithmicelse\ {#3}%
}
\algnewcommand\InlineIfElse[3]{%
  {#2}\ \algorithmicif\ {#1}\ \algorithmicelse\ {#3}%
}

\newcolumntype{M}[1]{>{\centering\arraybackslash}m{#1}}

\newcommand{\pred}{\ensuremath{P}}
\newcommand{\predi}[1]{\ensuremath{\pred_{#1}}}

\newcommand{\vset}{\ensuremath{D}}
\newcommand{\vseto}{\ensuremath{E}}

\newcommand{\vsetop}{\ensuremath{\vseto'}}
\newcommand{\vsetp}{\ensuremath{F}}
\newcommand{\vsetpp}{\ensuremath{\vsetp'}}
\newcommand{\vsetq}{\ensuremath{G}}
\newcommand{\vsetqp}{\ensuremath{\vsetq'}}

\newcommand{\vseti}[1]{\ensuremath{\vset_{#1}}}

\newcommand{\vsets}{\ensuremath{\mathbf{\vset}}}
\newcommand{\vsetsi}[1]{\ensuremath{\mathbf{\vset}_{#1}}}

\newcommand{\setopset}{\ensuremath{\mathbf{O}}}
\newcommand{\univ}{\ensuremath{U}}
\newcommand{\univi}[1]{\ensuremath{\univ_{#1}}}
\makeatletter
    \@ifdefinable{\univpvi}{\def\arnold/{Arnold Schwarzenegger}}
\makeatother
\newcommand{\univpvi}[1]{\ensuremath{\univi{i-1} \cup \{\pred(\vseti{i})\}}}

\newcommand{\vj}{\ensuremath{v_j}}

\newcommand{\boolformorig}{\ensuremath{P^*}}

\newcommand{\setformspace}{\ensuremath{\Psi}}

\newcommand{\critipos}{\vipos{}}
\newcommand{\critineg}{\vineg{}}
\newcommand{\vipos}{\ensuremath{v|_{i=T}}}
\newcommand{\vineg}{\ensuremath{v|_{i=F}}}

\newcommand{\uipos}{\ensuremath{u|_{i=T}}}
\newcommand{\uineg}{\ensuremath{u|_{i=F}}}

\newcommand{\node}{{\tt node}}

\newcommand{\child}{{\tt child}}

\newcommand{\refnode}{\ensuremath{\text{\tt ref}_{\text{\tt node}}}}
\newcommand{\refnodei}[1]{\ensuremath{\texttt{ref}_{#1}}}

\newcommand{\done}{\texttt{done}}
\newcommand{\donei}[1]{\ensuremath{\texttt{done}^{(#1)}}}
\newcommand{\tdo}{\texttt{todo}}

\newcommand{\rootnode}{\texttt{root}}
\newcommand{\nodepa}{\node{}.\text{applyPred}}

\newcommand{\childi}[1]{\ensuremath{\child_{#1}}}
\newcommand{\step}{\ensuremath{S}}
\newcommand{\stepi}[1]{\ensuremath{\step_{#1}}}
\newcommand{\stepouti}[1]{\ensuremath{X_{#1}}}

\newcommand{\anc}{\ensuremath{A}}
\newcommand{\anci}[1]{\ensuremath{\anc_{#1}}}
\newcommand{\ancp}{\ensuremath{\anc'}}

\newcommand{\selec}{\ensuremath{\gamma}}
\newcommand{\seleci}[1]{\ensuremath{\gamma_{#1}}}

\newcommand{\stepseq}{\ensuremath{\mathbf{S}}}
\newcommand{\predseq}{\ensuremath{\mathbf{P}}}

\newcommand{\act}{\ensuremath{O}}
\newcommand{\acti}[1]{\ensuremath{\act_{#1}}}
\newcommand{\sordering}{{\tt suborderings}}
\newcommand{\rank}{\ensuremath{r}}
\newcommand{\frank}{\ensuremath{r^*}}
\newcommand{\parent}{\text{\upshape parent}}

\newcommand{\jmax}{\ensuremath{j_{\max}}}

\newcommand{\chl}{\ensuremath{\text{\upshape children}}}
\newcommand{\chlp}{\ensuremath{\text{\upshape children}^{+}}}
\newcommand{\chln}{\ensuremath{\text{\upshape children}^-}}
\newcommand{\chlc}{\ensuremath{\text{\upshape children}^{\circ}}}
\newcommand{\cany}{\ensuremath{c}}
\newcommand{\cpos}{\ensuremath{c^+}}
\newcommand{\cneg}{\ensuremath{c^-}}
\newcommand{\cnegp}{\ensuremath{c^{-\prime}}}

\newcommand{\ccmp}{\ensuremath{c^{\circ}}}
\newcommand{\ccmpp}{\ensuremath{c^{\circ\prime}}}

\newcommand{\cspc}{\ensuremath{c^*}}

\newcommand{\canyp}{\ensuremath{c^{\prime}}}
\newcommand{\cposp}{\ensuremath{c^{+\prime}}}

\newcommand{\true}{\ensuremath{T}}
\newcommand{\false}{\ensuremath{F}}

\newcommand{\system}{\textsc{Cham}}

\newcommand{\algo}{\textsc{EvalPred}}

\newcommand{\bestd}{\textsc{BestD}}
\newcommand{\bestdp}{\textsc{BestD}\ensuremath{^{\prime}}}
\newcommand{\bestdf}{\textsc{BestDFlipped}}

\newcommand{\bestp}{\textsc{OrderP}}
\newcommand{\hanani}{\textsc{Hanani}}

\newcommand{\combinedp}{\textsc{Combined}}
\newcommand{\combineda}{\textsc{CombinedAll}}
\newcommand{\bdc}{\textsc{BDC}}
\newcommand{\deep}{\textsc{GreedyD3}}

\newcommand{\deeporderp}{\textsc{One\-Look\-aheadP}}

\newcommand{\update}{\textsc{Update}}
\newcommand{\updatep}{\textsc{Update}\ensuremath{^{\prime}}}

\newcommand{\estchild}{\textsc{OrderNode}}

\newcommand{\orderhelper}{\textsc{OrderNodeHelper}}
\newcommand{\sortfand}{\textsc{GetAndWeight}}
\newcommand{\sortfor}{\textsc{GetOrWeight}}

\newcommand{\byp}{\textsc{Byp}}
\newcommand{\tdacb}{\textsc{Tdacb}}
\newcommand{\noopt}{\textsc{NoOrOpt}}
\newcommand{\evalnode}{\textsc{EvalNode}}

\newcommand{\opteval}{\textsc{GeneralEvalPred}}

\newcommand{\mystate}{{\tt state}}
\newcommand{\mapping}{\texttt{leafNodes}}
\newcommand{\nil}{\textsc{Nil}}

\newcommand{\cmp}[1][]{%
  \begingroup%
  \ifthenelse{\isin{!}{#1}}{%
    \def\x{incomplete}%
  }{%
    \ifthenelse{\isin{n}{#1}}{%
      \def\x{completion}%
    }{%
      \ifthenelse{\isin{d}{#1}}{%
        \def\x{completed}%
      }{%
        \ifthenelse{\isin{s}{#1}}{%
          \def\x{completes}%
        }{%
          \ifthenelse{\isin{v}{#1}}{%
            \def\x{{\tt cmp}}%
          }{%
            \def\x{complete}%
          }%
        }%
      }%
    }%
  }%
  {\x}%
  \endgroup%
}

\newcommand{\detpos}[1][]{%
  \begingroup%
  \ifthenelse{\isin{v}{#1}}{%
    \def\x{\ensuremath{{\tt det}^{\texttt{+}}}}%
  }{%
    \def\x{\ensuremath{\text{determ}^{+}}}%
  }%
  {\x}%
  \endgroup%
}

\newcommand{\detneg}[1][]{%
  \begingroup%
  \ifthenelse{\isin{v}{#1}}{%
    \def\x{\ensuremath{{\tt det}^{\texttt{-}}}}%
  }{%
    \def\x{\ensuremath{\text{determ}^{-}}}%
  }%
  {\x}%
  \endgroup%
}

\newcommand{\vatom}[1][]{%
  \begingroup%
  \ifthenelse{\isin{m}{#1}}{%
    \ensuremath{\Gamma}%
    }{%
    \ifthenelse{\isin{c}{#1}}{%
      \def\x{Vertex group}%
      }{%
      \ifthenelse{\isin{C}{#1}}{%
        \def\x{Vertex Group}%
        }{%
        \def\x{vertex group}%
      }%
    }%
    \ifthenelse{\isin{s}{#1}}{%
      {\x}s%
      }{%
      {\x}%
    }%
  }%
  \endgroup%
}

\newcommand{\determ}[1][]{%
  \begingroup%
  \ifthenelse{\isin{!}{#1}}{%
    \def\x{undeterminable}%
    }{%
    \def\x{determinable}%
  }%
  {\x}%
  \endgroup%
}

\newcommand{\vatomvi}{\ensuremath{\vatom[m](v,i)}}

\newcommand{\patom}[1][]{%
  \begingroup%
  \ifthenelse{\isin{c}{#1}}{%
    \def\x{Predicate}%
  }{%
    \ifthenelse{\isin{C}{#1}}{%
      \def\x{Predicate}%
    }{%
      \ifthenelse{\isin{a}{#1}}{%
        \def\x{P}%
      }{%
        \def\x{predicate}%
      }%
    }%
  }%
  \ifthenelse{\isin{s}{#1}}{%
    {\x}s%
    }{%
    {\x}%
  }%
  \endgroup%
}

\newcommand{\makeref}[1][]{%
  \begingroup%
  \ifthenelse{\isempty{#1}}{%
    \def\x{???}%
  }{%
    \def\x{#1}%
  }%
  \textcolor{red}{[\x]}%
  \endgroup%
}

\makeatletter
\g@addto@macro\@floatboxreset{\centering}
\makeatother

\iftoggle{paper}{\setlength{\textfloatsep}{10pt plus 5pt minus 5pt}}{}

\iftoggle{paper}{%
  \newcommand\vldbdoi{XX.XX/XXX.XX}
  \newcommand\vldbpages{XXX-XXX}
  \newcommand\vldbvolume{17}
  \newcommand\vldbissue{1}
  \newcommand\vldbyear{2024}
  \newcommand\vldbauthors{\authors}
  \newcommand\vldbtitle{\shorttitle}
  \newcommand\vldbavailabilityurl{https://github.com/alkim0/disjunct-opt}
  \newcommand\vldbpagestyle{plain}

  \author{Albert Kim}
  \affiliation{%
    \institution{MIT}
  }
  \email{alkim@csail.mit.edu}

  \author{Atalay Mert Ileri}
  \affiliation{%
    \institution{Kansas State University}
  }
  \email{atalay@k-state.edu}

  \author{Sam Madden}
  \affiliation{%
    \institution{MIT}
  }
  \email{madden@csail.mit.edu}
}{
  \author{%
    Albert Kim\\
    MIT \\
    {\tt alkim@csail.mit.edu} \and
    Atalay Mert Ileri \\
    Kansas State University \\
    {\tt atalay@k-state.edu} \and
    Sam Madden \\
    MIT \\
    {\tt madden@csail.mit.edu}
  }
  \date{}
}

\title{Optimizing Query Predicates with Disjunctions for Column-Oriented Engines}

\begin{document}

\iftoggle{paper}{%
\begin{abstract}
Despite decades of work on query optimization, database research has given limited attention to optimizing predicates with disjunctions.  What little past work there is, has mostly focused on optimizations for traditional row-oriented databases.
  However, a key difference between how row-oriented and column-oriented engines evaluate predicates is that while row-oriented engines apply predicates to a \emph{single} tuple at a time, column-oriented engines apply predicates to \emph{sets} of tuples, adding another dimension to the problem.
  As such, row-oriented engines are focused only on the best order to apply predicates in to ``short-circuit'' the overall predicate expression, but column-oriented engines must additionally decide on the input sets of tuples for each predicate application.
  This is important, since smaller inputs lead to faster runtimes, and nontrivial, since the results of earlier predicates can be used to reduce the inputs to later predicates and predicates may be combined via disjunctions in the predicate expression. 
  In this work, we formally analyze the predicate evaluation problem for column-oriented engines and present \bestd{}/\update{}, the first ever polynomial-time, \emph{provably
optimal} algorithms to deduce the minimum input sets for each predicate application.
  \bestd{}/\update{}'s optimality is guaranteed under a wide range of cost models, representing different real-world scenarios. 
  Furthermore, when combined with the predicate ordering algorithm \hanani{}, \bestd{}/\update{} reduce into \algo{}, a simple $O(n \log^2 n )$ algorithm, which we recommend for practical use and optimal for all predicate expressions of nested depth 2 or less.
  Our evaluation shows, thanks to its optimality and polynomial planning time, \algo{} outperforms not implementing any disjunction optimizations and exiting optimal algorithms by up to 2.6$\times$ and 28$\times$ respectively for synthetic workloads and by up to 1.3$\times$ and 100$\times$ respectively for queries from TPC-H and the CH-benchmark.
\end{abstract}

  \maketitle

  \pagestyle{\vldbpagestyle}
  \begingroup\small\noindent\raggedright\textbf{PVLDB Reference Format:}\\
  \vldbauthors. \vldbtitle. PVLDB, \vldbvolume(\vldbissue): \vldbpages, \vldbyear.\\
  \href{https://doi.org/\vldbdoi}{doi:\vldbdoi}
  \endgroup
  \begingroup
  \renewcommand\thefootnote{}\footnote{\noindent
    This work is licensed under the Creative Commons BY-NC-ND 4.0 International License. Visit \url{https://creativecommons.org/licenses/by-nc-nd/4.0/} to view a copy of this license. For any use beyond those covered by this license, obtain permission by emailing \href{mailto:info@vldb.org}{info@vldb.org}. Copyright is held by the owner/author(s). Publication rights licensed to the VLDB Endowment. \\
    \raggedright Proceedings of the VLDB Endowment, Vol. \vldbvolume, No. \vldbissue\ %
    ISSN 2150-8097. \\
    \href{https://doi.org/\vldbdoi}{doi:\vldbdoi} \\
  }\addtocounter{footnote}{-1}\endgroup

  \ifdefempty{\vldbavailabilityurl}{}{
      \vspace{.3cm}
      \begingroup\small\noindent\raggedright\textbf{PVLDB Artifact Availability:}\\
      The source code, data, and/or other artifacts have been made available at \url{\vldbavailabilityurl}.
      \endgroup
    }
}{%
  \maketitle

}

\section{Introduction}
\label{sec:intro}

Despite being a core topic of database research for decades, query optimization paid surprisingly little attention to optimizing queries with disjunctions.
What little work there is, has mostly focused on optimizations for traditional row-oriented databases.
However, there is a critical difference in how row-oriented and column-oriented execution engines handle predicates.
In row-oriented engines, the input to a predicate is a \emph{single} tuple and the output is a true/false value.
Thus, given a predicate expression, the main focus is on finding the best order to apply the base predicates in to maximize the chances of ``short-circuiting'' the predicate expression as soon as possible (i.e., derive a true/false value for the predicate expression as soon as possible).
On the other hand, both the inputs and outputs to predicates in column-oriented engines are \emph{sets} of tuples, and evaluating a predicate expression must return the set of all tuples which satisfy that predicate expression.
This adds another dimension to the problem.
Not only must column-oriented engines determine the best order to apply predicate in, but they must also determine what input sets to use for each predicate application and how to combine the output sets to form the final result; we call this the \emph{set management} problem.
Smaller inputs lead to less time spent on predicate evaluation, and this is particularly important if predicates include expensive components, such as JSON processing and unoptimized user-defined functions, or if the I/O to retrieve the values for predicate evaluation is slow.
Thus, the engine must strive to select the smallest possible input sets for each predicate application, while still ensuring it can form the final result from the output sets.
As a simple example, consider:

\begin{query}
  Query with both conjunction and disjunction.
  \begin{quoting}[leftmargin=0pt]
    \centering
    {\tt SELECT * FROM $T$ WHERE ((\predi{1} AND \predi{2}) OR \predi{3}) AND \predi{4}}
  \end{quoting}
  \label{query:ex}
\end{query}

\noindent
Here, a row-oriented engine would only be interested in finding the best ordering for the base predicates \predi{1}, \predi{2}, \predi{3}, and \predi{4}.
However, a column-oriented engine must go one step further and manage the input and output sets of each predicate application.
Specifically, if the engine determines the best ordering is $[\predi{4}, \predi{3}, \predi{1}, \predi{2}]$, to ensure the smallest possible input to each predicate, it would:
\begin{enumerate}[(1)]
  \item Evaluate \predi{4} on the set of all tuples in $T$.
  \item Evaluate \predi{3} on the resulting tuples from the previous step.
  \item Evaluate \predi{1} on the resulting tuples from step 1 minus the resulting tuples from step 2.
  \item Evaluate \predi{2} on the resulting tuples from the previous step.
  \item Return the union of the resulting tuples from steps 2 and 4.
\end{enumerate}
How can we be sure this is the best we can do, and how does the ordering of predicates affect set management?
What happens when the predicate expression becomes more complex and has more predicates?
These are the questions which we seek to address in this work.

\subsection{Our Work}
\label{sec:intro:our-work}
In our work, we formally analyze the predicate evaluation problem for column-oriented engines.
We show through our analysis that the problem can be split into the two aforementioned components:
\begin{enumerate*}
  \item predicate ordering
  \item set management
\end{enumerate*}.
This split allows us to reuse existing work~\cite{hanani_optimal_1977}~\cite{kemper_optimizing_1992} from traditional row-oriented databases for the predicate ordering component.
For set management, we present \bestd{}/\update{}, the first polynomial-time algorithms to produce \emph{provably optimal} set management plans for any ordering of predicates.
Here, we define optimality as incurring the fewest number of \patom[] evaluations possible.
Our proofs show that \bestd{}/\update{} deduce the smallest possible input set for each predicate application, and this analysis holds under a wide range of cost models, representing different real-world situations, as long as the cost model obeys a few properties. 
Furthermore, when combined with the predicate ordering algorithm \hanani{}~\cite{hanani_optimal_1977}, \bestd{}/\update{} reduce into \algo{}.
\algo{} is a simple $O(n \log^2 n )$ algorithm which generates optimal predicate evaluation plans for all predicate expressions of nested depth 2 or less (e.g., ANDs of ORs) and is the algorithm which we recommend for practical use; note $n$ here is the number of predicates.
Compared to other optimal algorithms~\cite{kastrati_generating_2018} which iterate over an exponential space, \algo{}'s strong theoretical backing allows it to achieve optimality while remaining almost trivially simple.
Even for predicate expressions of depth 3 or greater (e.g., ANDs of ORs or ANDs), we show via experimentation that \algo{} comes quite close to optimal; for 92\% of the queries, the number of \patom[] applications invoked was within within 5\% of minimum required.

\noindent
\textbf{Contributions.} In short, our contributions are:
\begin{enumerate}
  \item Our algorithms. \algo{}, a simple, practical algorithm which generates optimal predicate evaluation plans for all predicate expression of depth 2 or less, and \bestd{}/\update{}, which generate the best set management plans for any predicate ordering.
  \item Our various theoretical results, including a formal analysis of the predicate evaluation problem for column-oriented engines and proofs of optimality for \bestd{}/\update{} and \algo{}.
  \item Evaluation of our algorithms, which outperform not implementing any disjunction optimizations and existing optimal algorithms by up to 2.6$\times$ and 28$\times$ respectively for synthetic workloads and by up to 1.3$\times$ and 100$\times$ respectively for queries from TPC-H and the CH-benchmark.
\end{enumerate}

The remainder of this paper is structured as follows.
We start by presenting related work in Section~\ref{sec:rel-work}.
We formulate the predicate evaluation problem in Section~\ref{sec:problem} and present the practical \algo{} first in Section~\ref{sec:algo}.
The more theoretical parts of the paper, including the problem analysis, \bestd{}/\update{}, and the proofs of optimality are all presented later in Section~\ref{sec:optimal}.
Section~\ref{sec:eval} shows our evaluation, and we conclude with Section~\ref{sec:conc}.

\section{Related Work}
\label{sec:rel-work}

{\noindent \bf \tdacb/\byp.}
As far as the authors are aware, Kastrati and Moerkotte's work on \tdacb{}~\cite{kastrati_generating_2018} and \byp{}~\cite{kastrati_optimization_2017} are the only other works which focus on optimizing queries with disjunctions for column-oriented engines.
\tdacb{} also stands unique from other related works in that it too generates optimal predicate evaluation plans.
As such, it is the main competitor of our work.
However, in contrast to the formal approach taken by our work, \tdacb{} achieves its optimality by searching over the exponential space of all query plans.
Despite several clever optimizations to prune the search space, such as branch-and-bound and memoization, \tdacb{} still incurs a time complexity of $O(n 3^n)$, in which $n$ is the number of \patom[s].
Given that Fontoura et al. report a real workload (from Yahoo Inc.) of up to 150k \patom[s] with hundreds of unique attributes~\cite{fontoura_efficiently_2010}, clearly this solution is not scalable.
In fact, our evaluation shows the planning time for \tdacb{} can be quite expensive for even as few as 16 \patom[s].
In comparison, \algo{} is able to produce the optimal predicate evaluation plan in $O(n \log^2 n)$ time, and our experiments show a 28$\times$ speedup in planning time over \tdacb{}.
\byp{} is Kastrati and Moerkotte's previous work which they improved upon with \tdacb{}.
\byp{} requires its input to be in DNF and is only locally optimal while still searching the entire search space of predicate evaluation plans.
Thus, \byp{} is strictly worse than \tdacb{}.
\iftoggle{paper}{}{%
  Finally, both \byp{} and \tdacb{} are meant for in-memory systems, and they assume little to no overhead in fetching the actual data.
Thus, their guarantees of optimality do not extend to situations in which data is not stored in memory.
In contrast, \algo{} makes no such assumption.
}

{\noindent \bf \hanani{}.}
In 1977, Hanani introduced a \patom[] ordering algorithm~\cite{hanani_optimal_1977}, which we refer to as \hanani{} henceforth.
Although \hanani{} was developed for row-oriented databases, the same algorithm can be used for column-oriented engines thanks to our split between \patom[] ordering and set management.
As mentioned, combining \hanani{} with \bestd{}/\update{} gives us the practical \algo{}.
However, while \hanani{} (and thus \algo{}) is optimal for all predicate expressions of nested depth 2 or less, this is no longer the case for predicate expressions of nested depth 3 or greater due to an implicit assumption of a depth-first search (DFS) traversal of the predicate expression.
In Section~\ref{sec:optimal}, we present a property called ``determinability'' which identifies this issue.

{\noindent \bf Boolean Difference Calculus.}
Kemper et al.~\cite{kemper_optimizing_1992} propose an alternative ordering for \patom[s] based on Boolean Difference Calculus (BDC).
In BDC, each \patom[]'s ``importance'' is measured based on their likelihood of affecting the overall result, and \patom[s] are ordered based on decreasing importance.
In our evaluation, we compare the performance of using BDC with \bestd{}/\update{} against \algo{} (\hanani{} and \bestd{}/\update{}).

{\noindent \bf Bypass.}
Some of the most seminal works in optimizing disjunctions are the works regarding the bypass technique~\cite{kemper_optimizing_1994}~\cite{steinbrunn_bypassing_1995}~\cite{sen_optimization_2000}.
This technique splits the incoming tuples of a predicate into ``true'' and ``false'' streams based on the results, and only the tuples in the false stream are evaluated further, while tuples in the true stream are allowed to bypass the other predicates.
While this technique was developed for row-oriented databases, Kastrati and Moerkotte's \byp{} adapt the main ideas for column-oriented engines.

{\noindent \bf Factorization.}
Chaudhuri et al.~\cite{chaudhuriFactorizingComplexPredicates2003} present a work on factorizing predicate expressions with disjunctions to take advantage of existing indexes.
As we discuss later, a key assumption to the optimality of our algorithms is the uniqueness of \patom[s].
If the input predicate expression has reappearing \patom[s], we can use the algorithm presented by Chaudhuri et al. to minimize the number of reappearing \patom[s] before applying our algorithms.

{\noindent \bf Dewey Evaluation.}
Fontoura et al.~\cite{fontoura_efficiently_2010} present a way to efficiently encode predicate expressions with disjunctions and tens of thousands of \patom[s] into simple labels.
This encoding scheme could be used with our algorithms to reduce memory footprint.

{\noindent \bf CNF/DNF.}
The remaining works on disjunction optimization all focus on first trying to convert the predicate expression into either conjunctive normal form (CNF) or disjunctive normal form (DNF) and optimizing the execution from there~\cite{jarkematthias_query_1984} \cite{straubeQueriesQueryProcessing1990} \cite{chang_optimization_1997} \cite{muralikrishna_optimization_1988}.
Unfortunately, the conversion process into CNF/DNF can result in an exponential number of terms~\cite{russellArtificialIntelligenceModern2016}, so just transforming the input into the correct form could be quite expensive.
In addition, Kastrati and Moerkotte report that CNF/DNF-based evaluations generally tend to produce very poor plans and should be avoided~\cite{kastrati_optimization_2017}.

{\noindent \bf In Practice.}
Other than simple factorization and reduction via Boolean implication, existing, well-used database systems implement few, if any, optimizations for disjunctions.
PostgreSQL~\cite{stonebrakerDesignPostgres1986}, for example, does partial transformations to CNF if separate clauses contain accesses to the same table to minimize the number of tuples for joins.
MonetDB~\cite{idreos_monetdb_2012} and Spark SQL~\cite{armbrustSparkSqlRelational2015} do not seem to implement any additional optimizations, and based on conversations with developers from Vertica~\cite{lamb2012vertica}, Vertica ``doesn't do anything special for ORs''.
The lack of exploration into this area has affected these systems, and we hope \algo{} will serve as a good ``rule-of-thumb'' for systems to follow.

{\noindent \bf Number of \patom[Cs].}
Some readers may doubt whether queries have enough \patom[s] to warrant such optimizing efforts.
However, as Fontoura et al.~\cite{fontoura_efficiently_2010} report, advertising exchanges and automatic query builders can result in queries with tens of thousands of \patom[s].
In addition, Chaudhuri et al.~\cite{chaudhuriFactorizingComplexPredicates2003} discuss a workload that ranges up to over a hundred \patom[s].
Johnson et al.~\cite{johnson_row-wise_2008} also evaluate their work on up to 40 \patom[s] and report that for in-memory systems, the cost of evaluating predicates outweighs the cost of \iftoggle{paper}{}{performing} joins.
\iftoggle{paper}{}{All of this suggests that further research into this area can be highly beneficial.}


\section{Problem Formulation}
\label{sec:problem}

This section presents a formulation of the exact problem which we wish to solve.
We discuss the setting in which our work is applicable in Section~\ref{sec:problem:setup} and describe the expected query properties in Section~\ref{sec:problem:query}.
Section~\ref{sec:problem:system} presents the model of the predicate evaluation system that we use in our analysis, and Section~\ref{sec:problem:cost} lists the cost model properties required for our algorithms' optimality, as well as a few example cost models.

\subsection{Setting}
\label{sec:problem:setup}
We first discuss the setting in which this work is applicable.
Our analysis and algorithms are applicable to both column-oriented traditional databases and column-oriented distributed computing systems (together which we call column-oriented engines).
As is common with column-oriented engines, we assume intermediate representations of tuples during query execution are simple indices into the tables which they come from and that the values of tuples are only materialized when needed.
We assume sets of tuples are represented using lightweight data structures, such as bitmaps, and although no strong assumptions are made about the data storage, we assume that the time to fetch and process the data is significantly greater than the time to manipulate the lightweight indices of tuples in memory.
Note that data storages such as hard disk drives, flash drives, and distributed file storages all fit within this criteria.

The primary resource we are concerned with is time.
We assume we either have enough memory to hold the data needed to process the query or that the query can be executed off disk using a conventional buffer pool design.
If indexing structures to accelerate query execution exist, our algorithms can take advantage of them, but the absence of them does not affect the correctness of our algorithms.

Finally, our algorithms produce plans at the \emph{logical} layer.
They dictate which tuples should be evaluated by which \patom[s] to minimize the total number of evaluations.
When actually executing the predicate evaluation, the system may choose to implement additional \emph{physical} layer optimizations, such as retrieving extra tuples in the case of block-based data storages or even applying the \patom[] on additional tuples for vector processing.
While these optimizations may blunt some of the benefits of our algorithms, it does not render our work meaningless.
Many workloads contain \patom[s] whose evaluation times overshadow the time it takes to fetch the data~\cite{johnson_row-wise_2008}, and common \patom[s] involving components such as regular expression matching and JSON processing cannot be vector processed.
In fact, our own evaluation system fetches data in blocks, but we still see speedups of up to 2.6$\times$ over not implementing any disjunction optimization.

\subsection{Query Properties}
\label{sec:problem:query}
We assume that the queries given to us are selection queries with complex predicate expressions.
Predicate expressions may be composed of any number and depth of conjunctions and disjunctions of Boolean-result \patom[s].
\patom[cs] may be expensive (perhaps user-defined), and different \patom[s] may have different costs.
In our proofs, we assume a query's \patom[s] are unique and that each appears only once in the predicate expression.
If the same \patom[] occurs multiple times in the given predicate expression, one of the many common subexpression elimination algorithms from either research~\cite{chaudhuriFactorizingComplexPredicates2003} or from practice~\cite{stonebrakerDesignPostgres1986}~\cite{grund_hyrise:_2010} may be applied to reduce the number of reoccurring \patom[s].
Note that without uniqueness, our algorithms are no longer guaranteed to be optimal, but they will still return correct results.
For clarity's sake, this paper assumes that we are working with single-table queries, but this is not a strict requirement.
As long as each tuple in the source (single or joined) table has a global id (real or virtual), our
algorithms are applicable.
Even with pushdown optimizations, our algorithms still apply to pushed-down \patom[s] at the individual table level.

\subsection{Predicate Evaluation System}
\label{sec:problem:system}

In our formulation, we model the predicate evaluation system as follows.
The system initially starts with the set of all tuples (in the table), and its goal is to find the set of all tuples which satisfy the given predicate expression.
To accomplish this, the system can perform two possible types of actions as many times as it wants:
\begin{enumerate}
  \item The system can \emph{apply} a \patom[] to a set of tuples to obtain the subset of tuples which satisfy that \patom[].
  \item The system can perform a \emph{set operation} (i.e., union, intersection, difference) on known sets of tuples to derive a new set of tuples.
\end{enumerate}
The application of a \patom[] corresponds to fetching the values of the input tuples and actually performing the \patom[] evaluation.
Since this involves both fetching the data from storage and applying potentially expensive \patom[s] (e.g., regular expression matching, JSON processing, and user-defined functions), we assume this action to be much more costly than performing a set operation action, which can be performed using bitwise operations on bitmaps in memory.
To be called optimal, our algorithms should be able to find the sequence of actions which deduce the set of all tuples that satisfy the given predicate expression, while minimizing the costs of the actions in the sequence.

\subsection{Cost Model}
\label{sec:problem:cost}

The cost model assigns a cost to each possible action the predicate evaluation system can take, as previously introduced.
Our analysis and proofs of optimality hold for any cost model as long as they exhibit following three properties:
\begin{enumerate}
  \item The cost of a \patom[] application is significantly greater than the cost of a set operation.
  \item The cost of a \patom[] application increases with the input size.
  \item When applying the same \patom[] to two different input sets, it is cheaper to apply the \patom[] once on the union of those sets than it is to apply the \patom[] separately on each set.
\end{enumerate}
The first property comes from the basic setting under which this work is applicable, and the second property is often trivially true.
The third property is a ``triangle-inequality''-like property; if $C_P(D)$ measures the cost of applying \patom[] $P$ to input set $D$, then the property can be stated as:
\[
  C_P(D \cup E) < C_P(D) + C_P(E)
\]

Our algorithms are optimal for any cost model which satisfies the above three properties.
A simple example cost model might be:
\begin{costmodel}
  \begin{gather*}
    C_P(D) = F_P |D| + \kappa \\
    C_{\star}(D, E) = \epsilon \left(|D| + |E| + |D \star E| + \kappa'\right)
  \end{gather*}
  \label{cost:1}
\end{costmodel}
\iftoggle{paper}{\vspace{-15pt}}{}
\noindent
The first line shows the cost of applying \patom[] $P$ to input set $D$.
Here, $F_P$ is some cost constant cost factor specific to $P$ (depending on what type of \patom[] $P$ is), $|D|$ is the number of tuples in $D$, and $\kappa$ is some constant overhead cost.
The second line shows the cost of applying set operation $\star$ to input sets $D$ and $E$.
Here, $\star$ is one of $\cup$~(union), $\cap$~(intersection), or $\setminus$~(difference), $\epsilon$ is the ratio of costs between set operations and \patom[] applications, and $\kappa$' is some other constant overhead cost.
The cost includes the sizes of the input sets as well as the size of the output set.
With respect to $\epsilon$, depending on the environment, a \patom[] application can easily cost $30\times\sim100000\times$ more than a set operation.
In our experimental environment, simply reading 10M 4-byte \texttt{int} values from a RAID5 setup of HDDs took 274ms, while ANDing together two bitmaps of 10M elements in memory to produce another bitmap only took 8ms.
Thus, $\epsilon$ can be thought of as approaching 0 in most cases.
The above cost model satisfies all three properties needed for our analysis.
A small $\epsilon$ value ensures the first property, and the costs of \patom[] applications grow proportionally with the sizes of the input sets, ensuring the second property.
The inclusion of the constant overhead cost $\kappa$ ensures the third property.

However, aside from the simple cost model presented above, more complex cost models representing different real-world scenarios also fit under our analysis.
For example, for spinning disk drives, random I/O is often significantly more expensive than sequential I/O, so if the selectivity of the input set is above a certain threshold (e.g., 20\%), then it is cheaper to simply scan the entire data column than it is to retrieve only the relevant tuples using random I/O.
A different cost model that takes this into account would be:
\begin{costmodel}
\begin{gather*}
  C_P(D) = \begin{cases}
    F_{scan} + F_P |D| + \kappa & \text{if selectivity > 20\%}\\
    F_{rnd} \text{\upshape blocks}(D) + F_P |D| + \kappa & \text{otherwise}
  \end{cases} \\
  C_{\star}(D, E) = \epsilon \left(|D| + |E| + |D \star E| + \kappa'\right)
\end{gather*}
\end{costmodel}
\noindent
Here, the set operation cost is the same as the previous cost model, but the cost the of the \patom[] application changes with respect to the selectivity of the input set $D$.
If the input set contains more than 20\% of all tuples in the table, the cost model chooses to scan the entire data column and incurs cost $F_{scan}$.
If the input set contains fewer tuples, then it retrieves only the blocks which contains tuples in $D$ using random I/O.
Accordingly, $\text{blocks}(D)$ returns the number of blocks, and $F_{rnd}$ is the cost factor associated with random I/O.
After fetching the data, the \patom[] is applied to the input tuples, incurring the cost $F_P |D|$.
Note that despite its complexity, this cost model still satisfies our three properties, so our algorithms would be optimal for this case as well.
Other cost models, representing other real-world scenarios, may also be used.
As long as the three properties are satisfies, our analysis holds.

\section{\algo{}}
\label{sec:algo}

\iftoggle{paper}{%
\begin{algorithm}[t]
  \begin{algorithmic}[1]
  \Require Predicate expression $\boolformorig{}$, set of all tuples $R$
  \Ensure The set of all tuples which satisfy \boolformorig{}
  \State $[\predi{1},...,\predi{n}] \gets \hanani{}(\boolformorig{})$ \label{line:algo:orderp} \label{line:algo:already-imp-beg}
  \State \rootnode{} $\gets \text{makeTree}(\boolformorig, [\predi{1},...,\predi{n}])$ \label{line:algo:maketree}
  \State \Return \evalnode(\rootnode{}, $R$) \label{line:algo:call}
  \Statex
  \Function{\evalnode}{\node{}, $D$}
  \If{isLeafNode(\node{})}
  \State \Return \nodepa{}($D$) \label{line:algo:apply}
  \ElsIf{isAndNode(\node)}
  \For{\child{} in children(\node)}
  \If{$D = \varnothing$} \label{line:algo:early-and-beg}
  \State \Return $D$ \label{line:algo:early-and-end}
  \EndIf
  \State $D \gets$ \evalnode(\child, $D$) \label{line:algo:update-and}
  \EndFor
  \State \Return $D$ \label{line:algo:and-return}
  \Else \label{line:algo:already-imp-end}
  \State \done{} $\gets \varnothing$ \label{line:algo:new-imp-beg}
  \For{\child{} in children(\node)}
  \State \tdo{} $\gets D \setminus \done{}$ \label{line:algo:set-todo}
  \If{$\tdo{} = \varnothing$} \label{line:algo:early-or-beg}
  \State \Return \done{} \label{line:algo:early-or-end}
  \EndIf
  \State \done{} $\gets$ \done{} $\cup$ \evalnode(\child, \tdo{}) \label{line:algo:update-or}
  \EndFor
  \State \Return \done \label{line:algo:or-return}
  \EndIf
  \EndFunction \label{line:algo:new-imp-end}
\end{algorithmic}

  \caption{\algo{}}
  \label{alg:algo}
\end{algorithm}
}{%
\begin{algorithm}[H]
  \begin{algorithmic}[1]
  \Require Predicate expression $\boolformorig{}$, set of all tuples $R$
  \Ensure The set of all tuples which satisfy \boolformorig{}
  \State $[\predi{1},...,\predi{n}] \gets \hanani{}(\boolformorig{})$ \label{line:algo:orderp} \label{line:algo:already-imp-beg}
  \State \rootnode{} $\gets \text{makeTree}(\boolformorig, [\predi{1},...,\predi{n}])$ \label{line:algo:maketree}
  \State \Return \evalnode(\rootnode{}, $R$) \label{line:algo:call}
  \Statex
  \Function{\evalnode}{\node{}, $D$}
  \If{isLeafNode(\node{})}
  \State \Return \nodepa{}($D$) \label{line:algo:apply}
  \ElsIf{isAndNode(\node)}
  \For{\child{} in children(\node)}
  \If{$D = \varnothing$} \label{line:algo:early-and-beg}
  \State \Return $D$ \label{line:algo:early-and-end}
  \EndIf
  \State $D \gets$ \evalnode(\child, $D$) \label{line:algo:update-and}
  \EndFor
  \State \Return $D$ \label{line:algo:and-return}
  \Else \label{line:algo:already-imp-end}
  \State \done{} $\gets \varnothing$ \label{line:algo:new-imp-beg}
  \For{\child{} in children(\node)}
  \State \tdo{} $\gets D \setminus \done{}$ \label{line:algo:set-todo}
  \If{$\tdo{} = \varnothing$} \label{line:algo:early-or-beg}
  \State \Return \done{} \label{line:algo:early-or-end}
  \EndIf
  \State \done{} $\gets$ \done{} $\cup$ \evalnode(\child, \tdo{}) \label{line:algo:update-or}
  \EndFor
  \State \Return \done \label{line:algo:or-return}
  \EndIf
  \EndFunction \label{line:algo:new-imp-end}
\end{algorithmic}

  \caption{\algo{}}
  \label{alg:algo}
\end{algorithm}
}

\begin{figure}
  \includegraphics[width=0.6\linewidth]{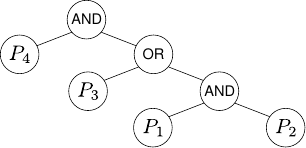}
  \caption{Predicate tree of Query~\ref{query:ex} in $[P_4, P_3, P_1, P_2]$ order.}
  \label{fig:pred-tree}
\end{figure}

We now introduce \algo{}, our simple predicate evaluation algorithm, optimal for all predicate expressions of depth 2 or less, and what we recommend practitioners implement for most cases.
Algorithm~\ref{alg:algo} presents \algo{}.
The algorithm first orders \patom[s] from the given predicate expression\footnote{We expect the given predicate expression to be in negation normal form (NOT operators are pushed inwards until they appear only in front of \patom[s]); if not, this conversion can be done in linear time. Also, all negative \patom[s] are replaced with positive \patom[s]: $\pred' = \neg \pred$.} according to \hanani{}\footnote{We present a compatible version of \hanani{} in \iftoggle{paper}{our technical report~\cite{techreport}}{Appendix~\ref{apx:hanani}}.} and creates a \emph{predicate tree} with that ordering (using makeTree).
In the predicate tree, each node represents a predicate subexpression: leaf nodes represent base \patom[s], and intermediate AND/OR nodes combine children subexpression nodes; the root node refers to the entire predicate expression.
The tree is also normalized such that the parent of each AND node is an OR node and vice versa, resulting in an interleaving of ANDs and ORs across different levels of the predicate tree.
Figure~\ref{fig:pred-tree} depicts the predicate tree for the predicate expression in Query~\ref{query:ex} for the ordering $[P_4, P_3, P_1, P_2]$.

\evalnode{} then uses this predicate tree to determine the input set to each \patom[] and performs the actual evaluation.
Conceptually, \evalnode{}(\node{}, $D$) returns the subset of tuples in $D$ which satisfy the predicate subexpression represented by \node{}.
Thus, assuming $R$ represents the set of all tuples in the table, $\evalnode{}(\rootnode{}, R)$ returns the set of all tuples which satisfy the given predicate expression \boolformorig{}.
As for the algorithm itself, \evalnode{} traverses the predicate tree in a depth-first search (DFS) manner and reduces the input set sizes using various set operations.
For leaf nodes, \evalnode{} simply applies \node{}'s associated \patom[] to $D$.
For AND nodes, the tuples that \evalnode{} returns must satisfy every one of \node{}'s children, so each child updates $D$ with its results, and later children are evaluated on only the set of tuples which have satisfied all previous children.
For OR nodes, each tuple that \evalnode{} returns can satisfy any one of \node{}'s children, so the tuples that satisfy one child are removed from the set of tuples which still have to be checked, and later children are evaluated on only the set of tuples which have not satisfied any previous children.
Also, both AND and OR nodes return early if there are no tuples left to check (Lines~\ref{line:algo:early-and-beg}-\ref{line:algo:early-and-end} and~\ref{line:algo:early-or-beg}-\ref{line:algo:early-or-end}).
An example of running \algo{} with the ordering $[P_4, P_3, P_1, P_2]$ was provided in the introduction.

We hope Algorithm~\ref{alg:algo} appears simple and maybe somewhat even obvious to the reader.
In fact, most column-oriented engines implement some variation of \algo{}.
However, they do not remove the \done{} tuples from the \tdo{} tuples as shown in Line~\ref{line:algo:set-todo} and instead call $\evalnode(\child{}, D)$ for each OR node \child{} without any filtration.
Furthermore, rather than its novelty, \algo{}'s true compelling point is that despite its simplicity, it is \emph{optimal} for all predicate expressions of depth 2 or less, as we show later in Section~\ref{sec:optimal}.
Thus, we can use this polynomial-time $O(n \log^2 n)$ algorithm to generate optimal predicate evaluation plans, instead of the existing exponential-time algorithms.

\noindent
\textbf{Implementation.}
Although \algo{} is presented as one algorithm here, in a real system, Lines~\ref{line:algo:orderp} and~\ref{line:algo:maketree} would be done during planning time, and only the call to \evalnode{} in Line~\ref{line:algo:call} would be performed during execution time.
Furthermore, a real system would most likely use bitmaps to represent sets of tuples and all set operations would be replaced with bitwise operations.

\noindent
\textbf{Time Complexity.}
\algo{}'s plan time complexity is bounded by \hanani{}, which takes $O(n \log^2 n)$ time.
The other operation (makeTree) simply creates the predicate tree according to the given \patom[] ordering and can be done in $O(n)$ time.
Thus, the overall plan time complexity of \algo{} is $O(n \log^2 n)$.

\iftoggle{paper}{}{%
  \noindent
  \textbf{Correctness.}
  The correctness of \algo{} can be proven by showing $\evalnode{}(\node, \vset)$ returns the subset of tuples in \vset{} which satisfy the predicate subexpression represented by \node{} (the full proof is in Appendix~\ref{apx:correct:algo}).
}

\section{Theory}
\label{sec:optimal}

This section presents the theory backing \algo{}.
We first formalize the problem and show that a predicate evaluation system can be modeled as a sequence of operator/operand steps in Section~\ref{sec:optimal:formal}.
We constrain the solution space and show that the only operators we need to worry about are \patom[] applications in Section~\ref{sec:optimal:reduction}.
We present \bestd{}/\update{} and prove their optimality for the set management problem in Section~\ref{sec:optimal:opteval}.
Finally, in Section~\ref{sec:optimal:bestp}, we discuss the \patom[] ordering problem, the conditions to reduce \bestd{}/\update{} into \algo{}, and \hanani{}'s issues with depth-3+ predicate expressions.
\iftoggle{paper}{%
  Due to space constraints, this paper only contains proofs for some of the main theorems.
  Proofs of lemmas can be found in our technical report~\cite{techreport}.
}{%
  Note that this section contains proofs for only the main theorems.
  The proofs for lemmas can be found in Appendix~\ref{apx:lemmas}.
}

\subsection{Formalization}
\label{sec:optimal:formal}
\subsubsection{Valuations}
Rather than reasoning directly about sets of tuples, it is easier for us to instead borrow from propositional logic~\cite{buning1999propositional} and reason about \emph{valuations} (or true/false assignments) instead.
\begin{restatable}{definition}{defval}
  A \emph{valuation} is a tuple of true/false values.
\end{restatable}
\noindent
Given a predicate expression with $n$ unique \patom[s] $(\predi{1}, \predi{2}, ..., \predi{n})$ and a (data) tuple $r$, an $n$-length valuation can be constructed by evaluating each \patom[] on that tuple: $(\predi{1}(r), \predi{2}(r), ..., \predi{n}(r))$.
The resulting valuation is an $n$-length tuple of true/false values (e.g., $(T, F, ..., T)$).
In our analysis, we can use a valuation to represent the set of all data tuples which evaluate to it\footnote{Note that the mapping from valuations to data tuples is merely logical and only used in our reasoning, so no actual evaluations have to be performed.}.
Furthermore, we can use a set of valuations (or a \emph{valuation set}) to represent the union of the tuples represented by each valuation.
For example, the set of all $2^n$ valuations represents the set of all tuples.
Valuation sets allow us to reason about tuples based on their \patom[] results and are sufficient to prove the optimality of our algorithms.

\begin{table}[b]
  \begin{tabular}{|c|c|c|}
    \hline
    \textbf{name} & \textbf{age} & \textbf{height} \\ \hline
    Alice & 25 & 160 \\ \hline
    Bob & 32 &  180 \\ \hline
    Charlie & 48 & 170 \\ \hline
    Dave & 56 & 178 \\ \hline
    Eve & 30 & 165 \\ \hline
  \end{tabular}
  \vspace{5pt}
  \caption{Example relation $T$}
  \label{tab:val-ex}
\end{table}

As an example, consider the relation from \autoref{tab:val-ex} and the predicate expression: (age < 35 OR height > 175).
Let \predi{1} = age < 35 and \predi{2} = height >  175.
The valuation $(T, F)$ represents the set of tuples containing Alice and Eve, since both are under the age of 35 and do not have a height greater than 175.
The set of all valuations which satisfy the predicate expression is $\{(T, T), (T, F), (F, T)\}$, and this represents the set of tuples containing Alice, Bob, Dave, and Eve.


\subsubsection{Predicate Evaluation System}
As mentioned in Section~\ref{sec:problem:system}, a predicate evaluation system starts with the set of all valuations (i.e, the set of all tuples), and its goal is to derive the set of all valuations that satisfy the predicate expression using two types of operators: \patom[] applications and set operations. We define them as:
\begin{restatable}{definition}{defpredapp}
The \emph{\patom[] application} of \patom[] $P$ on valuation set $D$ is denoted $P(D)$ and returns the subset of valuations in $D$ which satisfy $P$.
\end{restatable}
\begin{restatable}{definition}{defsetop}
A \emph{set operation} is one of union ($\cup)$, intersection ($\cap$), or difference ($\setminus$).
\end{restatable}
\noindent
The system performs these actions in a sequence of \emph{steps}:
\begin{restatable}{definition}{defstep}
  A \emph{step} is an (operator, operand) pair: $\step{} = (\act, \vsets)$, in which operator \act{} is either a \patom[] or a set operation, and operand $\vsets${} is the input valuations set(s).
  Applying the operator to operand results in the step's output and is denoted $\act(\vsets)$.
  \label{def:step}
\end{restatable}
\noindent
Assuming $\stepi{i}$ denotes the $i$th step, $\stepi{1} = (P, \vset)$ states that the first step is the \patom[] application of $P$ on the valuation set $D$.
Each step outputs a new valuation set, and later steps may use the outputs of previous steps.
To keep track of all output valuation sets:
\begin{restatable}{definition}{defuniv}
  Given the sequence of steps $[(\acti{1}, \vsetsi{1}), (\acti{2}, \vsetsi{2}), ...,\allowbreak (\acti{m}, \vsetsi{m})]$, \emph{universe} \univi{i} is defined by the recurrence relation:
  \[
    \univi{i} = \univi{i-1} \cup \{\acti{i}(\vsetsi{i})\} \\
  \]
  with the initial universe: $\univi{0} = \{\{T, F\}^n\}$.
\end{restatable}
\begin{restatable}{definition}{defvalid}
  A sequence of steps $[(\acti{1}, \vsetsi{1}), (\acti{2}, \vsetsi{2}), ...,\allowbreak (\acti{m}, \vsetsi{m})]$ is \emph{valid} if, for each step $(\acti{i}, \vsetsi{i})$, all valuation sets in \vsetsi{i} are members of \univi{i-1}.
\end{restatable}
\noindent
Universe \univi{i} contains all output valuations sets from the first $i$ steps, and the $(i+1)$th step can select any valuation set(s) from it to use as its input(s).
The original universe \univi{0} contains only the set of all valuations, since the predicate evaluation system must consider all tuples for its first step/\patom[] application.
The system can be said to have achieved its goal after the $m$th step if universe \univi{m} contains the set of all valuations which satisfy the predicate expression.

\iftoggle{paper}{}{%
  \begin{table}[H]
    \begin{tabular}{|c|c|c|M{.5\linewidth}|}
      \hline
      $i$ & Step (\acti{i}, \vsetsi{i}) & Output (\stepouti{i}) & Universe (\univi{i}) \\ \hline
      1 & (\predi{1}, $\{T,F\}^n$) & $\{(T, T), (T, F)\}$ & $\{\{T,F\}^n, \; \{(T, T), (T, F)\}\}$ \\ \hline
      2 & $\left(\setminus, [\{T,F\}^n, \stepouti{1}]\right)$ & $\{(F, F), (F, T)\}$ & $\{\{T,F\}^n, \; \{(T, T), (T, F)\}, \; \{(F, F), (F, T)\}\}$ \\ \hline
      3 & $\left(\predi{2}, \stepouti{2}\right)$ & $\{(F, T)\}$ & $\{\{T,F\}^n, \; \{(T, T), (T, F)\}, \; \{(F, F), (F, T)\}, \; \{(F, T)\} \}$ \\ \hline
      4 & $\left(\cup, [\stepouti{1}, \stepouti{3}]\right)$ & $\{(T, T), (T, F), (F, T)\}$ & $\{\{T,F\}^n, \; \{(T, T), (T, F)\}, \; \{(F, F), (F, T)\}, \; \{(F, T)\},\allowbreak \{(T, T), (T, F), (F, T)\} \}$ \\ \hline
    \end{tabular}
    \caption{Example sequence of steps for the predicate expression (\predi{1} OR \predi{2}).}
    \label{tab:steps-ex}
  \end{table}

  As an example, consider once again the relation from \autoref{tab:val-ex} and the predicate expression: (age < 35 OR height > 175), in which \predi{1} = (age < 35) and \predi{2} = (height >  175).
  \autoref{tab:steps-ex} shows an example sequence of steps to deduce the set of all valuations which satisfy this predicate expression.
  The table shows, for each step, the operator and operand pair, the step's output (denoted \stepouti{i}), and the universe after that step.
  For example, the first step applies \predi{1} to the set of all valuations $\{T,F\}^n$.
  This results in the valuation set $\{(T, T), (T, F)\}$, and this valuation set is added to the universe to form \univi{1}.
  The second step is a set operation step and evaluates $\{T,F\}^n \setminus \stepouti{1}$, in which \stepouti{1} is the output valuation set from the first step (i.e., $\{(T, T), (T, F)\}$).
  After the last step, the universe contains the valuation set $\{(T, T), (T, F), (F, T)\}$, which is the set of all valuations that satisfies the predicate expression (\predi{1} OR \predi{2}), so the predicate evaluation system is done after this step.
  As another point of reference, \autoref{tab:steps-tuples-ex} presents the same sequence of steps as \autoref{tab:steps-ex}, but uses sets of tuples from \autoref{tab:val-ex} rather than valuation sets.
  Note that the valuation sets in \autoref{tab:steps-ex} and the sets of tuples in \autoref{tab:steps-tuples-ex} form a one-to-one relationship, so the output of the first step in \autoref{tab:steps-ex} (i.e., $\{(T, T), (T, F)\}$) represents the output of the first step in \autoref{tab:steps-tuples-ex} (i.e., $\{\text{Alice}, \text{Bob}, \text{Eve}\}$), and the set of all valuations which satisfy the predicate expression (i.e., $\{(T, T), (T, F), (F, T)\}$) represents the set of all tuples which satisfy the predicate expression (i.e., $\{\text{Alice}, \text{Bob}, \text{Dave}, \text{Eve}\}$).

  \begin{table}[H]
    \begin{tabular}{|c|c|c|M{.5\linewidth}|}
      \hline
      $i$ & Step (\acti{i}, \vsetsi{i}) & Output (in tuples) & Universe (in tuples) \\ \hline
      1 & (\predi{1}, $\{T,F\}^n$) & $\{\text{Alice}, \text{Bob}, \text{Eve}\}$ & $\{\{\text{Alice}, \text{Bob}, \text{Charlie}, \text{Dave}, \text{Eve}\}, \; \{\text{Alice}, \text{Bob}, \text{Eve}\}\}$ \\ \hline
      2 & $\left(\setminus, [\{T,F\}^n, \stepouti{1}]\right)$ & $\{\text{Charlie}, \text{Dave}\}$ & $\{\{\text{Alice}, \text{Bob}, \text{Charlie}, \text{Dave}, \text{Eve}\}, \; \{\text{Alice}, \text{Bob}, \text{Eve}\}, \allowbreak \{\text{Charlie}, \text{Dave}\}\}$ \\ \hline
      3 & $\left(\predi{2}, \stepouti{2}\right)$ & $\{\text{Dave}\}$ & $\{\{\text{Alice}, \text{Bob}, \text{Charlie}, \text{Dave}, \text{Eve}\}, \; \{\text{Alice}, \text{Bob}, \text{Eve}\}, \allowbreak \{\text{Charlie}, \text{Dave}\}, \; \{\text{Dave}\} \}$ \\ \hline
      4 & $\left(\cup, [\stepouti{1}, \stepouti{3}]\right)$ & $\{\text{Alice}, \text{Bob}, \text{Dave}, \text{Eve}\}$ & $\{\{\text{Alice}, \text{Bob}, \text{Charlie}, \text{Dave}, \text{Eve}\}, \; \{\text{Alice}, \text{Bob}, \text{Eve}\}, \allowbreak \{\text{Charlie}, \text{Dave}\}, \; \{\text{Dave}\}, \; \{\text{Alice}, \text{Bob}, \text{Dave}, \text{Eve}\} \}$ \\ \hline
    \end{tabular}
    \caption{Example sequence of steps for the predicate expression (\predi{1} OR \predi{2}) using tuples from \autoref{tab:val-ex}.}
    \label{tab:steps-tuples-ex}
  \end{table}
}



\subsubsection{Cost Model}
Given the definitions above, we can now formally state the cost model properties from Section~\ref{sec:problem:cost}.
\begin{restatable}{definition}{defcost}
  The cost of a step $(\act, \vsets)$ is denoted as $C_{\act}(\vsets)$, and the cost of a sequence of steps $[(\acti{1}, \vsets{1}), \allowbreak (\acti{2}, \vsets{2}), ... (\acti{m}, \vsets{m})]$ is the sum of the costs of its individual steps: $\sum_{i=1}^{m} C_{\acti{i}}(\vsetsi{i})$.
\end{restatable}
\begin{restatable}{property}{propcosta}
  The cost of any \patom[] application step $(\pred, \vset)$ is significantly greater than the cost of any set operation step $(O, \vsets')$:
  \[
    C_O(\vsets') / C_P(\vset) \approx 0
  \]
  \label{prop:cost:1}
\end{restatable}
\begin{restatable}{property}{propcostb}
  For any \patom[] $P$, if step $(\pred, \vset)$ is more costly than step $(\pred, \vset')$, then $\vset'$ cannot be a superset of $\vset$.
  \[
    C_{\pred}(\vset) > C_{\pred}(\vset') \implies \vset' \not\supseteq \vset
  \]
  Similarly, if $\vset'$ is a strict superset of \vset{}, then step $(\pred, \vset')$ is more costly than step $(\pred, \vset)$:
  \[
    \vset' \supset \vset \implies C_{\pred}(\vset') > C_{\pred}(\vset)
  \]
  \label{prop:cost:2}
\end{restatable}
\begin{restatable}{property}{propcostc}
  For any \patom[] $P$ and valuation sets $D$ and $E$, applying $P$ once to $D \cup E$ is less costly than applying $P$ to $D$ and $E$ individually:
  \[
    C_{\pred}(D \cup E) < C_{\pred}(D) + C_P(E)
  \]
  \label{prop:cost:3}
\end{restatable}
\iftoggle{paper}{\vspace{-15pt}}{}
\noindent
Note that \autoref{prop:cost:2} has a more relaxed condition than the one stated in Section~\ref{sec:problem:cost}.
With these properties, we can formally state:
\begin{problem}
  Given a predicate expression \boolformorig{}, find the lowest cost, valid sequence of steps such that universe after the sequence contains the set of all valuations which satisfy \boolformorig{}.
  \label{prob:1}
\end{problem}

\subsection{Problem Reduction}
\label{sec:optimal:reduction}

We now leverage the cost model properties to reduce the solution space.
The reduction allows us to think about the problem using only \patom[] applications and constrains the total number of steps.

\subsubsection{\texorpdfstring{\patom[C]-only Sequences}{Predicate Atom-only Sequences}}
According to \autoref{prop:cost:1}, set operations are practically free compared to \patom[] applications.
Rather than worrying about what set operation steps can be added at which points to create a lower cost sequence, let us instead consider that after each \patom[] application step, all possible set operation steps are performed before the next \patom[] application step.
This way, we only need to consider the order of \patom[] application steps in finding our optimal solution.
Once the optimal sequence of \patom[] application steps has been found, we can reconstruct the intermediate set operation steps with bookkeeping.
Thus, instead of considering sequences of steps, in which each step can be either a set operation or a \patom[] application, let us now consider sequences of only \patom[] application steps.
We provide the following definitions to formalize this idea.
\begin{restatable}{definition}{defsetform}
  A \emph{set formula} is any combination of sets (as operands) using any number of set operations and \patom[s] (as operators).
  The \emph{result} of a set formula is its value after evaluation.
\end{restatable}
\begin{restatable}{definition}{defsetformspace}
  Let \setopset{} be a set of operators, and \vsets{} be a set of sets.
  The \emph{set formula space} $\setformspace_{\setopset{}} (\vsets)$ is the set of all possible formula results that can be constructed using only the operators in \setopset{} as the operators and the sets in \vsets{} as the operands.
\end{restatable}
\noindent
For example, $\pred(\vset)$ and $\vseti{1} \cap (\vseti{2} \setminus \vseti{3})$ are both examples of set formulas, and the set formula space $\setformspace_{\{\cap,\cup\}} (\{\vseti{1}, \vseti{2}\})$ is $\{\vseti{1}, \vseti{2}, \vseti{1} \cap \vseti{2}, \vseti{1} \cup \vseti{2}\}$.
Although an infinite number of set formulas can be constructed from a set of operators and operands, the number of unique results, and thus the set formula space itself, is finite.
\begin{restatable}{definition}{defderive}
  Set $\vset$ is said to be \emph{derived} from set of sets \vsets{} if there exists some set formula which only uses the elements of \vsets{} as operands and results in \vset{}.
  More explicitly, set \vset{} can be derived from the \emph{origin set} \vsets{} using the operators in \setopset{} if $\vset \in \setformspace_{\setopset}(\vsets)$.
\end{restatable}
\noindent
For example, both $\{(T,T)\}$ and $\{(T,F)\}$ can be derived from $\{\{(T,T),\allowbreak (T,F)\}, \{(T,T),(F,T)\}\}$ with the operators $\{\cap, \setminus\}$, but $\{(F,F)\}$ can never be derived from this set of sets using the same operators.
With these definitions, we can extend the idea of a universe to contain all valuation sets which can be derived from the first $i$ \patom[] application-only steps using any number of set operations.
\begin{restatable}{definition}{defextuniv}
  Given the sequence of \patom[] application steps $[(\predi{1}, \vseti{1}), (\predi{2}, \vseti{2}),..., (\predi{m}, \vseti{m})]$, the \emph{extended universe} $\univi{i}'$ is defined by the recurrence relation:
  \[
    \univi{i}' = \setformspace_{\{\cap,\cup,\setminus\}}(\univi{i-1}' \cup \{\predi{i}(\vseti{i})\})
  \]
  with the initial extended universe: $\univi{0}' = \{\{T, F\}^n\}$.
\end{restatable}
\noindent
Although the initial extended universe once again contains only the set of all valuations, the $(i+1)$th step of a \patom[] application-only sequence may now select any valuation set(s) from the extended universe $\univi{i}'$ as its input(s), rather than the regular universe \univi{i}.
For each solution to Problem~\ref{prob:1}, there must exist a corresponding solution in this setup\footnote{%
  Note the extended universe is an abstract concept used only in our analysis and not actually realized in memory by any of our algorithms.
}, so our problem can be restated with:
\begin{restatable}{definition}{defsolseq}
  Given a predicate expression, a \emph{solution sequence} is a valid sequence of \patom[] application steps, after which the extended universe contains the set of all valuations which satisfy the predicate expression.
\end{restatable}
\begin{problem}
  Given a predicate expression, find the lowest cost solution sequence.
  \label{prob:2}
\end{problem}
\noindent

\subsubsection{\texorpdfstring{Length of \patom[C]-only Sequences}{Length of Predicate Atom-only Sequences}}
It turns out that the lowest cost sequence of \patom[] application-only steps must apply each unique \patom[] exactly once.
The proof for this is encapsulated in two theorems.
First, \autoref{thm:need-pred} states each \patom[] must appear as part of a \patom[] application step at least once in the sequence to ensure correctness\iftoggle{paper}{ (proven in our technical report~\cite{techreport})}.
Next, \autoref{thm:one-pred} states that in the lowest cost solution sequence, each \patom[] appears as part of a \patom[] application step exactly once.
\begin{restatable}{theorem}{thmneedpred}
  Given a predicate expression and a solution sequence \stepseq{}, for each \patom[] \pred{} in the predicate expression, there must be a step \step{} in \stepseq{} such that $\step = (\pred, \vset)$ for some valuation set \vset{}.
  \label{thm:need-pred}
\end{restatable}
\iftoggle{paper}{}{%
\begin{proof}
  We first introduce some notation.
  For any valuation $v$, let $v_i$ refer to the $i$th value of $v$ (i.e., the result of \patom[] of \predi{i}).
  In addition, given valuation $v$, let the valuation \vipos{} be the same valuation as $v$ except the $i$th value is true.
  Similarly, \vineg{} is the same valuation as $v$ except that the $i$th value is false.
  Note that one of \vipos{} and \vineg{} must be equivalent to $v$.
  Multiple values may be assigned as well; the valuation $v|_{i=T,j=F}$ is the same as $v$ except the $i$th value is true and the $j$th value is false.

  Next, we introduce Lemmas~\ref{lem:crit-i} and~\ref{lem:nosplit-pi}.
  Lemma~\ref{lem:crit-i} states that for each \patom[] \predi{i}, there must exist some valuation for which only the result of \predi{i} (i.e., the $i$th element of the valuation) can determine whether the valuation satisfies the overall given predicate expression.
  In addition, Lemma~\ref{lem:nosplit-pi} states that before applying \patom[] \predi{i}, there is no way to split a set of valuations based on the $i$th value of a valuation.
  \begin{restatable}{lemma}{lemcriti}
    Let \predi{i} refer to the $i$th \patom[] of the given predicate expression.
    There exists a valuation $v$ such that $\vipos$ satisfies the overall predicate expression, but \vineg{} does not.
    \label{lem:crit-i}
  \end{restatable}

  \begin{restatable}{lemma}{lemnosplitpi}
    Given a predicate expression with \patom[s] $\{\predi{1},...,\predi{n}\}$, some valuation $v \in \{T,F\}^n$, and some set of valuation sets \vsets{},
    if every valuation set $\vset \in \vsets$ contains either both or neither \vipos{} and \vineg{}, then every valuation set which can be derived from \vsets{} without using \predi{i} as an operator must also contain either both or neither \vipos{} and \vineg{}. In other words:
    \begin{align*}
    &\forall \vset \in \vsets, (\vipos \in \vset \land \vineg \in \vset) \lor (\vineg \not\in \vset \land \vineg \not\in \vset)\\
    & \mspace{-10mu} \implies \forall \setopset \subseteq \{\cup,\cap,\setminus,\predi{1},...,\predi{i-1},\predi{i+1},...,\predi{n}\}, \forall \vset' \in \setformspace_{\setopset}(\vsets),\\
    & \quad \quad \; (\vipos \in \vset' \land \vineg \in \vset') \lor (\vipos \not\in \vset' \land \vineg \not\in \vset')\\
    \end{align*}
    \label{lem:nosplit-pi}
  \end{restatable}

  Assume to the contrary, sequence $[\stepi{1},...,\stepi{m}]$ does not contain any step with $\predi{i}$.
  Based on Lemma~\ref{lem:crit-i}, let $v$ be the valuation such that \vipos{} satisfies the overall predicate expression and \vineg{} does not.
  For $[\stepi{1},...,\stepi{m}]$ to be a solution sequence, there must exist a valuation set
  in the final extended universe $\univi{m}'$ which contains \vipos{} but does not contain
  \vineg{}.
  However, based on our definition of the extended universe, $\univi{m}' \subseteq \setformspace_{\{\cap,\cup,\setminus,\predi{1},...,\predi{i-1},\predi{i+1},...,\predi{n}\}}(\{\{T,F\}^n\})$.
  Since the origin set only contains the set of all valuations, and this obviously contains both \vipos{} and \vineg{}, the precondition of Lemma~\ref{lem:nosplit-pi} is met.
  Thus, every valuation set in the extended universe $\univi{m}'$ must contain either both \vipos{} and \vineg{} or neither \vipos{} nor \vineg{}.
  This is a contradiction, so the sequence $[\stepi{1},...,\stepi{n}]$ must contain \patom[] \predi{i} as an operator.
\end{proof}
}

\begin{restatable}{theorem}{thmonepred}
  Given a predicate expression with $n$ unique \patom[s], the lowest cost solution sequence must have exactly $n$ steps.
  \label{thm:one-pred}
\end{restatable}

\begin{proof}
  Assume to the contrary that the lowest cost solution sequence is $n+k$ steps long for some $k \ge 1$ (note the sequence must be at least $n$ thanks to \autoref{thm:need-pred}).
  By the pigeonhole principle, at least one \patom[] must appear at least twice in the sequence.
  Let \pred{} be the first \patom[] to do so, and let indices $i$ and $j$, for some $i < j$, be the first two steps that \pred{} appears in: $\stepi{i} = (\pred, \vseti{i})$ and $\stepi{j} = (\pred, \vseti{j})$.
  We show that we can always construct a new, less costly sequence which does not include both \stepi{i} and \stepi{j} as steps, leading to a contradiction.
  There are two major cases:

  \noindent $\underline{\vseti{i} \cap \vseti{j} \ne \varnothing:}$
  We can replace \stepi{j} with $\stepi{j}' = (\pred, \vseti{j} \setminus \vseti{i})$ and still construct every valuation set derived from $\univi{j-1} \cup \{\pred(\vseti{j})\}$, since we can derive $\pred(\vseti{j}) = \pred(\vseti{j} \setminus \vseti{i}) \cup (\vseti{j} \cap \pred(\vseti{i}))$.
  Based on \autoref{prop:cost:2}, $\stepi{j}'$ is cheaper than \stepi{j}, so replacing \stepi{j} with $\stepi{j}'$ gives us a less costly sequence while maintaining equivalence, leading to a contradiction.

  \noindent $\underline{\vseti{i} \cap \vseti{j} = \varnothing:}$
    In this case, instead of applying
    \pred{} separately, we can combine steps \stepi{i} and \stepi{j} and replace
    them with a single step $\step' = (\pred, \vseti{i} \cup \vseti{j})$.
    With step $\step'$, we can directly calculate both
    $\pred(\vseti{i}) = \pred(\vseti{i} \cup \vseti{j}) \cap \vseti{i}$ and
    $\pred(\vseti{j}) = \pred(\vseti{i} \cup
    \vseti{j}) \cap \vseti{j}$.
    \autoref{prop:cost:3} states that performing a single $\step'$ is less costly than performing both \stepi{i} and \stepi{j}, so by replacing \stepi{i} with $\step'$ and removing
    \stepi{j}, we have found a less costly sequence
    while maintaining
    equivalence, leading to a
  contradiction%
  \footnote{\scriptsize%
    There is a caveat when $\vseti{i} \cap \vseti{j} = \varnothing$ but \vseti{j} is derived from $\univi{i-1} \cup \{\pred(\vseti{i})\}$.
    Calculating $\vseti{i} \cup \vseti{j}$ requires $\vseti{j}$, which in turn requires $P(\vseti{i})$ which is not available before the $i$th step.
    Fortunately, we show that this situation cannot arise in \iftoggle{paper}{out technical report~\cite{techreport}}{Appendix~\ref{sec:caveat}}.
  }.
\end{proof}

\subsection{\bestd{} and \update{}}
\label{sec:optimal:opteval}
\autoref{thm:one-pred} states that for a predicate expression with $n$ \patom[s], the lowest cost solution sequence has exactly $n$ steps, so a predicate evaluation system should be looking for sequences of the form $[(\predi{1}, \vseti{1}),...,(\predi{n},\vseti{n})]$.
This search can be be split into two:
\begin{enumerate}[(1)]
  \item \patom[c] ordering: Searching for the best ordering of \patom[s] $[\predi{1},...,\predi{n}]$.
  \item Set management: Given an ordering of \patom[s], searching for the operands $[\vseti{1},...,\vseti{n}]$ which lead to the lowest overall cost, while still ensuring the final output set can be constructed.
\end{enumerate}
As mentioned, we can reuse past work for \patom[] ordering.
For set management, we present \bestd{}/\update{}.

\subsubsection{Algorithms}
To understand \bestd{} and \update{}, we first provide some definitions.
In the following definitions,
the expression $\node.\patom[a]$ denotes the \patom[] associated with leaf \node{}.
\iftoggle{paper}{%
\begin{restatable}{definition}{defcmp}
  Given a predicate expression and some ordering of its \patom[s] \patom[s] $[\predi{1},...,\predi{n}]$, node $\node$ of the associated predicate tree is \emph{\cmp[]} on the $i$th step if:
  \begin{equation*}
    \text{\cmp}(\node, i) = \begin{cases}
      \node.\patom[a] = \predi{j} \text{ and } j < i & \text{if leaf node} \\
      \bigwedge_{c \in \text{children}(\node)} \text{\cmp}(c, i) & \text{otherwise} \\
    \end{cases}
  \end{equation*}
  Node $\node$ is \emph{positively determinable} on the $i$th step if:
  \begin{equation*}
    \text{\detpos[]}(\node, i) = \begin{cases}
      \node.\patom[a] = \predi{j} \text{ and } j < i & \text{if leaf node} \\
      \bigwedge_{c \in \text{children}(\node)} \text{\detpos[]}(c, i) & \text{if AND node} \\
    \bigvee_{c \in \text{children}(\node)} \text{\detpos}(c, i) & \text{if OR node} \\
    \end{cases}
  \end{equation*}
  Node $\node$ is \emph{negatively determinable} on the $i$th step if:
  \begin{equation*}
    \text{\detneg}(\node, i) = \begin{cases}
      \node.\patom[a] = \predi{j} \text{ and } j < i & \text{if leaf node} \\
      \bigvee_{c \in \text{children}(\node)} \text{\detneg[]}(c, i) & \text{if AND node} \\
      \bigwedge_{c \in \text{children}(\node)} \text{\detneg[]}(c, i) & \text{if OR node} \\
    \end{cases}
  \end{equation*}
  \label{def:cmp}
\end{restatable}
}{%
\begin{restatable}{definition}{defcmp}
  Given a predicate expression and some ordering of its \patom[s] \patom[s] $[\predi{1},...,\predi{n}]$, node $\node$ of the associated predicate tree is \emph{\cmp[]} on the $i$th step, if:
  \begin{equation*}
    \text{\cmp}(\node, i) = \begin{cases}
      \node.\patom[a] = \predi{j} \text{ and } j < i & \text{if leaf node} \\
      \bigwedge_{c \in \text{children}(\node)} \text{\cmp}(c, i) & \text{otherwise} \\
    \end{cases}
  \end{equation*}
  \label{def:cmp}
\end{restatable}
\begin{restatable}{definition}{defdetpos}
  Given a predicate expression and some ordering of its \patom[s] \patom[s] $[\predi{1},...,\predi{n}]$, node $\node$ of the associated predicate tree is \emph{positively determinable} on the $i$th step, if:
  \begin{equation*}
    \text{\detpos[]}(\node, i) = \begin{cases}
      \node.\patom[a] = \predi{j} \text{ and } j < i & \text{if leaf node} \\
      \bigwedge_{c \in \text{children}(\node)} \text{\detpos[]}(c, i) & \text{if AND node} \\
    \bigvee_{c \in \text{children}(\node)} \text{\detpos}(c, i) & \text{if OR node} \\
    \end{cases}
  \end{equation*}
  \label{def:detpos}
\end{restatable}
\begin{restatable}{definition}{defdetneg}
  Given a predicate expression and some ordering of its \patom[s] \patom[s] $[\predi{1},...,\predi{n}]$, node $\node$ of the associated predicate tree is \emph{negatively determinable} on the $i$th step, if:
  \begin{equation*}
    \text{\detneg}(\node, i) = \begin{cases}
      \node.\patom[a] = \predi{j} \text{ and } j < i & \text{if leaf node} \\
      \bigvee_{c \in \text{children}(\node)} \text{\detneg[]}(c, i) & \text{if AND node} \\
      \bigwedge_{c \in \text{children}(\node)} \text{\detneg[]}(c, i) & \text{if OR node} \\
    \end{cases}
  \end{equation*}
  \label{def:detneg}
\end{restatable}
}

%

\noindent
If $\node$ is complete on step $i$, then every \patom[] leaf node descendant of $\node$ has already been applied by step $i$, and the extended universe $\univi{i-1}'$ contains the set of all valuations which satisfy the predicate subexpression represented by $\node$.
Positively (negatively) determinable nodes, on the other hand, can determine a \emph{subset} of valuations which satisfy (do not satisfy) the predicate subexpression represented by $\node$.

\begin{algorithm}
  \begin{algorithmic}[1]
    \Require Predicate tree node \node{}, algorithm state \mystate{}, \patom[] \predi{i}, step index $i$
    \Ensure Optimal operand \vseti{i} to apply \predi{i} to
    \If{$\node  = \nil$}
    \State \Return $\{T,F\}^n$
    \ElsIf{\text{isAndNode}(\node)}
    \State $X \gets \bestd(\text{parent}(\node), \mystate, i)$\label{line:bestd:and-bestd}
    \For{$\child \in \text{children}(\node)$}
    \If{$\text{\cmp[]}(\child, i)$}
    \State $X \gets X \cap \mystate[\child].\cmp[v]$ \label{line:bestd:and-cmp}
    \ElsIf{$\detneg[](\child, i)$ and $\neg$isAnc(\child, \predi{i})}
    \State $X \gets X \setminus \mystate[\child].\detneg[v]$ \label{line:bestd:and-detneg}
    \EndIf
    \EndFor
    \State \Return $X$
    \Else
    \State $X \gets \bestd(\text{parent}(\node), \mystate, i)$\label{line:bestd:or-bestd}
    \State $Y \gets \{\}$
    \For{$\child \in \text{children}(\node)$}
    \If{$\text{\cmp[]}(\child, i)$}
    \State $Y \gets Y \cup \mystate[\child].\cmp[v]$ \label{line:bestd:or-cmp}
  \ElsIf{$\detpos[](\child, i)$ and $\neg$isAnc(\child, \predi{i})}
    \State $Y \gets Y \cup \mystate[\child].\detpos[v]$ \label{line:bestd:or-detpos}
    \EndIf
    \EndFor
    \State \Return $X \setminus Y$
    \EndIf
  \end{algorithmic}
  \caption{\bestd{}}
  \label{alg:bestd}
\end{algorithm}

\bestd{} and \update{} are presented in Algorithm~\ref{alg:bestd} and~\ref{alg:update} (next page) respectively.
We also provide pseudocode of how to use these algorithms with some ordering algorithm \bestp{} in Algorithm~\ref{alg:opteval} (next page).
As shown in Algorithm~\ref{alg:opteval}, \bestd{} determines the optimal operands $[\vseti{1},...,\vseti{n}]$, while \update{} applies \patom[s] to these operands and uses the \mystate{} variable to keep track of valuation sets for complete and positively/negatively determinable nodes.
The valuation sets in \mystate{} are used by \bestd{} to reduce the size of generated operands.

\iftoggle{paper}{}{%
  \begin{algorithm}
  \begin{algorithmic}[1]
    \Require Predicate tree node \node{}, algorithm state \mystate{}, \patom[] \predi{i}, set of valuations \vseti{i}, step index $i$
    \If{$\node = \nil$}
    \State \Return
    \ElsIf{$\text{isLeafNode}(\node)$}
    \State $\mystate[\node].\cmp[v] \gets \predi{i}(\vseti{i})$\label{line:update:apply}
    \State $\mystate[\node].\detpos[v] \gets \predi{i}(\vseti{i})$
    \State $\mystate[\node].\detneg[v] \gets \vseti{i} \setminus \predi{i}(\vseti{i})$
    \ElsIf{\text{isAndNode}(\node)}
    \State $X \gets \bestd(\text{parent}(\node), \mystate, i)$\label{line:update:and-bestd}
    \If{$\text{\cmp[]}(\node, i+1)$}
    \State $\mystate[\node].\cmp[v] \gets \bigcap_{\child} \mystate[\child].\cmp[v] \cap X$ \label{line:update:and-cmp}
    \EndIf
    \If{$\detpos[](\node,i+1)$}
    \State $\mystate[\node].\detpos[v] \gets \bigcap_{\child} \mystate[\child].\detpos[v] \cap X$\label{line:update:and-detpos}
    \EndIf
    \If{$\detneg[](\node,i+1)$}
    \State $\mystate[\node].\detneg[v] \gets \bigcup_{\child} \mystate[\child].\detneg[v] \cap X$\label{line:update:and-detneg}
    \EndIf
    \Else
    \State $X \gets \bestd(\text{parent}(\node), \mystate, i)$\label{line:update:or-bestd}
    \If{$\text{\cmp[]}(\node, i+1)$}
    \State $\mystate[\node].\cmp[v] \gets \bigcup_{\child} \mystate[\child].\cmp[v] \cap X$ \label{line:update:or-cmp}
    \EndIf
    \If{$\detpos[](\node,i+1)$}
    \State $\mystate[\node].\detpos[v] \gets \bigcup_{\child} \mystate[\child].\detpos[v] \cap X$\label{line:update:or-detpos}
    \EndIf
    \If{$\detneg[](\node,i+1)$}
    \State $\mystate[\node].\detneg[v] \gets \bigcap_{\child} \mystate[\child].\detneg[v] \cap X$\label{line:update:or-detneg}
    \EndIf
    \EndIf
    \State $\update(\text{parent}(\node), \mystate, \predi{i}, \vseti{i}, i)$
  \end{algorithmic}
  \caption{\update{}}
  \label{alg:update}
\end{algorithm}

\begin{algorithm}[!ht]
  \begin{algorithmic}[1]
    \Require Predicate expression \boolformorig{}
    \Ensure The set of all valuations which satisfy \boolformorig{}
    \State $[\predi{1},...,\predi{n}] \gets \bestp(\boolformorig)$
    \State \rootnode{} $\gets \text{makeTree}(\boolformorig, [\predi{1},...,\predi{n}])$
    \State $\mapping \gets \text{createMapping}(\rootnode, [\predi{1},...,\predi{n}])$
    \State $\mystate \gets \text{initState}(\rootnode)$
    \For{$i \gets 1,...,n$}
    \State $\vseti{i} \gets \bestd(\text{parent}(\mapping[\predi{i}]), \mystate, \predi{i}, i)$ \label{line:opteval:1}
    \State $\update(\mapping[\predi{i}], \mystate, \predi{i}, \vseti{i}, i)$
    \EndFor
    \State \Return \mystate[\rootnode].\cmp[v]
  \end{algorithmic}
  \caption{\opteval{}}
  \label{alg:opteval}
\end{algorithm}
}

In Algorithm~\ref{alg:opteval}, \mapping{} is a mapping from \patom[] \predi{i} to the leaf node containing that \patom[].
Variable \mystate{} is an object which has three attributes: \cmp[v], \detpos[v], and \detneg[v].
Each of these attributes is a mapping from a predicate tree node to the valuation sets that can be determined by completion, positive determinability, and negative determinability respectively.
The ``initState'' function initializes each of these attributes to the empty set for every node in the predicate tree.
The expression $\neg$isAnc(\node, \predi{i}) is true if \predi{i} is not a descendant of \node{}.
Finally, the ``parent'' function returns \nil{} if called on the root node.

\bestd{} determines the optimal operand for a \patom[] \predi{i} by traversing the predicate tree in a bottom-up manner.
In the beginning, no nodes are complete nor determinable, so \bestd{} has no choice but to return the set of all possible valuations $\{T,F\}^n$.
As \mystate{} gets updated, the operand is trimmed according to the completeness and determinability of sibling nodes.
The intersection is taken for complete AND children, and additional sets are added to the set subtraction for complete OR children.
Additionally, for AND (OR) children which are negatively (positively) determinable, the sets in \detneg[v] (\detpos[v]) are removed from the returned operand.
\update{} applies \patom[] \predi{i} to the generated operand \vseti{i}.
Then, it traverses up the ancestors of the applied \patom[] leaf node and updates the \cmp[v], \detpos[v], and \detneg[v] values for complete, positively determinable, and negatively determinable nodes respectively.

\iftoggle{paper}{%
\begin{algorithm}
  \begin{algorithmic}[1]
    \Require Predicate tree node \node{}, algorithm state \mystate{}, \patom[] \predi{i}, set of valuations \vseti{i}, step index $i$
    \If{$\node = \nil$}
    \State \Return
    \ElsIf{$\text{isLeafNode}(\node)$}
    \State $\mystate[\node].\cmp[v] \gets \predi{i}(\vseti{i})$\label{line:update:apply}
    \State $\mystate[\node].\detpos[v] \gets \predi{i}(\vseti{i})$
    \State $\mystate[\node].\detneg[v] \gets \vseti{i} \setminus \predi{i}(\vseti{i})$
    \ElsIf{\text{isAndNode}(\node)}
    \State $X \gets \bestd(\text{parent}(\node), \mystate, i)$\label{line:update:and-bestd}
    \If{$\text{\cmp[]}(\node, i+1)$}
    \State $\mystate[\node].\cmp[v] \gets \bigcap_{\child} \mystate[\child].\cmp[v] \cap X$ \label{line:update:and-cmp}
    \EndIf
    \If{$\detpos[](\node,i+1)$}
    \State $\mystate[\node].\detpos[v] \gets \bigcap_{\child} \mystate[\child].\detpos[v] \cap X$\label{line:update:and-detpos}
    \EndIf
    \If{$\detneg[](\node,i+1)$}
    \State $\mystate[\node].\detneg[v] \gets \bigcup_{\child} \mystate[\child].\detneg[v] \cap X$\label{line:update:and-detneg}
    \EndIf
    \Else
    \State $X \gets \bestd(\text{parent}(\node), \mystate, i)$\label{line:update:or-bestd}
    \If{$\text{\cmp[]}(\node, i+1)$}
    \State $\mystate[\node].\cmp[v] \gets \bigcup_{\child} \mystate[\child].\cmp[v] \cap X$ \label{line:update:or-cmp}
    \EndIf
    \If{$\detpos[](\node,i+1)$}
    \State $\mystate[\node].\detpos[v] \gets \bigcup_{\child} \mystate[\child].\detpos[v] \cap X$\label{line:update:or-detpos}
    \EndIf
    \If{$\detneg[](\node,i+1)$}
    \State $\mystate[\node].\detneg[v] \gets \bigcap_{\child} \mystate[\child].\detneg[v] \cap X$\label{line:update:or-detneg}
    \EndIf
    \EndIf
    \State $\update(\text{parent}(\node), \mystate, \predi{i}, \vseti{i}, i)$
  \end{algorithmic}
  \caption{\update{}}
  \label{alg:update}
\end{algorithm}

\begin{algorithm}
  \begin{algorithmic}[1]
    \Require Predicate expression \boolformorig{}
    \Ensure The set of all valuations which satisfy \boolformorig{}
    \State $[\predi{1},...,\predi{n}] \gets \bestp(\boolformorig)$
    \State \rootnode{} $\gets \text{makeTree}(\boolformorig, [\predi{1},...,\predi{n}])$
    \State $\mapping \gets \text{createMapping}(\rootnode, [\predi{1},...,\predi{n}])$
    \State $\mystate \gets \text{initState}(\rootnode)$
    \For{$i \gets 1,...,n$}
    \State $\vseti{i} \gets \bestd(\text{parent}(\mapping[\predi{i}]), \mystate, \predi{i}, i)$ \label{line:opteval:1}
    \State $\update(\mapping[\predi{i}], \mystate, \predi{i}, \vseti{i}, i)$
    \EndFor
    \State \Return \mystate[\rootnode].\cmp[v]
  \end{algorithmic}
  \caption{\opteval{}}
  \label{alg:opteval}
\end{algorithm}
}{}

\subsubsection{Optimality}
We now prove the optimality of \bestd{}/\update{}.
\begin{restatable}{theorem}{thmoptimal}
  For a given predicate expression and some ordering of its \patom[s] $[\predi{1},...,\predi{n}]$, let $[\vseti{1},...,\vseti{n}]$ be the sequence of operands generated by \bestd{}/\update{}.
  The sequence of steps $[(\predi{1}, \vseti{1}),...,\allowbreak (\predi{n}, \vseti{n})]$ must have the lowest cost of any solution sequence.
\end{restatable}
\begin{proof}
Assume to the contrary that there exists a better sequence of operands $[\vseti{1}',...,\vseti{n}']$.
This must mean that there is at least one step $i$ for which $C_{\predi{i}}(\vseti{i}') < C_{\predi{i}}(\vseti{i})$.
Let $i$ be the index of the first such step.
To incur a cheaper cost, based on \autoref{prop:cost:2}, $\vseti{i}'$ must be missing at least one valuation $v$ that is present in \vseti{i}.
However, we show that if we apply $\predi{i}$ to the operand $\vseti{i}'$ that is missing $v$, we are not able to generate the set of all valuations which satisfy the given predicate expression, leading to a contradiction.

\iftoggle{paper}{%
Central to this argument is the idea that before we apply \patom[] \predi{i}, there is no way to split a valuation set based on the $i$th value of a valuation.
Lemma~\ref{lem:nosplit-pi}, shown below, formally states this.
We first introduce some notation.
For any valuation $v$, let $v_i$ refer to the $i$th value of $v$, and
let the valuation \vipos{} be the same valuation as $v$ except the $i$th value is true.
Similarly, \vineg{} is the same valuation as $v$ except that the $i$th value is false.
Multiple values may be assigned as well; the valuation $v|_{i=T,j=F}$ is the same as $v$ except the $i$th value is true and the $j$th value is false.

\begin{restatable}{lemma}{lemnosplitpi}
    Given a predicate expression with \patom[s] $\{\predi{1},...,\predi{n}\}$, some valuation $v \in \{T,F\}^n$, and some set of valuation sets \vsets{},
  if every valuation set $\vset \in \vsets$ contains either both or neither \vipos{} and \vineg{}, then every valuation set which can be derived from \vsets{} without using \predi{i} as an operator must also contain either both or neither \vipos{} and \vineg{}. In other words:
  \begin{align*}
    &\forall \vset \in \vsets, (\vipos \in \vset \land \vineg \in \vset) \lor (\vineg \not\in \vset \land \vineg \not\in \vset)\\
    & \mspace{-10mu} \implies \forall \setopset \subseteq \{\cup,\cap,\setminus,\predi{1},...,\predi{i-1},\predi{i+1},...,\predi{n}\}, \forall \vset' \in \setformspace_{\setopset}(\vsets),\\
    & \quad \quad \; (\vipos \in \vset' \land \vineg \in \vset') \lor (\vipos \not\in \vset' \land \vineg \not\in \vset')\\
  \end{align*}
  \label{lem:nosplit-pi}
\end{restatable}
}{}

\iftoggle{paper}{\vspace{-30pt}}{}
Let us consider valuation set \vseti{i} from the extended universe $\univi{i-1}' \subseteq \setformspace_{\{\cap,\cup,\setminus,\predi{1},...,\predi{i-1}\}}(\{\{T,F\}^n\})$.
In this case, the set of all possible valuations $\{T,F\}^n$ obviously contains all valuations (satisfying the pre-condition),
thus the derived valuation set $\vseti{i} \in \univi{i-1}'$ must also contain either both or neither \vipos{} and \vineg{} for any valuation $v$ according to Lemma~\ref{lem:nosplit-pi}.
If valuation $v \in \vseti{i}$, then both \vipos{} and \vineg{} must be in \vseti{i}.
Furthermore, since \predi{i+1} is also not included as an operator in the construction of $\univi{i-1}'$ and $\vipos{} \in \vseti{i}$, $\{v|_{i=T,(i+1)=T}, v|_{i=T,(i+1)=F}, v|_{i=F,(i+1)=T}, v|_{i=F,(i+1)=F}\}$ must be a subset of \vseti{i} by Lemma~\ref{lem:nosplit-pi}.
%
This same logic can be applied recursively for all indices in $i,...,n$ and in converse for when $v \not\in \vseti{i}$.
\begin{restatable}{definition}{defvalgroup}
  The \emph{ith valuation group} of a valuation $v$ is the set of all valuations for which the first $i-1$ values of the valuation are equal to the first $i-1$ values of $v$.
  In other words, if $\vatom[m](v,i)$ represents the $i$th valuation group of $v$:
\iftoggle{paper}{\vspace{-5pt}}{}
  \[
    \vatom[m](v,i) = \{u \in \{T,F\}^n \mid \bigwedge\nolimits_{j=1}^{i-1} v_j = u_j \}
  \]
\end{restatable}
\iftoggle{paper}{\vspace{-5pt}}{}
\begin{corollary}
  Given a predicate expression, some ordering of its \patom[s] $[\predi{1},...,\predi{n}]$, and some valuation $v \in \{T,F\}^n$, any set derived from $\{\{T,F\}^n\}$ without using $\{\predi{i},...,\predi{n}\}$ as operators must either be a superset of the $i$th valuation group of $v$ or be disjoint from it:
  \begin{align*}
    \forall v \in & \{T,F\}^n, \forall \vset \in \setformspace_{\{\cap,\cup,\setminus,\predi{1},...,\predi{i-1}\}}(\{\{T,F\}^n\}), \\
                  & (\vatom[m](v,i) \subseteq \vset) \lor (\vatom[m](v,i) \cap \vset = \varnothing)
  \end{align*}
  \label{cor:vatom}
\end{corollary}

\iftoggle{paper}{\vspace{-15pt}}{}
In a similar vein, we present the following lemma which remarks that since each \patom[] is necessary for the final result, there exists at least one pair of valuations which must be distinguished by the $i$th value of valuation $v$ using the $i$th \patom[] \predi{i}:
\begin{restatable}{lemma}{lematomstatic}
  Given a predicate expression and some ordering of its \patom[s] $[\predi{1},...,\predi{n}]$,
  let valuation set $\vseti{i}$ be the $i$th operand generated by \bestd{}/\update{}.
  For all valuations $v \in \vseti{i}$, there exists a valuation $u \in \vatomvi$ such that only one of \uipos{} and \uineg{} satisfies the given predicate expression.
  \label{lem:atom-static}
\end{restatable}

As we initially stated, if the step $(\predi{i}, \vseti{i}')$ is cheaper than operand $(\predi{i}, \vseti{i})$,
then $\vseti{i}$ must contain some valuation $v$ that is not in $\vseti{i}'$.
Since \vseti{i} and $\vseti{i}'$ must both come from the extended universe $\univi{i-1}'$,
Corollary~\ref{cor:vatom} applies.
Valuation $v$'s $i$th valuation group must be a subset of \vseti{i}, and at the same time, cannot have any elements in common with $\vseti{i}'$: $(\vatom[m](v,i) \subseteq \vseti{i}) \land (\vatom[m](v,i) \cap \vseti{i}' = \varnothing)$.
By Lemma~\ref{lem:atom-static}, there must exist at least one valuation $u \in \vatom[m](v,i)$ such that only one of \uipos{} and \uineg{} satisfies the predicate expression.
However, since $\vatom[m](v,i)$ is mutually exclusive with $\vseti{i}'$, any derived valuation set from $\univi{i-1}' \cup \{\predi{i}(\vseti{i}')\}$ without \predi{i} as an operator must contain either both or neither \uipos{} and \uineg{} according to Lemma~\ref{lem:nosplit-pi}.
Since each \patom[] can be applied once at most, we have no more opportunities to apply \predi{i}, and every valuation set in the final extended universe will contain either both or neither \uipos{} and \uineg{}.
However, this is a contradiction since the set of all valuations which satisfy the given predicate expression must come from the final extended universe and contains only one of \uipos{} and \uineg{}.
Thus, the sequence $[\vseti{1},...,\vseti{n}]$ generated by \bestd{}/\update{} is optimal.
\end{proof}

\subsection{\texorpdfstring{\patom[C] Ordering}{Predicate Ordering}}
\label{sec:optimal:bestp}

We now turn our attention to the ordering of \patom[s].
We first discuss the reduction from \bestd{}/\update{} to \algo{}, then we explain why \hanani{} is not optimal for predicate expressions of nested depth 3 or greater.

\subsubsection{Reduction from \bestd{}/\update{} to \algo{}}
As mentioned, there already exist works which focus on \patom[] ordering~\cite{hanani_optimal_1977}~\cite{kemper_optimizing_1992}, and we can use any of these works as \bestp{} in Algorithm~\ref{alg:opteval}.
However, in the special case that $\bestp = \hanani{}$, \bestd{}/\update{} can be reduced to \algo{}.
This is because the ordering returned by \hanani{} is guaranteed to traverse the predicate tree in a DFS manner.
A DFS ordering ensures that a node is complete before moving on to its siblings, so no incomplete, positively/negatively determinable children ever exist.
Thus, for DFS orderings, all logic regarding determinability can be removed from \bestd{}/\update{}, and from this point on, the reduction is straightforward (see \iftoggle{paper}{our technical report~\cite{techreport}}{Appendix~\ref{apx:algo-reduction}} for details).

\subsubsection{Depth-3+ Predicate Expressions}
\label{sec:optimal:bestp:depth3}
\hanani{} always returns a DFS ordering.
For predicate expressions of depth 2 or less, this is not a problem since the optimal ordering is guaranteed to be in DFS as well due to the following lemma:
\begin{restatable}{theorem}{thmdfsopt}
  Given a predicate expression of depth 2 or less and an ordering of its \patom[s] $\predseq{} = [\predi{1}, ..., \predi{n}]$, let $[\vseti{1},...,\vseti{n}]$ be the sequence of operands generated by \bestd{}/\update{}.
  If the sequence of steps $[(\predi{1}, \vseti{1}), ..., \allowbreak (\predi{n}, \vseti{n})]$ has the lowest cost of any solution sequence, \predseq{} must be in DFS ordering.
  \label{thm:dfs-opt}
\end{restatable}
\noindent
\iftoggle{paper}{%
\begin{proof}[Proof Idea]
  Assume to the contrary that there exists a non-DFS \patom[] ordering $\predseq'$ which instead leads to the lowest cost solution sequence.
  Since $\predseq'$ has a non-DFS ordering, there must be a node $q$ for which on step $i$:
  \begin{enumerate*}
    \item $q$ is not complete on step $i$
    \item $q$ has some descendant \patom[] \predi{j} such that $j < i$
    \item \patom[] \predi{i} is not a descendant of $q$
  \end{enumerate*}.
  Let $k$ refer to the last step for which the above conditions hold for any node $q'$.
  We argue that we can find a \patom[] ordering which leads to a lower cost solution sequence by taking all the \patom[] descendants of $q'$ and placing them after \predi{k} in the same relative ordering.
  This is a contradiction, so the optimal ordering must be in DFS.
\end{proof}
}{%
  \begin{proof}
    Assume to the contrary that there exists a non-DFS \patom[] ordering $\predseq'$ which instead leads to the lowest cost solution sequence.
    Since $\predseq'$ has a non-DFS ordering, there must be a node $q$ and step index $i$, such that:
    \begin{enumerate}
      \item $q$ is not complete on the $i$th step
      \item $q$ has some descendant \patom[] \predi{j} such that $j < i$
      \item \patom[c] \predi{i} is not a descendant of $q$
    \end{enumerate}
    In other words, one of $q$'s \patom[] descendant (i.e., \predi{j}) comes before \predi{i}, but before completing $q$, \predi{i} is applied.
    Let $i$ be the last step index for which the above conditions hold for any node $q$.
    We argue that we can find a \patom[] ordering which leads to a lower cost solution sequence by taking all of $q$'s \patom[] descendants and placing them after \predi{i} in the same relative ordering, thus causing a contradiction.

    To make this argument, we first introduce \autoref{lem:depth2} (shown below), which states that, for predicate expressions of depth 2 or less, all negatively determinable children of AND nodes must also be complete, and all positively determinable children of OR nodes must also be complete.
    In other words, an incomplete node can be neither a negatively determinable child of an AND node nor a positively determinable child of an OR node.
    Based on Algorithms~\ref{alg:bestd} and~\ref{alg:update}, this means that \patom[] descendants of $q$ cannot affect the operands generated by \bestd{}/\update{} for any non-descendant \patom[] until $q$'s completion.
    Thus, regardless of whether $q$'s \patom[] descendants are all moved to be after \predi{i} or not, the non-descendant \patom[s] in $[\predi{1},...,\predi{i}]$ will generate the same \bestd{}/\update{} operands and have the same cost.
    On the other hand, since $q$ is the last node for which the above conditions hold, \predi{i} must complete a sibling of $q$; let us refer to this newly completed sibling as $q'$ (if \predi{i} does not complete $q'$, then there exists a later step index $k > i$ for which the above conditions hold for $q'$).
    As a completed node, $q'$ must have a non-empty $\cmp[v]$ value, and this \cmp[v] value must help reduce the \bestd{}/\update{}-generated operands of all \patom[s] which come after \predi{i}.
    Thus, moving the \patom[] descendants of $q$ to after \predi{i} reduces their respective operands and lowers the costs of the steps containing these \patom[s].
    Note that moving the \patom[] descendants to after \predi{i} does not change the \cmp[v] value of $q$ upon completion, so the cost of any \patom[s] after $q$'s completion remains the same as well.
    As such, taking the \patom[] descendants of $q$ and placing them after \predi{i} (in the same relative ordering) leads to a lower cost ordering, and a contradiction is reached.
  \end{proof}

  \begin{restatable}{lemma}{lemdepth}
    \noindent
    Given any ordering of \patom[s] for a predicate expression of depth 2 or less, if node \node{} of the associated predicate tree is the child of an AND node and negatively determinable, then \node{} must be complete.
    Similarly, if \node{} is the child of an OR node and positively determinable, then \node{} must be complete.
    \label{lem:depth2}
  \end{restatable}
}

However, for predicate expressions of depth 3 or greater, \autoref{thm:dfs-opt} no longer holds, and a non-DFS \patom[] ordering may lead to a lower cost solution sequence.
To take an example, let us once again consider the predicate expression from \autoref{query:ex}, and for this example let us use a simplified version of \autoref{cost:1}, in which we ignore the constant overhead $\kappa$ and the costs of set operations:
\begin{example}
  Consider the predicate expression be $((\predi{1} \land \predi{2}) \lor \predi{3}) \land \predi{4}$.
  Let the selectivities of \predi{1}, \predi{2}, \predi{3}, and \predi{4} be $\seleci{1} = 0.469$, $\seleci{2} = 0.984$, $\seleci{3} = 0.313$, and $\seleci{4} = 0.820$ respectively.
  Let the constant cost factors of all \patom[s] be 1: $F_1 = F_2 = F_3 = F_4 = 1$.
  \label{ex:depth3}
\end{example}
For this example, \hanani{} returns the ordering $[\predi{1}, \predi{2}, \predi{3}, \predi{4}]$.
When combined with the operands from \bestd{}/\update{}, the cost of the resulting sequence of steps can be calculated with the following (assuming the total number of tuples is $|R|$):
\begin{enumerate}
  \item \predi{1} is applied to all tuples, so it has a cost of $|R|$.
  \item \predi{2} is applied to all tuples for which \predi{1} is true, so the cost is $\seleci{1}|R| = 0.469|R|$.
  \item \predi{3} is then applied to all tuples for which either \predi{1} or \predi{2} is not satisfied. Using the independence assumption, $\seleci{1 \land 2}$ $= \seleci{1}\seleci{2} $ $= 0.461$. Thus, the cost is $(1 - 0.461)|R| = 0.539|R|$.
  \item Finally, \predi{1} is applied to all tuples for which $\predi{B} \lor (\predi{C} \land \predi{D})$ is satisfied.
    Using the independence assumption:
    $\seleci{B \lor (C \land D)}$ $ = \seleci{B} + \seleci{C \land D} - \seleci{B}\seleci{C \land D}$ $= 0.630$, so the cost is $0.630|R|$.
\end{enumerate}
Thus, this sequence has a total cost of $|R| + 0.469|R| + 0.539|R| + 0.630|R| = 2.638|R|$.

Let us now consider the non-DFS ordering $[\predi{3}, \predi{1}, \predi{4}, \predi{2}]$:
\begin{enumerate}
  \item \predi{3} is applied to all tuples, so it has a cost of $|R|$.
  \item \predi{1} is applied to all tuples for which \predi{3} is false, for a cost of $(1 - \seleci{3})|R| = 0.687|R|$.
  \item \predi{4} is applied to all tuples for which either \predi{1} or \predi{3} is true: $\seleci{1 \lor 3} = \seleci{1} + \seleci{3} - \seleci{1}\seleci{3} = 0.635$, so the cost is $0.635|R|$.
  \item Finally, \predi{2} is applied to all tuples for which $\predi{4} \land \neg \predi{3} \land \predi{1}$ is true: $\seleci{4 \land \neg 3 \land 1} = \seleci{4} (1-\seleci{3}) \seleci{1} = 0.264$, so the cost is $0.264|R|$.
\end{enumerate}
This leads to a total cost of $2.586|R|$, confirming the suboptimality of \hanani{} for depth-3 predicate expressions.
Even though \predi{1}'s parent was not complete during the third step, it was negatively determinable, and this determinability helped reduce the input set of tuples for \predi{4}'s application in step 3.
Thus, for predicate expressions of depth 3 or greater, determinability prevents us from dividing the problem into well-contained subproblems, so a DFS ordering is no longer always optimal.

As such, rather than using \algo{} for these predicate expressions, it may be beneficial to use a combination of \bestd{}/\update{} and another ordering algorithm instead.
For example, we could combine \bestd/\update{} with an ordering algorithm based on Boolean Difference Calculus (BDC)~\cite{kemper_optimizing_1992} as mentioned in Section~\ref{sec:rel-work}.
Another example might be to use the following greedy, one-step lookahead algorithm \deeporderp{}.
\deeporderp{} generates the \patom[] ordering one step at a time, and for each step, it selects the \patom[] which it estimates will lead to the lowest cost evaluation of the remaining \patom[s] (the estimate for each \patom[] $P$ is the cost of applying $P$ plus the cost of evaluating the remaining \patom[s] other than $P$ using \algo{}).
Although neither of these algorithms can guarantee optimality, as \hanani{} did for predicate expressions of 2 or less, they serve as alternatives which can sometimes generate better plans for predicate expressions of depth 3 or greater.
In fact, our experiments showed that although \hanani{} outperformed \deeporderp{} for most cases, for 6\% of the queries, \deeporderp{} led to better plans than \hanani{}.

\section{Evaluation}
\label{sec:eval}

\noindent
\textbf{Algorithms.}
We evaluated a total of five algorithms in our experiments: \algo{}, \noopt{} ,\tdacb{}~\cite{kastrati_generating_2018}, \deep{}, and \bdc{}.
\begin{enumerate}
  \item \algo{} (\bestd{}/\update{} + \hanani{}). Our recommended algorithm, with $O(n \log^2 n)$ planning time, and optimal for all predicate expressions of 2 or less.
  \item \noopt{}. The baseline algorithm with no optimizations for disjunctions; essentially the same as \algo{}, except the outputs of disjunctive clauses do not help reduce the inputs of other disjunctive clauses.
    Note this is the actual strategy employed by many real-world systems for disjunctions.
  \item \tdacb. Our main competitor and the current state-of-the-art algorithm for generating optimal predicate evaluation plans.
    It can generate optimal plans for predicate expressions of any depth, but its planning time complexity is exponential.
  \item \deep{} (\bestd{}/\update{} + \deeporderp{}). The combination of \bestd{}/\update{} and the greedy, one-step lookahead algorithm \deeporderp{} described at the end of Section~\ref{sec:optimal:bestp:depth3}.
    Used as a comparison point for depth-3+ predicate expressions.
  \item \bdc{} (\bestd{}/\update{} + BDC). Combination of \bestd{}/\update{} and BDC~\cite{kemper_optimizing_1992}, described in Section~\ref{sec:rel-work}.
    Used as a comparison point for depth-3+ predicate expressions.
\end{enumerate}

\noindent
\textbf{Workload and Metric.}
We evaluated our algorithms on both a synthetic workload using the Forest dataset~\cite{forest} and on Q19 (which has disjunctions) from TPC-H~\cite{TPCHHomepage} and the CH-benchmark~\cite{cole2011mixed} with the following metrics:
\begin{enumerate}[(1)]
  \item Runtime: The total time it takes to generate and execute a predicate evaluation plan. 
  \item Number of Evaluations: The sum total number of times a query's \patom[s] are evaluated (each evaluation of a \patom[] on a tuple counts as 1).
    This provides an implementation-agnostic measure of our algorithms.
\end{enumerate}
Note that for the synthetic workload, with the exception of \tdacb{}, the plan times for all other algorithms accounted for only $0.01\%$ of the total runtime, so we do not report the plan times separately there.
For TPC-H and the CH-benchmark, we present the plan times and execution times separately.

\noindent
\textbf{System.}
All experiments were performed on our column-oriented execution engine \system{}\footnote{A chameleon's eyes can move in complete ``disjunction'' with one another~\cite{ott2001chameleons}!}, which implements all five aforementioned algorithms.
\system{} is coded in $\sim$6500 lines of Rust, stores data on disk, and internally represents sets of tuples with Roaring bitmaps~\cite{chambi2016better}.
The hardware for our experiments was a server with 64 Intel(R) Xeon(R) CPU E7-4830 @ 2.13GHz processors, 256GB of memory, and a RAID5 configuration of 7200rpm HDDs.

\iftoggle{paper}{%
\begin{figure*}
  \centering
  \begin{subfigure}{.32\linewidth}
    \centering
    \includegraphics[clip, trim=10px 0 10px 0,width=\linewidth]{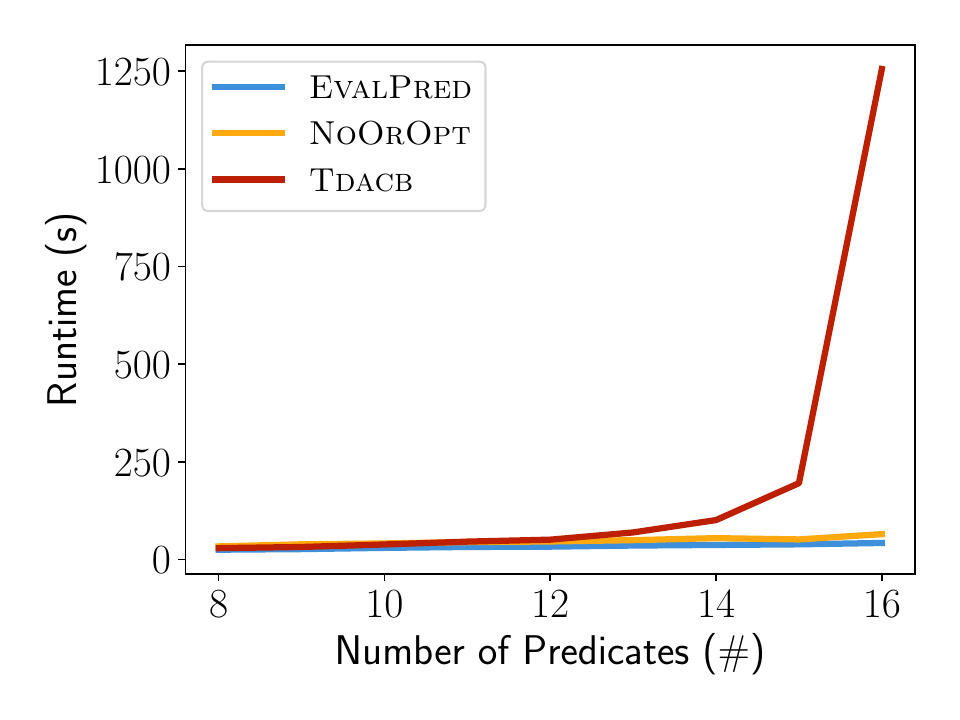}
    \caption{Runtimes}
    \label{fig:depth2-uniform-times-inc-tdacb}
  \end{subfigure}
  \begin{subfigure}{.32\linewidth}
    \centering
    \includegraphics[clip, trim=10px 0 10px 0,width=\linewidth]{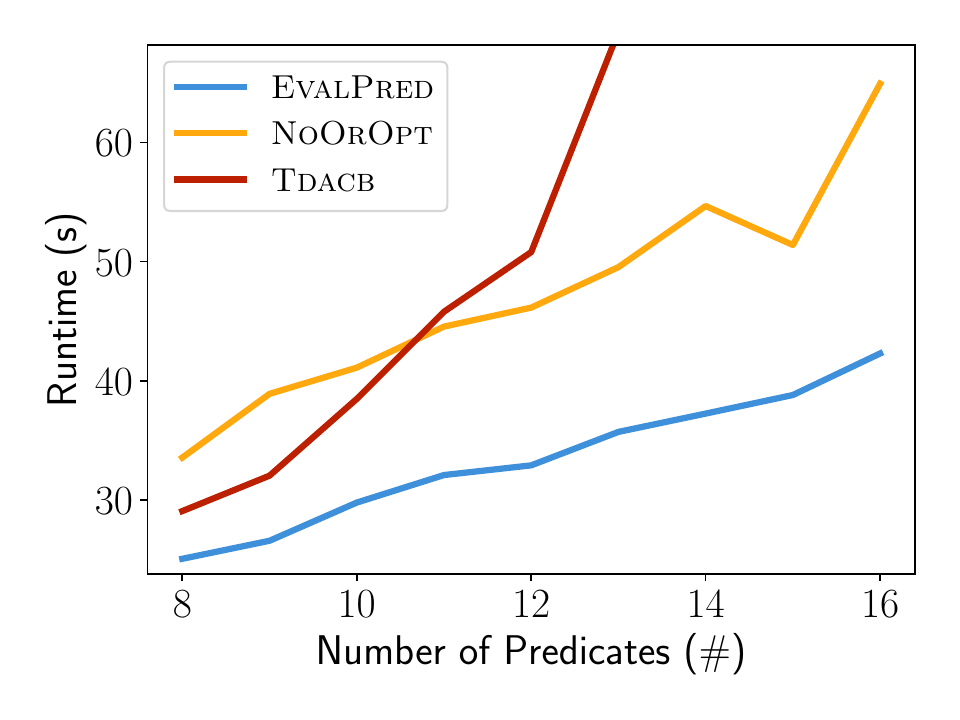}
    \caption{Runtimes (zoomed-in)}
    \label{fig:depth2-uniform-times-no-tdacb}
  \end{subfigure}
  \begin{subfigure}{.32\linewidth}
    \centering
    \includegraphics[clip, trim=10px 0 10px 0,width=\linewidth]{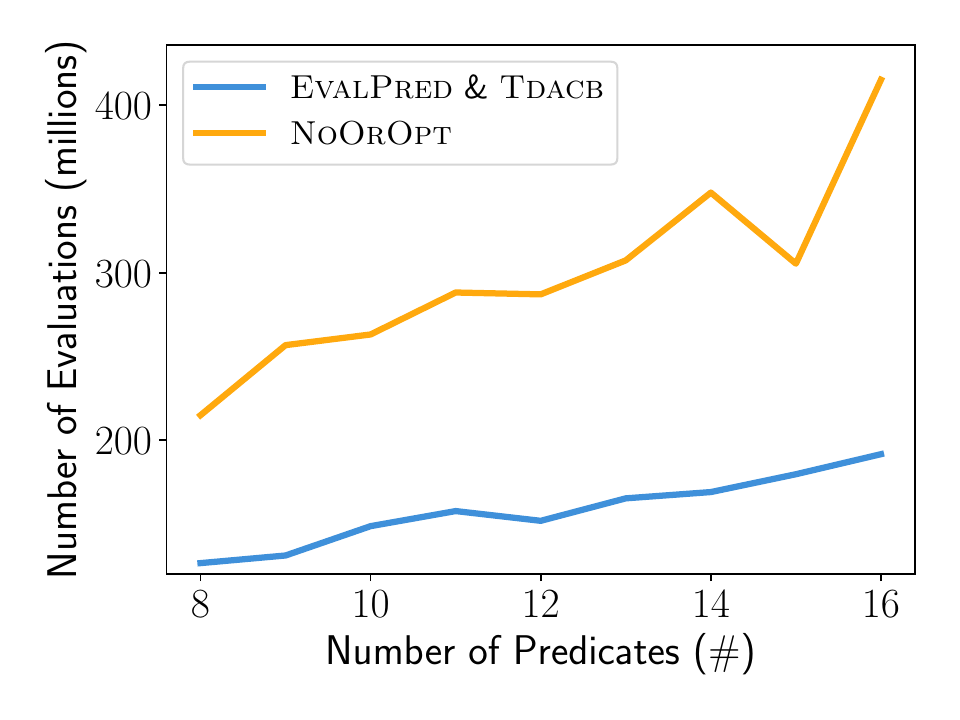}
    \caption{Number of evaluations}
    \label{fig:depth2-uniform-preds-inc-tdacb}
  \end{subfigure}
  \caption{%
  Runtimes and number of evaluations for depth-2 predicate expressions with uniform-cost \patom[s].
}
  \label{fig:depth2-uniform}
\end{figure*}
}{}

\smallskip
\begin{mdframed}
  \textbf{Key Takeaways:}
  In general, \algo{} performed the best out of any algorithm.
  For the synthetic workload, \algo{} had average speedups of up to 2.6$\times$ over \noopt{}, 1.4$\times$ over \deep{}, and 1.8$\times$ over \bdc{} for the top 10\% of queries.
  As for \tdacb{}, while executions times were consistently low, due to its exponential planning time, its total runtimes ended up being several orders of magnitude greater than even \noopt{} for the more complex expressions.
  For depth-2 predicate expressions, \algo{} had the same number of evaluations as \tdacb{}, verifying that it is optimal for the depth-2 case.
  For depth-3 predicate expressions, \algo{} still came quite close to the optimal number of evaluations, with 92\% of queries within 5\% of the minimum possible.
  For TPC-H and the CH-benchmark, \algo{} once again had the fastest runtimes.
  In comparison, \tdacb{}'s substantial planning times for TPC-H made its total runtime 100$\times$ slower than \algo{}'s for this benchmark.
  For the CH-benchmark, the number of unique \patom[s] in the predicate expression was fewer, so \tdacb{} had similar total runtimes to \algo{}.
\end{mdframed}

\iftoggle{paper}{\vspace{5pt}}{}
The remainder of this  section presents the results of the synthetic workload first, with Section~\ref{sec:eval:setup} describing the details of the workload, Section~\ref{sec:eval:depth2} presenting the results for depth-2 predicate expressions, and Section~\ref{sec:eval:depth3} presenting the results for depth-3+ predicate expressions.
The results for TPC-H and the CH-benchmark are presented later in Section~\ref{sec:eval:ch}.


\subsection{Synthetic Workload}
\label{sec:eval:setup}

{\noindent \bf Dataset.}
We used the Forest dataset~\cite{forest} that was used in the majority of the
experiments evaluating \tdacb{}~\cite{kastrati_generating_2018}.
The dataset has 10 quantitative attributes and 2 qualitative attributes.
However, because we wanted to evaluate our algorithms on more than just 12 independent \patom[s],
we duplicated the original dataset 40 times, shuffled the tuples of each duplicate dataset, and added the attributes of the
shuffled datasets as additional attributes of the original dataset,
giving us a total of 480 attributes.
Furthermore, the original dataset only contained 581K tuples, so we repeated the tuples 100 times for a total of 58M tuples.
Altogether, the size of the generated dataset was 208GB.

{\noindent \bf Queries.}
To mirror the evaluation of \tdacb{}~\cite{kastrati_generating_2018}, we used selection queries with randomly generated predicate expressions ranging up to 16 \patom[s].
  However, Kastrati and Moerkotte's evaluation of \tdacb{} included logically related \patom[s], and \tdacb{} leveraged Boolean implication (e.g., \predi{1} implies $\neg\predi{2}$) to significantly increase its runtime performance.
  Since we wanted to directly measure the impact of the number of \patom[s] on our algorithms, we chose to make every \patom[] independent of each other.
  Furthermore, \tdacb{} was only evaluated on CNF and DNF queries, meaning their predicate expressions had a maximum depth of 2.
In contrast, our predicate expressions were generated according to the following criteria.
The root of the predicate tree was randomly chosen to either be an AND or an OR.
Each predicate expression had a predetermined max depth of 2, 3, or 4.
Each non-leaf node had a randomly chosen number of children between 2 and 5, and each child had some chance to be a leaf node, ensuring that we did not only test on balanced trees.
For each of the quantitative attributes, we generated a simple $x < c$ comparison \patom[] where $c$ could one of 9 different constants, resulting in selectivities of $[0.1, 0.2, ..., 0.9]$.
For the two qualitative attributes, we had equality \patom[s] of the form $x = c$ where $c$ is one of
the possible values of that attribute (one had 4 possible values and the other had 7).
For experiments with variable-cost \patom[s], sleep times of 1-10ns were added per evaluation of a tuple to emulate variable cost.

{\noindent \bf Experiments.}
We conducted a total of 6 sets of experiments.
Each set of experiments had either a uniform or varying cost for all its \patom[s] and a fixed depth for its predicate expressions.
500 queries with randomly generated predicate expressions were used for each experiment.

\subsection{Depth-2 Predicate Expressions}
\label{sec:eval:depth2}

\iftoggle{paper}{}{%
\begin{figure*}
  \centering
  \begin{subfigure}{.32\linewidth}
    \centering
    \includegraphics[clip, trim=10px 0 10px 0,width=\linewidth]{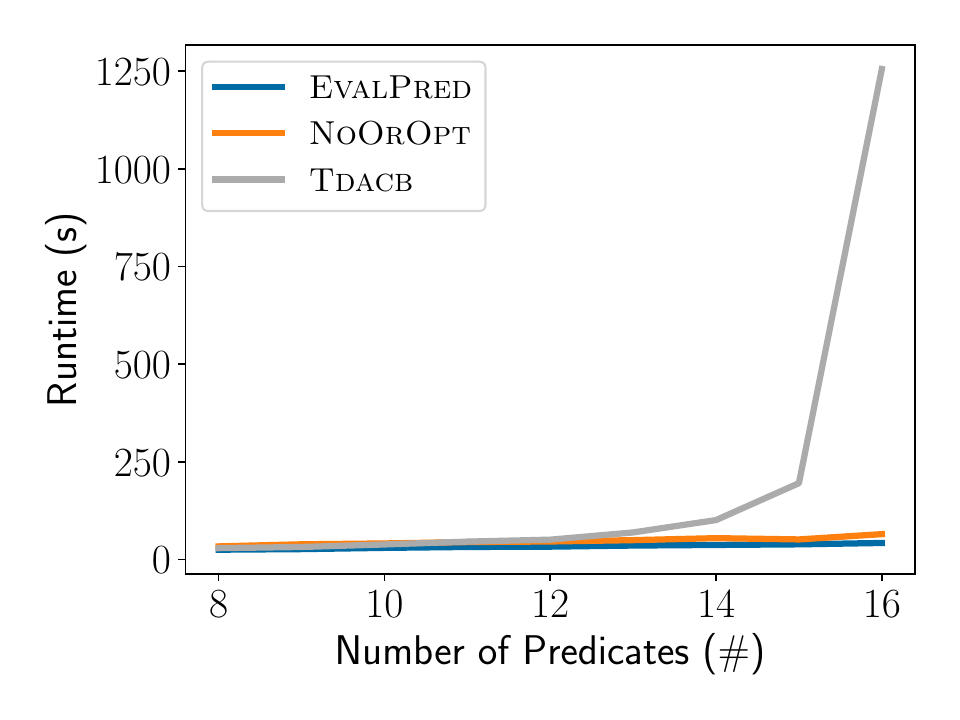,tdacb-with-plan,no-bdc,bdc-noplan,no-bdc-bestd,bdc-bestd-noplan.pdf}
    \caption{Runtimes}
    \label{fig:depth2-uniform-times-inc-tdacb}
  \end{subfigure}
  \begin{subfigure}{.32\linewidth}
    \centering
    \includegraphics[clip, trim=10px 0 10px 0,width=\linewidth]{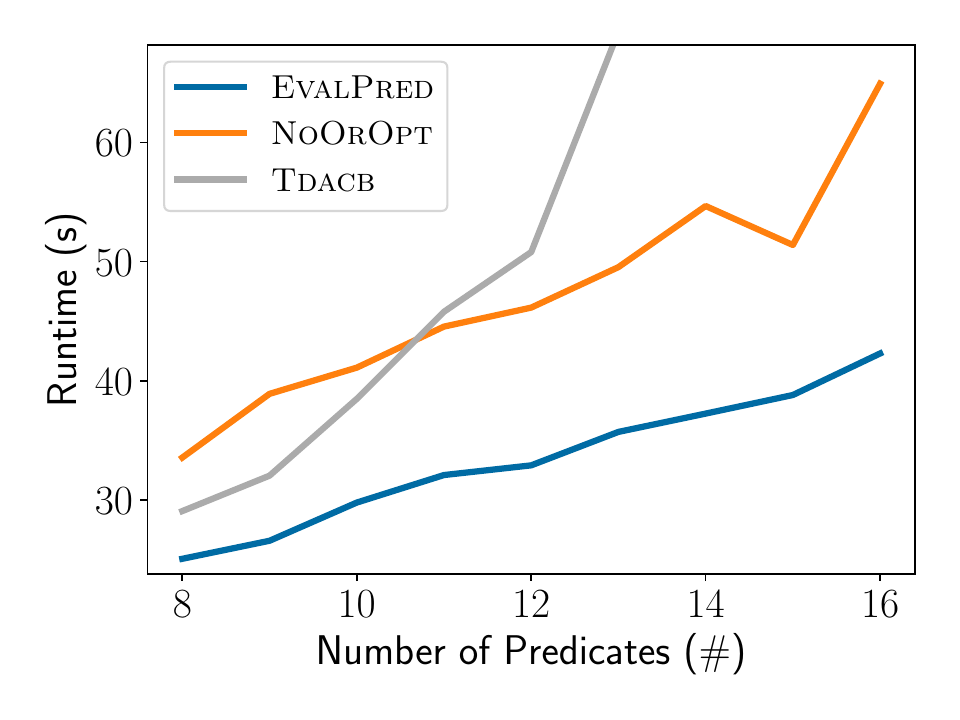,tdacb-with-plan,no-bdc,bdc-noplan,no-bdc-bestd,bdc-bestd-noplan.pdf}
    \caption{Runtimes (zoomed-in)}
    \label{fig:depth2-uniform-times-no-tdacb}
  \end{subfigure}
  \begin{subfigure}{.32\linewidth}
    \centering
    \includegraphics[clip, trim=10px 0 10px 0,width=\linewidth]{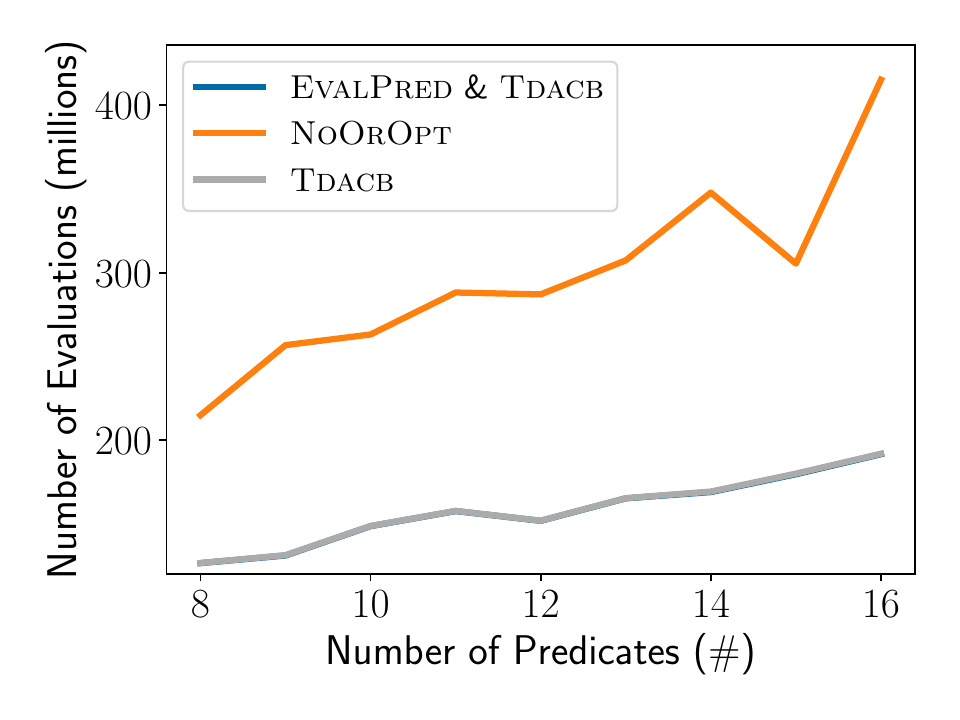,tdacb-with-plan,no-bdc,bdc-noplan,no-bdc-bestd,bdc-bestd-noplan.pdf}
    \caption{Number of evaluations}
    \label{fig:depth2-uniform-preds-inc-tdacb}
  \end{subfigure}
  \caption{%
  Runtimes and number of evaluations for depth-2 predicate expressions with uniform-cost \patom[s].
}
  \label{fig:depth2-uniform}
\end{figure*}
}

Figure~\ref{fig:depth2-uniform} shows the results of our experiments for depth-2 predicate expressions with uniform \patom[] cost.
We plot the average runtimes of \algo{}, \noopt{}, and \tdacb{} when grouped by the number of \patom[s] in Figure~\ref{fig:depth2-uniform-times-inc-tdacb}.
The same data is shown zoomed-in in Figure~\ref{fig:depth2-uniform-times-no-tdacb}.
Finally, we plot the average total number of evaluations incurred by each algorithm when grouped by the number of \patom[s] in Figure~\ref{fig:depth2-uniform-preds-inc-tdacb}.
As can be seen, \tdacb{} quickly becomes unviable for higher numbers of \patom[s] due to its exponential planning time, incurring total runtimes orders of magnitude greater than even \noopt{}.
Even for smaller numbers of \patom[s], \algo{} outperformed \tdacb{} with average speedups of 1.12$\times$ and 1.28$\times$ for 8 and 10 \patom[s] respectively.
Since Figure~\ref{fig:depth2-uniform-preds-inc-tdacb} shows that number of evaluations performed by \algo{} and \tdacb{} are the same, this disparity must come from the difference in planning time.
With respect to \noopt{}, \algo{} enjoyed a consistent 1.41$\times$ average speedup in runtime and a total average of 1.88$\times$ speedup in number of evaluations.
For the top 10\% of queries, \algo{} had an average speedup of 2.12$\times$ in runtime and 4.04$\times$ in number of evaluations over \noopt{}.

The same patterns arose for depth-2 predicate expressions with varying-cost \patom[s].
\tdacb{} exhibited the same exponential behavior, but \algo{} outperformed \tdacb{} for even smaller numbers of \patom[s] with the same average speedups of 1.12$\times$ and 1.28$\times$ for 8 and 10 \patom[s] respectively.
\algo{} also consistently outperformed \noopt{} with a total average speedup of 1.43$\times$ in runtime and a total average speedup of 1.86$\times$ in number of evaluations.

\subsection{Depth-3+ Predicate Expressions}
\label{sec:eval:depth3}

\begin{figure*}
  \centering
  \begin{subfigure}{.32\linewidth}
    \includegraphics[clip, trim=10px 0 10px 0,width=\linewidth]{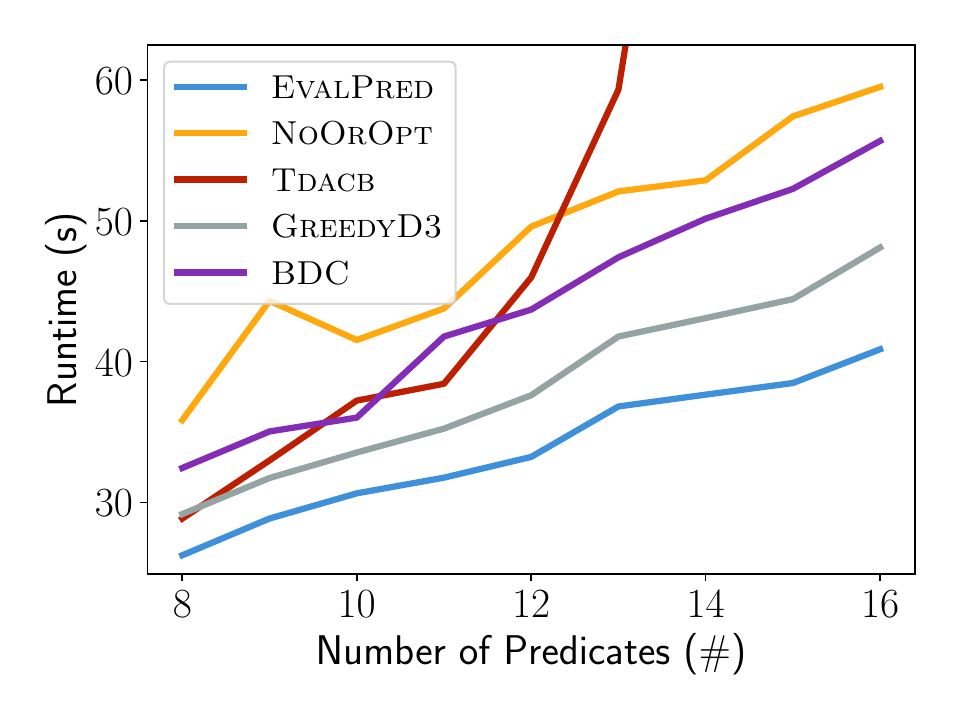}
    \caption{Runtimes}
    \label{fig:depth3-varcost-times-no-tdacb}
  \end{subfigure}
  \begin{subfigure}{.32\linewidth}
    \includegraphics[clip, trim=10px 0 10px 0,width=\linewidth]{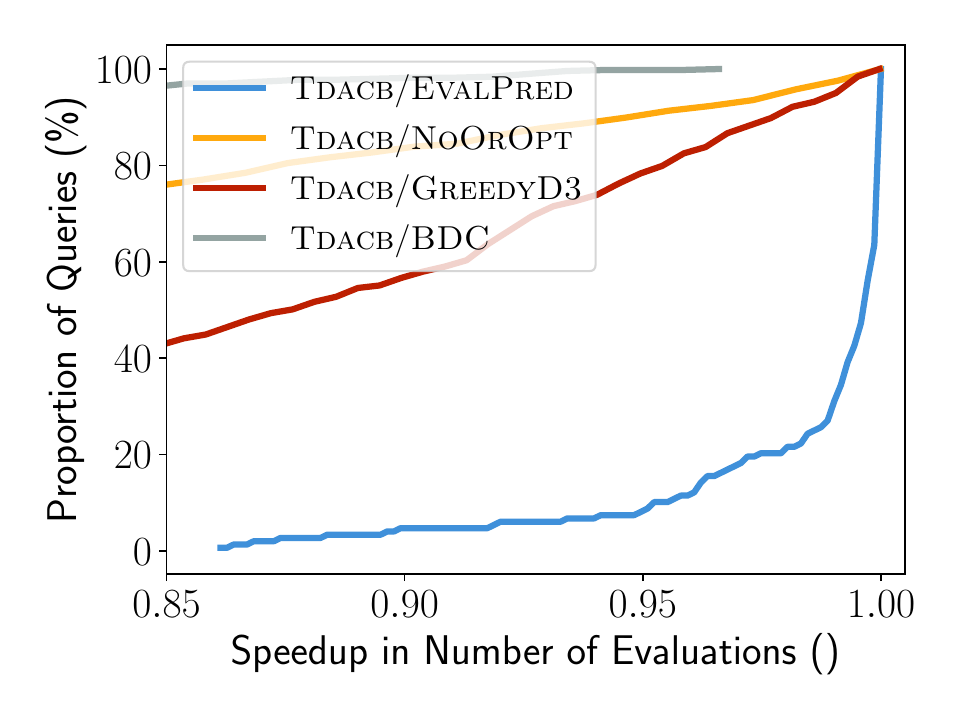}
    \caption{CDF of \tdacb{}/Others}
    \label{fig:depth3-tdacb-cdf}
  \end{subfigure}
  \begin{subfigure}{.32\linewidth}
    \includegraphics[clip, trim=10px 0 10px 0,width=\linewidth]{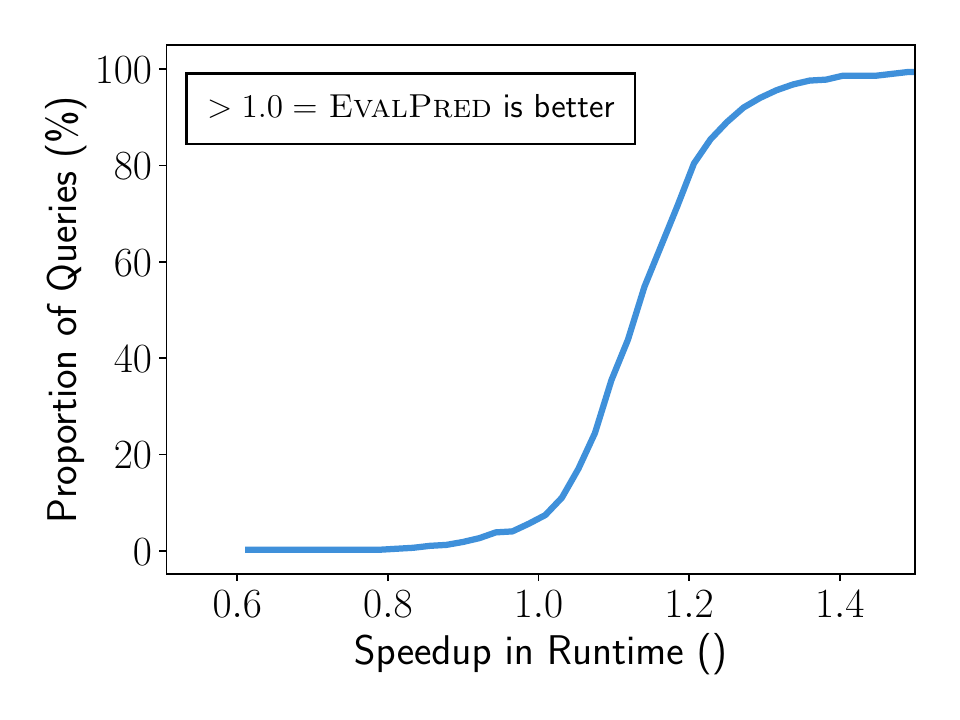}
    \caption{CDF of \deep{}/\algo{}}
    \label{fig:depth3-deep-cdf}
  \end{subfigure}
  \caption{%
    Runtimes and CDFs of speedups for depth-3 predicate expressions with varying-cost \patom[s].
  }
  \label{fig:depth3}
\end{figure*}

Figure~\ref{fig:depth3} (next page) shows the results of our experiments for depth-3 predicate expressions with varying-cost \patom[s].
We plot the average runtimes of \algo{}, \noopt{}, \deep{}, \bdc{}, and \tdacb{} when grouped by the number of \patom[s] in Figure~\ref{fig:depth3-varcost-times-no-tdacb}.
As Figure~\ref{fig:depth3-varcost-times-no-tdacb} shows, despite not being optimal, \algo{} still achieved the best runtime performance out of any algorithm, with total average speedups of 1.43$\times$ over \noopt{}, 1.13$\times$ over \deep{}, and 1.29$\times$ over \bdc{}.
For the top 10\% of queries, \algo{} had average speedups of 2.60$\times$ over \noopt{}, 1.36$\times$ over \deep{}, and 1.84$\times$ over \bdc{}.
\tdacb{} exhibited the same exponential behavior as before and had runtimes orders of magnitude greater than \noopt{} for 14-16 \patom[s].
Even for 8 and 10 \patom[s], \algo{} outperformed \tdacb{} with average speedups of 1.10$\times$ and 1.21$\times$ respectively.

Though \tdacb{}'s runtimes may be high, the plans that it generates are still optimal.
As such, the number of evaluations incurred by \tdacb{} can be seen as the minimum possible for any given query and serve as a reference for how close \algo{} comes to optimal for depth-3+ predicate expressions.
Figure~\ref{fig:depth3-tdacb-cdf} shows the CDF of speedups in number of evaluations of various algorithms over \tdacb{}.
It is apparent that \algo{} comes quite close to achieving optimal performance even for depth-3 predicate expressions, with 92\% of queries invoking within 5\% of the minimum number of evaluations possible.
In comparison, the other algorithms fare much worse.
To serve as another comparison point, Figure~\ref{fig:depth3-deep-cdf} shows the CDF of speedups in runtime of \algo{} over \deep{}.
\deep{} is able to generate orderings which are not DFS traversals, and this makes a difference for 6\% of queries, in which \deep{} had faster runtimes than \algo{}.
However, for the vast majority of queries, \algo{} still performed at least as well as, if not better than, \deep{}.
\bdc{} also had similar results with only 4\% of queries performing better than \algo{}.

Depth-3, uniform-cost predicate expressions and depth-4 predicate expressions exhibited very similar patterns to the depth-3, varying-cost case.
With less than a 2\% difference in total average speedups, the plotted figures were almost identical, so we do not show them separately here.

\subsection{TPC-H and CH-benchmark}
\label{sec:eval:ch}

\iftoggle{paper}{%
\begin{table}
  \begin{tabular}{c|c|c|c}
    Algorithm & Plan Time (s) & Exec Time (s) & \# of Evaluations \\ \hline
    \algo{} & 0 & 4.20 & 10.8M \\
    \noopt{} & 0 & 4.28 & 10.9M \\
    \tdacb{} & 416 & 4.21 & 10.8M \\
    \deep{} & 0 & 4.25 & 10.9M \\
    \bdc{} & 0.09 & 5.64 & 61.1M
  \end{tabular}
  \vspace{3pt}
  \caption{TPC-H results for Query 19.}
  \label{tab:tpc-h}
  \iftoggle{paper}{\vspace{-20pt}}{}
\end{table}
}{%
  \begin{table}[b]
  \begin{tabular}{c|c|c|c}
    Algorithm & Plan Time (s) & Exec Time (s) & \# of Evaluations \\ \hline
    \algo{} & 0 & 4.20 & 10.8M \\
    \noopt{} & 0 & 4.28 & 10.9M \\
    \tdacb{} & 416 & 4.21 & 10.8M \\
    \deep{} & 0 & 4.25 & 10.9M \\
    \bdc{} & 0.09 & 5.64 & 61.1M
  \end{tabular}
  \vspace{3pt}
  \caption{TPC-H results for Query 19.}
  \label{tab:tpc-h}
  \iftoggle{paper}{\vspace{-20pt}}{}
\end{table}
}

\begin{table}
  \begin{tabular}{c|c|c|c}
    Algorithm & Plan Time (ms) & Exec Time (s) & \# of Evaluations \\ \hline
    \algo{} & 0 & 12.6 & 77.3M \\
    \noopt{} & 0 & 16.1 & 102M \\
    \tdacb{} & 12 & 12.6 & 77.3M \\
    \deep{} & 0 & 13.5 & 89.2M \\
    \bdc{} & 0 & 16.6 & 118M
  \end{tabular}
  \vspace{3pt}
  \caption{CH-benchmark results for Query 19.}
  \label{tab:ch}
  \iftoggle{paper}{\vspace{-20pt}}{}
\end{table}

In addition to our synthetic workload, we looked at the queries from TPC-H and the CH-benchmark and found that Q19 from each benchmark has disjunctions.
TPC-H's joined table has 6M tuples and its predicate expression is in DNF with 18 total \patom[s], but only 14 unique \patom[s].
CH-benchmark's joined table has 15M tuples and its predicate expression is also in DNF with 15 total \patom[s], but only 9 unique \patom[s].
The results for TPC-H and CH-benchmark are shown in Tables~\ref{tab:tpc-h} and~\ref{tab:ch} respectively.
The tables show the average runtimes of executing each query 5 times for each algorithm.
For TPC-H, although \algo{} and \tdacb{} had similar execution times, \tdacb{}'s planning time took an enormous 416 seconds, so \algo{} had a 100$\times$ speedup over \tdacb{} in total runtime.
In comparison, \noopt{} exhibited similar runtimes to \algo{} for this query.
For the CH-benchmark, there were only 9 unique \patom[s].
Thus,\tdacb{}'s planning time was significantly reduced, and it had comparable total runtimes to \algo{}.
\noopt{}, on the other hand, was slower for this query, and \algo{} had a 1.3$\times$ speedup over it in runtime.

\section{Conclusion}
\label{sec:conc}

In this paper, we formally analyze the problem of predicate evaluation for column-oriented engines and present \algo{} as our recommended algorithm.
Our analysis and evaluation shows that \algo{} is optimal for all predicate expressions of depth 2 or less and still quite close to optimal for depth-3+ predicate expressions.
Thus, we hope that many
systems can take advantage of our work and implement these optimizations
to improve the performance of predicate evaluation.

{
  \iftoggle{paper}{%
    \bibliographystyle{ACM-Reference-Format}
  }{
    \bibliographystyle{plainnat}
  }
  \bibliography{main}

\begin{thebibliography}{30}
\providecommand{\natexlab}[1]{#1}
\providecommand{\url}[1]{\texttt{#1}}
\expandafter\ifx\csname urlstyle\endcsname\relax
  \providecommand{\doi}[1]{doi: #1}\else
  \providecommand{\doi}{doi: \begingroup \urlstyle{rm}\Url}\fi

\bibitem[for(1999)]{forest}
Forest dataset.
\newblock \url{http://kdd.ics.uci.edu/databases/covertype/covertype.data.html}, 1999.

\bibitem[TPC(2022)]{TPCHHomepage}
{{TPC-H Homepage}}, 2022.
\newblock URL \url{https://www.tpc.org/tpch/}.

\bibitem[Armbrust et~al.(2015)Armbrust, Xin, Lian, Huai, Liu, Bradley, Meng, Kaftan, Franklin, and Ghodsi]{armbrustSparkSqlRelational2015}
Michael Armbrust, Reynold~S. Xin, Cheng Lian, Yin Huai, Davies Liu, Joseph~K. Bradley, Xiangrui Meng, Tomer Kaftan, Michael~J. Franklin, and Ali Ghodsi.
\newblock Spark sql: {{Relational}} data processing in spark.
\newblock In \emph{Proceedings of the 2015 {{ACM SIGMOD}} International Conference on Management of Data}, pages 1383--1394. ACM, 2015.

\bibitem[B{\"u}ning and Lettmann(1999)]{buning1999propositional}
Hans~Kleine B{\"u}ning and Theodor Lettmann.
\newblock \emph{Propositional logic: deduction and algorithms}, volume~48.
\newblock Cambridge University Press, 1999.

\bibitem[Chambi et~al.(2016)Chambi, Lemire, Kaser, and Godin]{chambi2016better}
Samy Chambi, Daniel Lemire, Owen Kaser, and Robert Godin.
\newblock Better bitmap performance with roaring bitmaps.
\newblock \emph{Software: practice and experience}, 46\penalty0 (5):\penalty0 709--719, 2016.

\bibitem[Chang and Lee(1997)]{chang_optimization_1997}
Jae-young Chang and Sang-goo Lee.
\newblock An optimization of disjunctive queries: union-pushdown.
\newblock In \emph{{COMPSAC}}, pages 356--361. IEEE Computer Society, 1997.

\bibitem[Chaudhuri et~al.(2003)Chaudhuri, Ganesan, and Sarawagi]{chaudhuriFactorizingComplexPredicates2003}
Surajit Chaudhuri, Prasanna Ganesan, and Sunita Sarawagi.
\newblock Factorizing complex predicates in queries to exploit indexes.
\newblock In \emph{{{SIGMOD}} 2003}. {ACM}, 2003.

\bibitem[Claussen et~al.(2000)Claussen, Kemper, Moerkotte, Peithner, and Steinbrunn]{sen_optimization_2000}
Jens Claussen, Alfons Kemper, Guido Moerkotte, Klaus Peithner, and Michael Steinbrunn.
\newblock Optimization and {Evaluation} of {Disjunctive} {Queries}.
\newblock \emph{IEEE Trans. Knowl. Data Eng.}, 12\penalty0 (2):\penalty0 238--260, 2000.

\bibitem[Cole et~al.(2011)Cole, Funke, Giakoumakis, Guy, Kemper, Krompass, Kuno, Nambiar, Neumann, Poess, et~al.]{cole2011mixed}
Richard Cole, Florian Funke, Leo Giakoumakis, Wey Guy, Alfons Kemper, Stefan Krompass, Harumi Kuno, Raghunath Nambiar, Thomas Neumann, Meikel Poess, et~al.
\newblock The mixed workload ch-benchmark.
\newblock In \emph{Proceedings of the Fourth International Workshop on Testing Database Systems}, pages 1--6, 2011.

\bibitem[Csaszar(1978)]{alma990000570490106761}
Akos. Csaszar.
\newblock \emph{General topology}.
\newblock Disquisitiones mathematicae Hungaricae ; v. 9. A. Hilger, Bristol [Eng, 1978.
\newblock ISBN 0852742754.

\bibitem[Fontoura et~al.(2010)Fontoura, Sadanandan, Shanmugasundaram, Vassilvitskii, Vee, Venkatesan, and Zien]{fontoura_efficiently_2010}
Marcus Fontoura, Suhas Sadanandan, Jayavel Shanmugasundaram, Sergei Vassilvitskii, Erik Vee, Srihari Venkatesan, and Jason~Y. Zien.
\newblock Efficiently evaluating complex boolean expressions.
\newblock In \emph{{SIGMOD} Conference}, pages 3--14. {ACM}, 2010.

\bibitem[Grund et~al.(2010)Grund, Kr{\"{u}}ger, Plattner, Zeier, Cudr{\'{e}}{-}Mauroux, and Madden]{grund_hyrise:_2010}
Martin Grund, Jens Kr{\"{u}}ger, Hasso Plattner, Alexander Zeier, Philippe Cudr{\'{e}}{-}Mauroux, and Samuel Madden.
\newblock {HYRISE} - {A} main memory hybrid storage engine.
\newblock \emph{Proc. {VLDB} Endow.}, 2010.

\bibitem[Hanani(1977)]{hanani_optimal_1977}
Michael~Z. Hanani.
\newblock An {Optimal} {Evaluation} of {Boolean} {Expressions} in an {Online} {Query} {System}.
\newblock \emph{Commun. ACM}, 20\penalty0 (5):\penalty0 344--347, May 1977.
\newblock ISSN 0001-0782.
\newblock \doi{10.1145/359581.359600}.
\newblock URL \url{http://doi.acm.org/10.1145/359581.359600}.

\bibitem[Idreos et~al.(2012)Idreos, Groffen, Nes, Manegold, Mullender, and Kersten]{idreos_monetdb_2012}
S.~Idreos, F.~Groffen, N.~Nes, S.~Manegold, S.~Mullender, and M.~Kersten.
\newblock Monetdb: {Two} decades of research in column-oriented database.
\newblock \emph{IEEE Data Engineering Bulletin}, 35\penalty0 (1):\penalty0 40--45, 2012.
\newblock Publisher: Citeseer.

\bibitem[JarkeMatthias and KochJurgen(1984)]{jarkematthias_query_1984}
JarkeMatthias and KochJurgen.
\newblock Query {Optimization} in {Database} {Systems}.
\newblock \emph{ACM Computing Surveys (CSUR)}, June 1984.
\newblock URL \url{https://dl.acm.org/doi/abs/10.1145/356924.356928}.

\bibitem[Johnson et~al.(2008)Johnson, Raman, Sidle, and Swart]{johnson_row-wise_2008}
Ryan Johnson, Vijayshankar Raman, Richard Sidle, and Garret Swart.
\newblock Row-wise parallel predicate evaluation.
\newblock \emph{Proceedings of the VLDB Endowment}, 1\penalty0 (1):\penalty0 622--634, 2008.
\newblock Publisher: VLDB Endowment.

\bibitem[Kastrati and Moerkotte(2017)]{kastrati_optimization_2017}
Fisnik Kastrati and Guido Moerkotte.
\newblock Optimization of disjunctive predicates for main memory column stores.
\newblock In \emph{{SIGMOD} Conference}, pages 731--744. {ACM}, 2017.

\bibitem[Kastrati and Moerkotte(2018)]{kastrati_generating_2018}
Fisnik Kastrati and Guido Moerkotte.
\newblock Generating optimal plans for boolean expressions.
\newblock In \emph{{ICDE}}, pages 1013--1024. {IEEE} Computer Society, 2018.

\bibitem[Kemper et~al.(1992)Kemper, Moerkotte, and Steinbrunn]{kemper_optimizing_1992}
Alfons Kemper, Guido Moerkotte, and Michael Steinbrunn.
\newblock Optimizing boolean expressions in object-bases.
\newblock In \emph{{VLDB}}, pages 79--90. Morgan Kaufmann, 1992.

\bibitem[Kemper et~al.(1994)Kemper, Moerkotte, Peithner, and Steinbrunn]{kemper_optimizing_1994}
Alfons Kemper, Guido Moerkotte, Klaus Peithner, and Michael Steinbrunn.
\newblock Optimizing disjunctive queries with expensive predicates.
\newblock In \emph{{SIGMOD} Conference}. {ACM}, 1994.

\bibitem[Lamb et~al.(2012)Lamb, Fuller, Varadarajan, Tran, Vandier, Doshi, and Bear]{lamb2012vertica}
Andrew Lamb, Matt Fuller, Ramakrishna Varadarajan, Nga Tran, Ben Vandier, Lyric Doshi, and Chuck Bear.
\newblock The vertica analytic database: C-store 7 years later.
\newblock \emph{arXiv preprint arXiv:1208.4173}, 2012.

\bibitem[Monk(1973)]{monk1973introduction}
J~Donald Monk.
\newblock Introduction to set theory.
\newblock \emph{Journal of Symbolic Logic}, 38\penalty0 (1), 1973.

\bibitem[Muralikrishna(1988)]{muralikrishna_optimization_1988}
Murali Muralikrishna.
\newblock Optimization of multiple-disjunct queries in a relational database system.
\newblock Technical report, University of Wisconsin-Madison Department of Computer Sciences, 1988.

\bibitem[Ott(2001)]{ott2001chameleons}
Matthias Ott.
\newblock Chameleons have independent eye movements but synchronise both eyes during saccadic prey tracking.
\newblock \emph{Experimental Brain Research}, 139:\penalty0 173--179, 2001.

\bibitem[Russell and Norvig(2016)]{russellArtificialIntelligenceModern2016}
Stuart~J. Russell and Peter Norvig.
\newblock \emph{Artificial Intelligence: A Modern Approach}.
\newblock {Malaysia; Pearson Education Limited,}, 2016.

\bibitem[Steinbrunn et~al.(1995)Steinbrunn, Peithner, Moerkotte, and Kemper]{steinbrunn_bypassing_1995}
Michael Steinbrunn, Klaus Peithner, Guido Moerkotte, and Alfons Kemper.
\newblock Bypassing joins in disjunctive queries.
\newblock In \emph{{VLDB}}, volume~95, pages 11--15, 1995.

\bibitem[Stephen and Yusun(2014)]{stephenCountingInequivalentMonotone2014}
Tamon Stephen and Timothy Yusun.
\newblock Counting inequivalent monotone boolean functions.
\newblock \emph{Discrete Applied Mathematics}, 167:\penalty0 15--24, 2014.

\bibitem[Stoll(1979)]{stollSetTheoryLogic1979}
Robert~Roth Stoll.
\newblock \emph{Set {{Theory}} and {{Logic}}}.
\newblock {Courier Corporation}, October 1979.
\newblock ISBN 978-0-486-63829-4.

\bibitem[Stonebraker and Rowe(1986)]{stonebrakerDesignPostgres1986}
Michael Stonebraker and Lawrence~A. Rowe.
\newblock \emph{The Design of {{Postgres}}}, volume~15.
\newblock {ACM}, 1986.

\bibitem[Straube and {\"O}zsu(1990)]{straubeQueriesQueryProcessing1990}
David~D. Straube and M.~Tamer {\"O}zsu.
\newblock Queries and query processing in object-oriented database systems.
\newblock \emph{ACM Transactions on Information Systems (TOIS)}, 1990.

\end{thebibliography}
}

\iftoggle{paper}{}{%
  \clearpage
\appendix

\onecolumn



\section{\large \hanani{}}
\label{apx:hanani}

For the reader's convenience, we provide a modern, updated version of \hanani{} in Algorithm~\ref{alg:hanani} (next page).
As mentioned, this returns the optimal ordering of \patom[s] for any Boolean predicate expression of depth 2 or less.
The majority of the work is done by \estchild{}, which returns the best ordering of all descendent \patom[s] for a given \node{}.
If \node{} is a leaf node, $\node.P$ is used to refer to the \patom[] that \node{} points to.
$\selec_P$ is the selectivity of $P$ (i.e., the fraction of tuples which satisfy \pred), and $F_{\pred}$ is the constant cost factor discussed in Cost Model~\ref{cost:1}.
Since a leaf node only has one descendent \patom[], it is the only \patom[] in the returned sequence.
If \node{} is an AND/OR node, \estchild{} is called on each of its children, and the returned \sordering{} are sorted in increasing/decreasing selectivity respectively.
Once \sordering{} has been sorted, \orderhelper{} calculates the total selectivity and cost for \node{} (using the independence assumption), flattens \sordering{}, and returns the combination of these values to the parent.

\begin{algorithm}[H]
  \begin{algorithmic}[1]
    \Require Predicate expression \boolformorig{}
    \Ensure Ordered sequence of \patom[s] $[\predi{1},...,\predi{n}]$
    \State \rootnode{} $\gets$ makeUnorderedTree(\boolformorig{})
    \State $(\selec, \text{cost}, \mathbf{P}) \gets \estchild(\rootnode)$
    \State \Return $\mathbf{P}$
    \Statex
    \Function{\estchild}{$\node{}$}
    \If{isLeafNode(\node)}
    \State $P \gets \node.P$
    \State \Return $(\selec_{P}, F_{P}, [P])$
    \label{line:base-cost}
    \EndIf
    \State
    \State $\sordering \gets []$
    \For{$c \in \text{children}(\node)$}
    \State $(\selec, \text{cost}, \mathbf{P}) \gets \estchild(c)$
    \State $\sordering \gets \sordering + [(\selec, \text{cost}, \mathbf{P})]$
    \EndFor
    \If{\text{isAndNode}(\node)}
    \State $\sordering \gets \text{increasingSort}(\sordering, \sortfand)$
    \Else
    \State $\sordering \gets \text{increasingSort}(\sordering, \sortfor)$
    \EndIf
    \State \Return $\orderhelper(\sordering, \node)$
    \EndFunction
    \Statex
    \Function{\orderhelper}{$\sordering, \node$}
    \State totalCost $ \gets 0$
    \State $\selec_{\text{total}} \gets1$
    \State $\mathbf{P}_{\text{all}} \gets []$
    \For{$i \gets 1,...,|\sordering|$}
    \State $(\selec, \text{cost}, \mathbf{P}) \gets \sordering[i]$
    \If{isAndNode(\node)}
    \State totalCost $\gets \text{totalCost} + \selec_{\text{total}} \cdot \text{cost}$
    \label{line:and-total-cost}
    \State $\selec_{\text{total}} \gets \selec_{\text{total}} \cdot \selec$
    \Else
    \State totalCost $\gets \text{totalCost} + (1 -
    \selec_{\text{total}}) \cdot \text{cost}$
    \label{line:or-total-cost}
    \State $\selec_{\text{total}} \gets \selec + \selec_{\text{total}} \cdot
    (1-\selec)$
    \EndIf
    \State $\mathbf{P}_{\text{all}} \gets \mathbf{P}_{\text{all}} + \mathbf{P}$
    \EndFor
    \State \Return $(\selec_{\text{total}}, \text{totalCost}, \mathbf{P}_{\text{all}})$
    \EndFunction
    \Statex
    \Function{\sortfand}{$\selec, \text{cost}, \mathbf{P}$}
    \State \Return  $\text{cost} / (1 - \selec)$
    \EndFunction
    \Statex
    \Function{\sortfor}{$\selec, \text{cost}, \mathbf{P}$}
    \State \Return  $\text{cost} / \selec$
    \EndFunction
  \end{algorithmic}
  \caption{\hanani{}}
  \label{alg:hanani}
\end{algorithm}


\section{Correctness Proofs}
\label{apx:correct}

\subsection{\algo{}}
\label{apx:correct:algo}

\begin{algorithm}
  \renewcommand{\thealgorithm}{\ref{alg:algo}}
  \begin{algorithmic}[1]
    \Require Predicate expression $\boolformorig{}$, set of all tuples $R$
    \Ensure The set of all tuples which satisfy \boolformorig{}
    \State $[\predi{1},...,\predi{n}] \gets \hanani{}(\boolformorig{})$
    \State \rootnode{} $\gets \text{makeTree}(\boolformorig, [\predi{1},...,\predi{n}])$
    \State \Return \evalnode(\rootnode{}, $R$)
    \Statex
    \Function{\evalnode}{\node{}, $D$}
    \If{isLeafNode(\node{})}
    \State \Return \nodepa{}($D$)
    \ElsIf{isAndNode(\node)}
    \For{\child{} in children(\node)}
    \If{$D = \varnothing$}
    \State \Return $D$
    \EndIf
    \State $D \gets$ \evalnode(\child, $D$)
    \EndFor
    \State \Return $D$
    \Else
    \State \done{} $\gets \varnothing$
    \For{\child{} in children(\node)}
    \State \tdo{} $\gets D \setminus \done{}$
    \If{$\tdo{} = \varnothing$}
    \State \Return \done{}
    \EndIf
    \State \done{} $\gets$ \done{} $\cup$ \evalnode(\child, \tdo{})
    \EndFor
    \State \Return \done
    \EndIf
    \EndFunction
  \end{algorithmic}
  \caption{\algo{}}
  \addtocounter{algorithm}{-1}
\end{algorithm}

In this section, we prove the correctness of \algo{}.
For the reader's convenience, we restate Algorithm~\ref{alg:algo} above.
We first introduce Theorem~\ref{thm:evalnode}, shown below.
Theorem~\ref{thm:evalnode} states that, for any predicate tree node \node{} and set of tuples \vset{}, $\evalnode(\node,\vset)$ returns the subset of tuples in \vset{} which satisfy the predicate subexpression represented by \node{}.
Thus, if $R$ represents the set of all tuples (in the table), $\evalnode(\rootnode, R)$ in Line~\ref{line:algo:call} must return the set of all tuples which satisfy the given predicate expression.

\begin{restatable}{theorem}{thmevalnode}
  Given a predicate expression and an ordering of its \patom[s], let \node{} be any node in the associated predicate tree, and let \refnode{} be the set of all tuples which satisfy the predicate subexpression represented by \node{}.
  For any set of tuples \vset{}, $\evalnode{}(\node, \vset)$ returns the subset of \vset{} which satisfies the predicate subexpression represented by \node{}, or:
  \[
    \evalnode(\node, \vset) = \refnode \cap \vset
  \]
  \label{thm:evalnode}
\end{restatable}
\begin{proof}
  Our proof is by strong induction on the height of the tree referred to by \node{}.
  In the base case (height 1), \node{} refers to a leaf node.
  Thus, \node{} must be a \patom[] application node.
  By definition, $\nodepa(\node,\vset)$ returns the subset of tuples in \vset{} which satisfy \node{}'s \patom[], taking care of the base case:
  \[
    \nodepa(\node, \vset) = \refnode \cap \vset
  \]

  As the inductive hypothesis, we assume Theorem~\ref{thm:evalnode} applies to all predicate trees with height $k$ or less.
  If \node{} has a height of $k+1$, \node{} must be either an AND or an OR node.
  For the purpose of this proof, let us first ignore the early returns featured in Lines~\ref{line:algo:early-and-beg}-\ref{line:algo:early-and-end} and Lines~\ref{line:algo:early-or-beg}-\ref{line:algo:early-or-end}.
  We examine the AND and OR cases separately.



  {\bf If \node{} is an AND node,} the inductive hypothesis must hold for each child of \node{} since they must all be trees of height $k$ or less.
  Let $\vset^{(i)}$ be the updated value of \vset{} in Line~\ref{line:algo:update-and} after the \evalnode{} call to the $i$th child of \node{} (denoted \childi{i}).
  We can express $\vset^{(i)}$ with the recurrence relationship:
  \[
    \vset^{(i)} = \refnodei{\childi{i}} \cap \vset^{(i-1)}
  \]
  If we assume $\vset^{(0)}$ is the original value of \vset{} passed in as the argument to \evalnode{} and \node{} has $t$ children, the return value of Line~\ref{line:algo:and-return} will be:
  \[
    \vset^{(0)} \cap \refnodei{\childi{1}} \cap \refnodei{\childi{2}} \cap \ldots \cap \refnodei{\childi{t}}
  \]
  The set of tuples which satisfy an AND node is the intersection of its children's satisfying sets, thus the above can be written as:
  \[
    \vset^{(0)} \cap \refnodei{\node}
  \]

  {\bf If \node{} is an OR node,} the inductive hypothesis must again hold for each child of \node{}.
  Let \donei{i} be the updated value of \done{} in Line~\ref{line:algo:update-or} after the \evalnode{} call to the $i$th child of \node{} (denoted \childi{i}).
  We can express \donei{i} with the recurrence relationship:
  \[
    \donei{i} = \donei{i-1} \cup \left(\refnodei{\childi{i}} \cap \left(\vset \setminus \donei{i-1}\right)\right)
  \]
  For any sets $A$, $B$, and $C$: $A \cup (B \cap (C \setminus A)) = A \cup (B \cap C)$ (Appendix~\ref{apx:set-alg}).
  Thus, the recurrence relationship can be rewritten as:
  \[
    \donei{i} = \donei{i-1} \cup (\refnodei{\childi{i}} \cap \vset)
  \]
  Assuming once again \node{} has $t$ children, Line~\ref{line:algo:or-return}'s return value is:
  \[
    (\refnodei{\childi{1}} \cap \vset) \cup (\refnodei{\childi{2}} \cap \vset) \cup \ldots \cup (\refnodei{\childi{t}} \cap \vset)
  \]
  Using the distributive property, we can clean this up as:
  \[
    \vset \cap (\refnodei{\childi{1}} \cup \refnodei{\childi{2}} \cup \ldots \cup \refnodei{\childi{t}})
  \]
  The set of tuples which satisfy an OR node is the union of its children's satisfying sets, thus the above can be simplified to:
  \[
    \vset \cap \refnode
  \]

\end{proof}

\subsection{\bestd{} and \update{}}
\label{apx:correct:bestd}

\begin{algorithm}
  \renewcommand{\thealgorithm}{\ref{alg:bestd}}
  \begin{algorithmic}[1]
    \Require Predicate tree node \node{}, algorithm state \mystate{}, \patom[] \predi{i}, step index $i$
    \Ensure Optimal operand \vseti{i} to apply \predi{i} to
    \If{$\node  = \nil$}
    \State \Return $\{T,F\}^n$
    \ElsIf{\text{isAndNode}(\node)}
    \State $X \gets \bestd(\text{parent}(\node), \mystate, i)$
    \For{$\child \in \text{children}(\node)$}
    \If{$\text{\cmp[]}(\child, i)$}
    \State $X \gets X \cap \mystate[\child].\cmp[v]$
    \ElsIf{$\detneg[](\child, i)$ and $\neg$isAnc(\child, \predi{i})}
    \State $X \gets X \setminus \mystate[\child].\detneg[v]$
    \EndIf
    \EndFor
    \State \Return $X$
    \Else
    \State $X \gets \bestd(\text{parent}(\node), \mystate, i)$
    \State $Y \gets \{\}$
    \For{$\child \in \text{children}(\node)$}
    \If{$\text{\cmp[]}(\child, i)$}
    \State $Y \gets Y \cup \mystate[\child].\cmp[v]$
  \ElsIf{$\detpos[](\child, i)$ and $\neg$isAnc(\child, \predi{i})}
    \State $Y \gets Y \cup \mystate[\child].\detpos[v]$
    \EndIf
    \EndFor
    \State \Return $X \setminus Y$
    \EndIf
  \end{algorithmic}
  \caption{\bestd{}}
  \addtocounter{algorithm}{-1}
\end{algorithm}

\begin{algorithm}
  \renewcommand{\thealgorithm}{\ref{alg:update}}
  \begin{algorithmic}[1]
    \Require Predicate tree node \node{}, algorithm state \mystate{}, \patom[] \predi{i}, set of valuations \vseti{i}, step index $i$
    \If{$\node = \nil$}
    \State \Return
    \ElsIf{$\text{isLeafNode}(\node)$}
    \State $\mystate[\node].\cmp[v] \gets \predi{i}(\vseti{i})$
    \State $\mystate[\node].\detpos[v] \gets \predi{i}(\vseti{i})$
    \State $\mystate[\node].\detneg[v] \gets \vseti{i} \setminus \predi{i}(\vseti{i})$
    \ElsIf{\text{isAndNode}(\node)}
    \State $X \gets \bestd(\text{parent}(\node), \mystate, i)$
    \If{$\text{\cmp[]}(\node, i+1)$}
    \State $\mystate[\node].\cmp[v] \gets \bigcap_{\child} \mystate[\child].\cmp[v] \cap X$
    \EndIf
    \If{$\detpos[](\node,i+1)$}
    \State $\mystate[\node].\detpos[v] \gets \bigcap_{\child} \mystate[\child].\detpos[v] \cap X$
    \EndIf
    \If{$\detneg[](\node,i+1)$}
    \State $\mystate[\node].\detneg[v] \gets \bigcup_{\child} \mystate[\child].\detneg[v] \cap X$
    \EndIf
    \Else
    \State $X \gets \bestd(\text{parent}(\node), \mystate, i)$
    \If{$\text{\cmp[]}(\node, i+1)$}
    \State $\mystate[\node].\cmp[v] \gets \bigcup_{\child} \mystate[\child].\cmp[v] \cap X$
    \EndIf
    \If{$\detpos[](\node,i+1)$}
    \State $\mystate[\node].\detpos[v] \gets \bigcup_{\child} \mystate[\child].\detpos[v] \cap X$
    \EndIf
    \If{$\detneg[](\node,i+1)$}
    \State $\mystate[\node].\detneg[v] \gets \bigcap_{\child} \mystate[\child].\detneg[v] \cap X$
    \EndIf
    \EndIf
    \State $\update(\text{parent}(\node), \mystate, \predi{i}, \vseti{i}, i)$
  \end{algorithmic}
  \caption{\update{}}
  \addtocounter{algorithm}{-1}
\end{algorithm}

  Next, we prove the correctness of \bestd{} and \update{}.
  For the reader's convenience, we restate Algorithms~\ref{alg:bestd} (\bestd{}) and~\ref{alg:update} (\update{}).
  In addition, before diving into the proofs, we first provide some definitions, notation, and basic properties that will be used throughout these proofs.
\begin{restatable}{definition}{defrank}
    Given a predicate expression and an ordering of its \patom[s] $[\predi{1},...,\predi{n}]$, let \node{} be any node in the associated predicate tree.
    The \emph{rank} of \node{} on the $i$th step (denoted $\rank(\node,i)$) is the largest index $j$ such that $j < i$ and \predi{j} is a \patom[] leaf node descendant of \node{}.
    If no such $j$ exists, $\rank(\node,i) = 0$.
    The \emph{final rank} of \node{} (denoted $\frank(\node)$) is the largest index $j$ such that \predi{j} is a \patom[] leaf node descendant of \node{}.
    \label{def:rank}
  \end{restatable}
  \noindent
  Put another way, the rank of \node{} on the $i$th step is the latest step index on which  a \patom[] descendant of \node{} is applied.
  More explicitly, if \node{} has \patom[] \predi{j} as a descendant and $j < i$, and \node{} has no other \patom[] descendants \predi{k} such that $j < k < i$, then $\rank(\node{},i)=j$.
  If \node{} is a leaf node, it considers itself as a descendant, so if \node{} refers to \patom[] $\predi{j}$, $\rank(\node,i)=j$ if $j< i$ and $\rank(\node,i) = 0$ otherwise.
  The final rank of \node{} is the largest index among its \patom[] descendants and can be seen as the last step before \node{} becomes complete.
  Note that $\frank(\node)=\rank(\node,n+1)$, since all \patom[s] must be complete after $n$ steps.

  To reason about the \mystate{} variable whose value continuously changes throughout multiple steps in \bestd{}/\update{}, we denote $\mystate_i$ to be the value of \mystate{} at the beginning of the $i$th step (before any updates).
  We also need a way to refer to \node{}'s complete, positively determinable, and negatively determinable children on the $i$th step.
  For this, we introduce the functions: $\chlc(\node,i)$, $\chlp(\node,i)$, and $\chln(\node,i)$ which return \node{}'s complete, (incomplete) positively determinable, and (incomplete) negatively determinable children on the $i$th step respectively.
  We also use the convention of \ccmp{}, \cpos{}, and \cneg{} to refer to a complete, positively determinable, and negatively determinable child respectively.
  Throughout the section we may refer to leaf nodes which contains \patom[s] as the \patom[s] themselves.
  For example, if \node{} is a leaf node which contains \patom[] \predi{i}, the expression ``\predi{i}'s parent'' may be used to refer to \node{}'s parent.
  As another example, if we talk about all \patom[] descendants of a node, we are referring to the \patom[s] of the leaf node descendants of that node.
  Finally, in calls to \bestd{}, we omit the \mystate{} argument since it refers to the same object every time.

  Next, we present some properties which we use throughout the proofs.

\begin{restatable}{property}{propbasic}
    For any \patom[] \pred{} and valuation sets \vset{} and \vseto{}:
    \begin{align*}
      \pred(\vset \cap \vseto) &= \pred(\vset) \cap \vseto \\
      \pred(\vset \cup \vseto) &= \pred(\vset) \cup \pred(\vseto) \\
      \pred(\vset \setminus \vseto) &= \pred(\vset) \setminus \vseto
    \end{align*}
    \label{prop:basic}
\end{restatable}
\noindent
  If we imagine that each \patom[] \pred{} has a corresponding set \refnodei{\pred} which contains all possible valuations that satisfy \pred{}, then applying $\pred(\vset)$ is akin to taking the intersection of \refnodei{\pred} and \vset: $\pred(\vset) = \refnodei{\pred} \cap \vset$.
  Thus, any properties which pertain to normal intersection of sets also pertain to applied \patom[s].

\begin{restatable}{property}{propcmponce}
    If \node{} is not complete on the $i$th step but complete on the $(i+1)$th step, then for all $j > i$:
    \[
      \mystate_{i+1}[\node].\cmp[v] = \mystate_{j}[\node].\cmp[v]
    \]
    \label{prop:cmp-once}
\end{restatable}
\noindent
  This property simply states that \mystate[\node].\cmp[v] does not change once \node{} is complete.
  This is easy to see since the \cmp[v] attribute of \mystate[\node] is only ever updated once: right after \node{} becomes complete.
  Once a node is complete, all its \patom[] descendants have already been applied, and it can never again become an ancestor of an incomplete \patom[] node.
  Thus, it will never be updated again as an ancestor in \update{}.

\begin{restatable}{property}{propdetneg}
    For any \node{} and step indices $i$ and $j$:
    \[
      \mystate_i[\node].\cmp[v] \cap \mystate_j[\node].\detneg[v] = \varnothing
    \]
    and
    \[
      \mystate_i[\node].\detpos[v] \cap \mystate_j[\node].\detneg[v] = \varnothing
    \]
    \label{prop:detneg}
\end{restatable}
\noindent
  This should make intuitive sense since the \cmp[v]/\detpos[v] attributes for \node{} holds valuations which satisfy the subexpression represented by \node{}, while \detneg[v] hold valuations which do not satisfy the subexpression represented by \node{}.

\begin{restatable}{property}{propdetpos}
  If \node{} is complete on the $i$th step:
    \[
      \mystate_i[\node].\cmp[v] = \mystate_i[\node].\detpos[v]
    \]
    \label{prop:detpos}
\end{restatable}
\noindent
  This is easy to verify with induction.
  For leaf nodes, the \cmp[v] and \detpos[v] attributes are set to the same value, and the lines in \update{} updating \cmp[v] and \detpos[v] for non-leaf nodes become equivalent once they are complete.

  We prove the correctness of \bestd{}/\update{} and \opteval{} with Theorem~\ref{thm:bestd-correct}.
  Theorem~\ref{thm:bestd-correct} states that if \node{} is complete on the $i$th step, then $\mystate_i[\node].\cmp[v]$ must be the set of valuations in $\bestd(\allowbreak \text{parent}(\node), \allowbreak \mystate_{\frank(\node)}, \frank(\node))$ which satisfy the predicate subexpression represented by \node{}.
  \rootnode{} has $n$ \patom[] descendants, so it must be complete after $n$ steps.
  Given that parent(\rootnode) is \nil, according to Algorithm~\ref{alg:bestd}, $\bestd(\text{parent}(\rootnode), \star, \star)$ returns $\{T,F\}^n$.
  Since \rootnode{} represents the entire predicate expression, $\mystate_{n+1}[\rootnode].\cmp[v]$ must return the subset of $\{T,F\}^n$ which satisfies the predicate expression.

\begin{restatable}{theorem}{thmbestdcorrect}
  Given a predicate expression and an ordering of its \patom[s] $[\predi{1},...,\predi{n}]$, and let \node{} be a node in the associated predicate tree.
    Furthermore, let \refnode{} be the set of all valuations which satisfy the predicate subexpression represented by \node{}.
    On the $i$th step, if \node{} is complete:
  \[
    \mystate_i[\node].\cmp[v] = \refnode \cap \bestd(\parent(\node), \frank(\node))
  \]
  \label{thm:bestd-correct}
\end{restatable}

\begin{proof}
  Our proof for Theorem~\ref{thm:bestd-correct} is by strong induction on the height of the tree referred to by \node{}.
  As the base case, let \node{} be a leaf node which refers to \patom[] \predi{j}.
  Since \node{} is complete on the $i$th step, that must mean that $j < i$, and \node{} has been complete since the $(j+1)$th step. Thus:
  \begin{steps}
    \mystate_i[\node].\cmp[v] & = \mystate_{j+1}[\node].\cmp[v] \label{eq:thm4:1} \\
                            & =\predi{j}(\bestd(\parent(\node), j)) \label{eq:thm4:2} \\
                            & =\predi{j}(\bestd(\parent(\node), \frank(\node)))  \label{eq:thm4:3} \\
                            & =\refnode \cap \bestd(\parent(\node), \frank(\node)) \label{eq:thm4:4}
  \end{steps}
  \stepref{eq:thm4:1} is a direct application of \autoref{prop:cmp-once}.
  \stepref{eq:thm4:2} comes directly from Algorithm~\ref{alg:update}.
  \stepref{eq:thm4:3} is simply the definition of $\frank(\node)$.
  Finally, since \node{} is a leaf node pointing to \patom[] \predi{j}, $\refnode = \predi{j}(\{T,F\}^n)$, and by applying Property~\ref{prop:basic}, we get the last step, thereby proving the base case.

  As the inductive hypothesis, assume that the theorem applies for all nodes in the predicate tree with height $k$ or less.
  If \node{} has a height of $k+1$, \node{} must either be an AND or OR node.

  \textbf{If \node{} is an AND node,} and \node{} is complete on the $i$th step, Property~\ref{prop:cmp-once} tells us that $\mystate_i[\node].\cmp[v] = \mystate_{\frank(\node)+1}[\node].\cmp[v]$.
  Furthermore, Line~\ref{line:update:and-cmp} of Algorithm~\ref{alg:update} states:
  \begin{equation}
    \mystate_{\frank(\node)+1}[\node].\cmp[v] = \bigcap_{\cany \in \chl(\node)} \mystate_{\frank(\node)+1}[\cany].\cmp[v] \cap \bestd(\parent(\node), \frank(\node))
    \label{eq:bestd-correct:7}
  \end{equation}
  Since all children of \node{} must have height $k$ or less and must be complete by the time \node{} is complete, they must follow the inductive hypothesis. For any child $\cany$ of \node{}:
  \begin{equation}
    \mystate_{\frank(\node)+1}[\cany].\cmp[v] = \refnodei{\cany} \cap \bestd(\parent(\cany), \frank(\cany))
    \label{eq:bestd-correct:6}
  \end{equation}
  Based on Lines~\ref{line:bestd:and-cmp} and~\ref{line:bestd:and-detneg} of Algorithm~\ref{alg:bestd}:
  \begin{multline*}
    \bestd(\parent(\cany), \frank(\cany)) = \bestd(\parent(\node),  \frank(\cany)) \cap \left(\bigcap_{\ccmp \in \chlc(\node, \frank(\cany))} \mystate_{\frank(\cany)}[\ccmp].\cmp[v]\right) \\
                                            \setminus \left( \bigcup_{\cneg \in \chln(\node, \frank(\cany)) \setminus \{\cany\}} \mystate_{\frank(\cany)}[\cneg].\detneg[v] \right)
  \end{multline*}
  Thus, Equation~\ref{eq:bestd-correct:6} expands to:
  \begin{multline}
    \mystate_{\frank(\node)+1}[\cany].\cmp[v] = \refnodei{\cany} \cap\bestd(\parent(\node),  \frank(\cany))  \cap \left( \bigcap_{\ccmp \in \chlc(\node, \frank(\cany))} \mystate_{\frank(\cany)}[\ccmp].\cmp[v] \right)\\
                                      \setminus \left( \bigcup_{\cneg \in \chln(\node, \frank(\cany)) \setminus \{\cany \}} \mystate_{\frank(\cany)}[\cneg].\detneg[v] \right)
  \label{eq:bestd-correct:8}
  \end{multline}
 Since \node{} is only complete once all its children are complete, we can apply
  Lemma~\ref{lem:subset} (stated below) and assert $\bestd(\parent(\node),  \frank(\node)) \subseteq \bestd(\parent(\node),  \frank(\cany))$.
  Then, we substitute Equation~\ref{eq:bestd-correct:8} back into Equation~\ref{eq:bestd-correct:7} to derive:
  \begin{multline}
      \mystate_{\frank(\node)+1}[\node].\cmp[v] = \bigcap_{\cany \in \chl(\node)} \refnodei{\cany} \cap \bestd(\parent(\node), \frank(\node)) \\
                                      \bigcap_{\ccmp \in \chlc(\node, \frank(\cany))} \mystate_{\frank(\cany)}[\ccmp].\cmp[v] \\
                                      \setminus \left( \bigcup_{\cneg \in \chln(\node, \frank(\cany)) \setminus \{\cany\}} \mystate_{\frank(\cany)}[\cneg].\detneg[v] \right)
                                      \label{eq:bestd-correct:9}
    \end{multline}
We observe that the $\mystate_{\frank(\cany)}[\ccmp].\cmp[v]$ component in the second line of Equation~\ref{eq:bestd-correct:9} can be expanded using Equation~\ref{eq:bestd-correct:8}.
By repeatedly plugging Equation~\ref{eq:bestd-correct:8} into Equation~\ref{eq:bestd-correct:9}, using Lemma~\ref{lem:subset}, applying the identities
\begin{enumerate*}[(1)]
  \item $A \cap (B \setminus C) = (A \cap B) \setminus C$,
  \item $A \cap (B \setminus (C \cup D)) = (A \setminus C) \cap (B \setminus D)$,
  \item $(A \setminus B) \setminus C = A \setminus (B \cup C)$
\end{enumerate*}
for any sets $A$, $B$, $C$, and $D$ (Appendix~\ref{apx:set-alg}), and removing duplicates, we get:
  \begin{multline*}
      \mystate_{\frank(\node)+1}[\node].\cmp[v] = \bestd(\parent(\node), \frank(\node)) \cap \bigcap_{\cany \in \chl(\node)} \refnodei{\cany} \setminus \left( \bigcup_{i_{\cany}} \mystate_{i_{\cany}}[\cany].\detneg[v] \right)
  \end{multline*}
for some set of step indices $i_{\cany}$ for each child $\cany$.
However, Property~\ref{prop:detneg} states that \refnodei{\cany} and $\mystate_i[\cany].\detneg[v]$ must be mutually exclusive for any step index $i$, so this simplifies to:
  \begin{equation*}
      \mystate_{\frank(\node)+1}[\node].\cmp[v] = \bestd(\parent(\node), \frank(\node)) \cap \bigcap_{\cany \in \chl(\node)} \refnodei{\cany}
\end{equation*}
Since the $\refnode = \bigcap_{c \in \chl(\node)} \refnodei{c}$:
  \begin{equation*}
    \mystate_{\frank(\node)+1}[\node].\cmp[v] = \bestd(\parent(\node), \frank(\node)) \cap \refnode
\end{equation*}

\begin{restatable}{lemma}{lemsubset}
  Let $[\predi{1},...,\predi{n}]$ be any ordering of \patom[s] for the given predicate expression, and let \node{} be a non-leaf node in the associated predicate tree.
  For all pairs of \node{}'s \patom[] descendants \predi{i} and \predi{j} for which $i < j$:
    \[
      \bestd(\parent(\node),  j) \subseteq \bestd(\parent(\node),  i)
    \]
    \label{lem:subset}
  \end{restatable}

\textbf{If \node{} is an OR node,} and \node{} is complete on the $i$th step, Property~\ref{prop:cmp-once} tells us again that $\mystate_i[\node].\cmp[v] = \mystate_{\frank(\node)+1}[\node].\cmp[v]$.
  Line~\ref{line:update:or-cmp} of Algorithm~\ref{alg:update} states:
  \begin{equation}
    \mystate_{\frank(\node)+1}[\node].\cmp[v] = \bigcup_{\cany \in \chl(\node)} \mystate_{\frank(\node)+1}[\cany].\cmp[v] \cap \bestd(\parent(\node), \frank(\node))
    \label{eq:bestd-correct:10}
  \end{equation}
  Since all children of \node{} must have height $k$ or less and must be complete by the time \node{} is complete, they must follow the inductive hypothesis. For any child $\cany$ of \node{}:
  \begin{equation*}
    \mystate_{\frank(\node)+1}[\cany].\cmp[v] = \refnodei{\cany} \cap \bestd(\parent(\cany), \frank(\cany))
  \end{equation*}
  Based on Algorithm~\ref{alg:bestd}, this expands to:
  \begin{multline*}
    \mystate_{\frank(\node)+1}[\cany].\cmp[v] = \refnodei{\cany} \cap \bestd(\parent(\node), \frank(\cany))  \\
    \setminus \left( \bigcup_{\ccmp \in \chlc(\node,\frank(\cany))} \mystate_{\frank(\cany)}[\ccmp].\cmp[v] \bigcup_{\cpos \in \chlp(\node,\frank(\cany)) \setminus \{\cany\}} \mystate_{\frank(\cany)}[\cpos].\detpos[v] \right)
  \end{multline*}
  \autoref{prop:detpos} states that for all complete children \ccmp{}, $\mystate[\ccmp].\cmp[v] = \mystate[\ccmp].\detpos[v]$, so this reduces to:
  \begin{multline*}
    \mystate_{\frank(\node)+1}[\cany].\cmp[v] = \refnodei{\cany} \cap \bestd(\parent(\node), \frank(\cany))  \\
    \setminus \left( \bigcup_{\canyp \in (\chlc(\node,\frank(\cany)) \cup \chlp(\node, \frank(\cany)) \setminus \{\cany\})} \mystate_{\frank(\cany)}[\canyp].\detpos[v] \right)
  \end{multline*}
  We can once again apply \autoref{lem:subset} ($\bestd(\parent(\node), \frank(\node)) \subseteq \bestd(\parent(\node), \frank(\cany))$) and substitute back into \autoref{eq:bestd-correct:10} to get:
  \begin{multline*}
    \mystate_{\frank(\node)+1}[\node].\cmp[v] = \bigcup_{\cany \in \chl(\node)} \refnodei{\cany} \cap \bestd(\parent(\node), \frank(\node))  \\
    \setminus \left( \bigcup_{\canyp \in (\chlc(\node,\frank(\cany)) \cup \chlp(\node, \frank(\cany)) \setminus \{\cany\})} \mystate_{\frank(\cany)}[\canyp].\detpos[v] \right)
  \end{multline*}
  We can apply the distributive property to pull the \bestd{} outwards:
  \begin{multline}
    \mystate_{\frank(\node)+1}[\node].\cmp[v] = \bestd(\parent(\node), \frank(\node)) \\
    \cap \left( \bigcup_{\cany \in \chl(\node)} \refnodei{\cany}  \setminus \left( \bigcup_{\canyp \in (\chlc(\node,\frank(\cany)) \cup \chlp(\node, \frank(\cany)) \setminus \{\cany\})} \mystate_{\frank(\cany)}[\canyp].\detpos[v] \right) \right)
    \label{eq:bestd-correct:13}
  \end{multline}

  Next, we introduce Lemma~\ref{lem:orsetred}.
  \begin{restatable}{lemma}{lemorsetred}
    For any sets $A$, $B$, $C$, $D$, $E$, and $F$, if $F \subseteq B$ and $F \cap C = \varnothing$, then:
    \[
      A \cup (B \setminus C) \cup (D \setminus (E \cup F)) = A \cup (B \setminus C) \cup (D \setminus E)
    \]
    \label{lem:orsetred}
  \end{restatable}
  \noindent
  In our case, for two different children \cspc{} and \canyp{}:
  \begin{equation*}
    \begin{aligned}
    A &= \bigcup_{\cany \in \chl(\node) \setminus\{\cspc,\canyp\}} \refnodei{\cany} \setminus \left(\bigcup_{\cpos \in (\chlc(\node, \frank(\cany)) \cup \chlp(\node, \frank(\cany)) \setminus \{\cany\})} \mystate_{\frank(\cany)}[\cpos].\detpos[v] \right) \\
    B &= \refnodei{\cspc} \\
    C &= \bigcup_{\cpos \in (\chlc(\node, \frank(\cspc)) \cup \chlp(\node, \frank(\cspc)) \setminus \{\cspc\})} \mystate_{\frank(\cspc)}[\cpos].\detpos[v] \\
    D &= \refnodei{\canyp} \\
    E &= \bigcup_{\cpos \in (\chlc(\node, \frank(\canyp)) \cup \chlp(\node, \frank(\canyp)) \setminus \{\cspc,\canyp\})} \mystate_{\frank(\canyp)}[\cpos].\detpos[v] \\
    F &= \mystate_{\frank(\canyp)}[\cspc].\detpos[v]
  \end{aligned}
  \end{equation*}
  \autoref{lem:detpos} (shown below) states that $\mystate_i[\cany].\detpos[v] \subseteq \refnodei{\cany}$ for any child \cany{}.
  The $C$ term explicitly does not iterate over \cspc{}, and \autoref{lem:detpos2} (shown below) states $\mystate_i[\cany].\detpos[v] \cap \mystate_i[\cspc].\detpos[v] = \varnothing$ if $\cany$ and \cspc{} are different children.
  Thus, we satisfy the preconditions for \autoref{lem:orsetred}.
  If we apply this lemma to every positively determinable child $\cpos$ in the $E$ term, Equation~\ref{eq:bestd-correct:13} reduces to:
  \begin{multline*}
    \mystate_{\frank(\node)+1}[\node].\cmp[v] = \bestd(\parent(\node), \frank(\node)) \\
    \cap \left( \refnodei{\canyp} \cup \left(\bigcup_{\cany \in \chl(\node) \setminus \{\canyp\}} \refnodei{\cany} \setminus \left(\bigcup_{\cpos \in (\chlc(\node,\frank(\cany)) \cup\chlp(\node,\frank(\cany)) \setminus \{c\})} \mystate_{\frank(\cany)}[\cpos].\detpos[v] \right)\right)\right)
  \end{multline*}
  If we apply this trick to every child $\cany$ of \node{}, this simplifies to:
  \begin{equation*}
      \mystate_{\frank(\node)+1}[\node].\cmp[v] = \bestd(\parent(\node), \frank(\node)) \cap \left(\bigcup_{\cany \in \chl(\node)} \refnodei{\cany} \right)
  \end{equation*}
  For OR nodes, $\refnode = \bigcup_{\cany \in \chl(\node)} \refnodei{\cany}$, the above equation reduces to:
  \begin{equation*}
      \mystate_{\frank(\node)+1}[\node].\cmp[v] = \refnode \cap \bestd(\node, \frank(\node))
  \end{equation*}
\end{proof}

    \begin{restatable}{lemma}{lemdetpos}
  Let $[\predi{1},...,\predi{n}]$ be any ordering of \patom[s] for the given predicate expression, and let \node{} be a node in the associated predicate tree.
  For any step index $i$:
    \[
      \mystate_i[\node].\detpos[v] \subseteq \refnodei{\node}
  \]
    \label{lem:detpos}
  \end{restatable}

    \begin{restatable}{lemma}{lemdetpostwo}
    Let $[\predi{1},...,\predi{n}]$ be any ordering of \patom[s] for the given predicate expression, and let \node{} be a non-leaf node in the associated predicate tree.
    Furthermore, Let \cany{} and \canyp{} be two different children of \node{}.
    If \node{} is an OR node, for all step indices $i$ and $j$:
  \[
    \mystate_i[\cany].\detpos[v] \cap \mystate_j[\canyp].\detpos[v] = \varnothing
  \]
  If \node{} is an AND node, then for all step indices $i$ and $j$:
  \[
    \mystate_i[\cany].\detneg[v] \cap \mystate_j[\canyp].\detneg[v] = \varnothing
  \]
    \label{lem:detpos2}
  \end{restatable}

\section{Caveat in Proof of Theorem~\ref{thm:one-pred}}
\label{sec:caveat}

There is a caveat in the proof of \autoref{thm:one-pred} when $\vseti{i} \cap \vseti{j} = \varnothing$, but \vseti{j} is derived from $\univi{i-1} \cup \{\pred(\vseti{i})\}$.
    Calculating $\vseti{i} \cup \vseti{j}$ requires $\vseti{j}$, which in turn requires $P(\vseti{i})$ which is not available before the $i$th step.
    However, Lemma~\ref{lem:caveat} (stated below) explicitly states that either this situation cannot arise or \vseti{j} does not require $\pred(\vseti{i})$ when being derived.
    Thus, the contradiction is maintained.
    \begin{restatable}{lemma}{lemcaveat}
      Given a predicate expression with $n$ unique \patom[s], let $[(\predi{1},\vseti{1}),...,(\predi{m},\vseti{m})]$ be a solution sequence, for which $m > n$.
      If step indices $i$ and $j$ are the first two times that \patom[] \pred{} appears in such a sequence (i.e., $\predi{i} = \predi{j} = \pred$) and
      $\vseti{i} \cap \vseti{j} = \varnothing$, then either $\vseti{j}$ can be derived directly from \univi{i-1} without using \pred{} as an operator, or \vseti{j} cannot be derived at all from $\univi{i-1} \cup \{\pred(\vseti{i})\}$ without using \pred{} as an operator.
      \label{lem:caveat}
    \end{restatable}


\section{Reduction from \bestd{}/\update{} to {\algo{}}}
\label{apx:algo-reduction}

To show the reduction from \bestd{}/\update{} to \algo{} for DFS \patom[] orderings, we first introduce \autoref{lem:dfs} below.
The lemma basically states that for DFS orderings, all negatively determinable children of AND nodes must also be complete, and all positively determinable determinable children of OR nodes must also be complete.
\begin{restatable}{lemma}{lemdfs}
  Given a predicate expression and a DFS ordering of its \patom[s],
  if node \node{} of the associated predicate tree is the child of an AND node and negatively determinable on the $i$th step, then \node{} must also be complete on the $i$th step.
  Similarly, if \node{} is the child of an OR node and positively determinable on the $i$th step, then \node{} must also be complete on the $i$th step.
  \label{lem:dfs}
\end{restatable}
\noindent
  Lemma~\ref{lem:dfs} allows us to effectively ignore determinability in \bestd{}, since all relevant determinable nodes are also complete.
  Since determinability is never checked, both \detpos[v] and \detneg[v] mappings are unused, and we can ignore the lines updating them in \update{}.
  Algorithm~\ref{alg:bestdp} and~\ref{alg:updatep} on the next page (\bestdp{} and \updatep{} respectively) show what \bestd{} and \update{} would look like after removing these lines.

  \begin{algorithm}[H]
  \begin{algorithmic}[1]
    \Require Predicate tree node \node{}, algorithm state \mystate{}, \patom[] \predi{i}, step index $i$
    \Ensure Optimal operand \vseti{i} to apply \predi{i} to
    \If{$\node  = \nil$}
    \State \Return $\{T,F\}^n$
    \ElsIf{isLeafNode(\node)} \label{line:bestdp:1}
    \State \Return \bestd(\parent(\node), \mystate, \predi{i}, i) \label{line:bestdp:2}
    \ElsIf{\text{isAndNode}(\node)}
    \State $X \gets \bestdp(\text{parent}(\node), \mystate, i)$\label{line:bestdp:and-bestdp}
    \For{$\child \in \text{children}(\node)$}
    \If{$\text{\cmp[]}(\child, i)$}
    \State $X \gets X \cap \mystate[\child].\cmp[v]$ \label{line:bestdp:and-cmp}
    \EndIf
    \EndFor
    \State \Return $X$
    \Else
    \State $X \gets \bestdp(\text{parent}(\node), \mystate, i)$\label{line:bestdp:or-bestdp}
    \State $Y \gets \{\}$
    \For{$\child \in \text{children}(\node)$}
    \If{$\text{\cmp[]}(\child, i)$}
    \State $Y \gets Y \cup \mystate[\child].\cmp[v]$ \label{line:bestdp:or-cmp}
    \EndIf
    \EndFor
    \State \Return $X \setminus Y$
    \EndIf
  \end{algorithmic}
  \caption{\bestdp{}}
  \label{alg:bestdp}
\end{algorithm}

\begin{algorithm}
  \begin{algorithmic}[1]
    \Require Predicate tree node \node{}, algorithm state \mystate{}, \patom[] \predi{i}, set of valuations \vseti{i}, step index $i$
    \If{$\node = \nil$}
    \State \Return
    \ElsIf{$\text{isLeafNode}(\node)$}
    \State $\mystate[\node].\cmp[v] \gets \predi{i}(\vseti{i})$\label{line:updatep:apply}
    \ElsIf{\text{isAndNode}(\node)}
    \If{$\text{\cmp[]}(\node, i+1)$}
    \State $\mystate[\node].\cmp[v] \gets \bigcap_{\child} \mystate[\child].\cmp[v] \cap \bestdp(\text{parent}(\node), \mystate, i)$ \label{line:updatep:and-cmp}
    \EndIf
    \Else
    \If{$\text{\cmp[]}(\node, i+1)$}
    \State $\mystate[\node].\cmp[v] \gets \bigcup_{\child} \mystate[\child].\cmp[v] \cap \bestdp(\text{parent}(\node), \mystate, i)$ \label{line:updatep:or-cmp}
    \EndIf
    \EndIf
    \State $\updatep(\text{parent}(\node), \mystate, \predi{i}, \vseti{i}, i)$
  \end{algorithmic}
  \caption{\updatep{}}
  \label{alg:updatep}
\end{algorithm}

\begin{algorithm}
  \begin{algorithmic}[1]
    \Require Predicate tree node \node{}, filtered valuation set $X$, algorithm state \mystate{}, \patom[] \predi{i}, step index $i$
    \Ensure Equivalent to $\bestdp(\parent(\predi{i}), \mystate, i)$
    \If{$\text{isLeafNode}(\node)$}
    \State \Return $X$
    \ElsIf{\text{isAndNode}(\node)}
    \For{$\child \in \chl(\node)$}
    \If{\cmp[]$(\child, i)$}
    \State $X \gets X \cap \mystate[\child].\cmp[v]$
    \EndIf
    \EndFor
    \State $\child \gets \text{childOfAndAncestorOf}(\node, \predi{i})$
    \State \Return $\bestdf(\child, X, \mystate, \predi{i}, i)$
    \Else
    \State $Y \gets \{\}$
    \For{$\child \in \chl(\node)$}
    \If{\cmp[]$(\child, i)$}
    \State $Y \gets Y \cup \mystate[\child].\cmp[v]$
    \EndIf
    \EndFor
    \State $\child \gets \text{childOfAndAncestorOf}(\node, \predi{i})$
    \State \Return $\bestdf(\child, X \setminus Y, \mystate, \predi{i}, i)$
    \EndIf
  \end{algorithmic}
  \caption{\bestdf{}}
  \label{alg:bestdf}
\end{algorithm}

\begin{algorithm}[!ht]
  \begin{algorithmic}[1]
    \Require Predicate tree node \node{}, filtered valuation set $X$, algorithm state \mystate{}, \patom[] \predi{i}, step index $i$
    \If{$\text{isLeafNode}(\node)$}
    \State $\mystate[\node].\cmp[v] \gets \predi{i}(X)$
    \ElsIf{\text{isAndNode}(\node)}
    \For{$\child \in \chl(\node)$}
    \If{\cmp[]$(\child, i)$}
    \State $X \gets X \cap \mystate[\child].\cmp[v]$
    \EndIf
    \EndFor
    \State $\child \gets \text{childOfAndAncestorOf}(\node, \predi{i})$
    \State $\combinedp(\child, X, \mystate, \predi{i}, i)$
    \If{\cmp[]$(\node, i+1)$}
    \State $\mystate[\node].\cmp[v] \gets \bigcap_{\child} \mystate[\child].\cmp[v] \cap X$
    \EndIf
    \Else
    \State $Y \gets \{\}$
    \For{$\child \in \chl(\node)$}
    \If{\cmp[]$(\child, i)$}
    \State $Y \gets Y \cup \mystate[\child].\cmp[v]$
    \EndIf
    \EndFor
    \State $\child \gets \text{childOfAndAncestorOf}(\node, \predi{i})$
    \State $\combinedp(\child, X \setminus Y, \mystate, \predi{i}, i)$
    \If{$\text{\cmp[]}(\node, i+1)$}
    \State $\mystate[\node].\cmp[v] \gets \bigcup_{\child} \mystate[\child].\cmp[v] \cap X$
    \EndIf
    \EndIf
  \end{algorithmic}
  \caption{\combinedp{}}
  \label{alg:combined}
\end{algorithm}

  Algorithm~\ref{alg:bestdp} works by traversing the predicate tree in a bottom-up fashion.
  Each node calls \bestdp{} on its parent, takes the returned value, modify it based on completed children, and return the filtered value to the child which called it.
  We can instead imagine constructing a dual algorithm which traverses the predicate top-down for any specific \patom[]; this is presented in Algorithm~\ref{alg:bestdf} (next page).
  For any \patom[] \predi{i}:
  \[
    \bestdf(\rootnode, \mystate, \predi{i}, i, \{T,F\}^n) = \bestdp(\mapping[\predi{i}], \mystate, \predi{i}, i)
  \]
  Note that the expression childOfAndAncestorOf(\node, \predi{i}) returns the child of \node{} which is the ancestor of the node referring to \predi{i}.

  Next, we can combine \bestdf{} with \updatep{}.
  This will traverse top-down from the root node to \predi{i} for the \bestdp{} value and then traversing back up to the root node, updating \mystate{} for complete nodes.
  Algorithm~\ref{alg:combined} presents this combined algorithm (\combinedp{}).
  A return value is no longer needed because they are stored in the \cmp[v] attributes of \mystate{}.
  \combinedp{} is used to traverse down and update only a single \patom[] \predi{i}.
  If multiple children are being updated with \combinedp{}, many extra checks for completeness can be avoided, since a DFS traversal guarantees completeness for any traversed child.
  Algorithm~\ref{alg:combineda} shows what it would look like if \combinedp{} calls to separate \patom[s] were combined together into one algorithm.
  We assume that the nodes of the predicate tree already sorted in the ordering given by $[\predi{1},...,\predi{n}]$.
  Also, since everything is done in one pass, instead of a separate \mystate{} variable, we simply use the returned values of each child.

  \begin{algorithm}[H]
  \begin{algorithmic}[1]
    \Require Predicate tree node \node{}, filtered valuation set $X$
    \If{$\text{isLeafNode}(\node)$}
    \State \Return $\predi{i}(X)$
    \ElsIf{\text{isAndNode}(\node)}
    \For{$\child \in \chl(\node)$}
    \State $X \gets X \cap \combineda(\child, X)$
    \EndFor
    \State \Return $X$
    \Else
    \State $Y \gets \{\}$
    \For{$\child \in \chl(\node)$}
    \State $Y \gets Y \cup \combineda(\child, X \setminus Y)$
    \EndFor
    \State \Return $Y$
    \EndIf
  \end{algorithmic}
  \caption{\combineda{}}
  \label{alg:combineda}
\end{algorithm}

  As we can see, there is a 1-to-1 correspondence between Algorithm~\ref{alg:combineda} and \algo{}.
  The only difference is that Algorithm~\ref{alg:combineda} does not feature any early returns.

\section{Supplementary Proofs}
\label{apx:lemmas}

\subsection{Additional Definitions}

We first provide some additional definitions that we will use throughout this section.



\begin{restatable}{definition}{deflineage}
  For any predicate tree node \node{}, we define the \emph{lineage} of \node{} as the sequence of ancestor nodes $[\anci{1},...,\anci{m}]$ in which \anci{1} is the parent of \node{}, and $\anci{i+1}$ is the parent of \anci{i} for all $i \in \{1,...,m-1\}$, and \anci{m} is the root node.
\end{restatable}
\noindent
Simply put, the lineage of \node{} is the sequence of its ancestors starting with the parent up to the root.

\begin{restatable}{definition}{defeval}
  Let \node{} be a node in the predicate tree, and let \refnode{} refer to the set of all valuations which satisfy the subexpression represented by \node{}.
  Evaluating a valuation $v$ with respect to the subexpression represented by \node{}, denoted $\node[v]$, is true if and only if $v \in \refnode$.
\end{restatable}
\noindent
For example, if \node{} a leaf node referring to the \patom[] $\predi{i}$, $\node[v] = \true$ if and only if $v_i = \true$.
In the following text, we may say that $v$ is ``evaluated against the subexpression represented by \node{}'' or ``evaluated against \node{}'' to refer to the same thing.

\begin{restatable}{definition}{defstat}
  Let \node{} be a node in the predicate tree.
  We say that \node{} is \emph{statically true} with respect to a valuation set \vset{} if $\vset \subseteq \refnode$ and \emph{statically false} with respect to \vset{} if $\vset \cap \refnode = \varnothing$.
  If \node{} is either statically true or statically false with respect to a valuation set, it is said to be \emph{static} with respect to that valuation set.
  Otherwise, \node{} is \emph{dynamic} with respect to that valuation set.
\end{restatable}
\noindent
More colloquially, \node{} is statically true (false) with respect to a valuation set \vset{} if every valuation in \vset{} evaluates to true (false) against \node{}, and \node{} is static with respect to valuation set \vset{} if all valuations in \vset{} evaluate to the same value against \node{} (otherwise, the \node{} is dynamic with respect to \vset{}).

\subsection{Reference}

For the reader's convenience, before diving into the proofs, we first list both \bestd{} and \update{} algorithms and all definitions, properties, and lemmas used in this paper.
The reader may refer to this section when reading through the proofs.
In addition, Table~\ref{tab:notation} provides a table of some notation that we think would be helpful for the reader to remember when reading this section.
Good luck.


\begin{algorithm}[H]
  \begin{algorithmic}[1]
    \Require Predicate tree node \node{}, algorithm state \mystate{}, \patom[] \predi{i}, step index $i$
    \Ensure Optimal operand \vseti{i} to apply \predi{i} to
    \If{$\node  = \nil$}
    \State \Return $\{T,F\}^n$
    \ElsIf{\text{isAndNode}(\node)}
    \State $X \gets \bestd(\text{parent}(\node), \mystate, i)$
    \For{$\child \in \text{children}(\node)$}
    \If{$\text{\cmp[]}(\child, i)$}
    \State $X \gets X \cap \mystate[\child].\cmp[v]$
    \ElsIf{$\detneg[](\child, i)$ and isNotAncestor(\child, \predi{i})}
    \State $X \gets X \setminus \mystate[\child].\detneg[v]$
    \EndIf
    \EndFor
    \State \Return $X$
    \Else
    \State $X \gets \bestd(\text{parent}(\node), \mystate, i)$
    \State $Y \gets \{\}$
    \For{$\child \in \text{children}(\node)$}
    \If{$\text{\cmp[]}(\child, i)$}
    \State $Y \gets Y \cup \mystate[\child].\cmp[v]$
  \ElsIf{$\detpos[](\child, i)$ and isNotAncestor(\child, \predi{i})}
    \State $Y \gets Y \cup \mystate[\child].\detpos[v]$
    \EndIf
    \EndFor
    \State \Return $X \setminus Y$
    \EndIf
  \end{algorithmic}
  \renewcommand{\thealgorithm}{\ref{alg:bestd}}
  \caption{\bestd{}}
  \addtocounter{algorithm}{-1}
\end{algorithm}

\begin{algorithm}[H]
  \begin{algorithmic}[1]
    \Require Predicate tree node \node{}, algorithm state \mystate{}, \patom[] \predi{i}, set of valuations \vseti{i}, step index $i$
    \If{$\node = \nil$}
    \State \Return
    \ElsIf{$\text{isLeafNode}(\node)$}
    \State $\mystate[\node].\cmp[v] \gets \predi{i}(\vseti{i})$
    \State $\mystate[\node].\detpos[v] \gets \predi{i}(\vseti{i})$
    \State $\mystate[\node].\detneg[v] \gets \vseti{i} \setminus \predi{i}(\vseti{i})$
    \ElsIf{\text{isAndNode}(\node)}
    \State $X \gets \bestd(\text{parent}(\node), \mystate, i)$
    \If{$\text{\cmp[]}(\node, i+1)$}
    \State $\mystate[\node].\cmp[v] \gets \bigcap_{\child} \mystate[\child].\cmp[v] \cap X$
    \EndIf
    \If{$\detpos[](\node,i+1)$}
    \State $\mystate[\node].\detpos[v] \gets \bigcap_{\child} \mystate[\child].\detpos[v] \cap X$
    \EndIf
    \If{$\detneg[](\node,i+1)$}
    \State $\mystate[\node].\detneg[v] \gets \bigcup_{\child} \mystate[\child].\detneg[v] \cap X$
    \EndIf
    \Else
    \State $X \gets \bestd(\text{parent}(\node), \mystate, i)$
    \If{$\text{\cmp[]}(\node, i+1)$}
    \State $\mystate[\node].\cmp[v] \gets \bigcup_{\child} \mystate[\child].\cmp[v] \cap X$
    \EndIf
    \If{$\detpos[](\node,i+1)$}
    \State $\mystate[\node].\detpos[v] \gets \bigcup_{\child} \mystate[\child].\detpos[v] \cap X$
    \EndIf
    \If{$\detneg[](\node,i+1)$}
    \State $\mystate[\node].\detneg[v] \gets \bigcap_{\child} \mystate[\child].\detneg[v] \cap X$
    \EndIf
    \EndIf
    \State $\update(\text{parent}(\node), \mystate, \predi{i}, \vseti{i}, i)$
  \end{algorithmic}
  \renewcommand{\thealgorithm}{\ref{alg:update}}
  \caption{\update{}}
  \addtocounter{algorithm}{-1}
\end{algorithm}

\begin{table}[H]
  \centering
  \begin{tabular}{|c|l|}
    \hline
    {\bf Symbol} & {\bf Meaning} \\ \hline
    $v_i$ & The value of the $i$th element of valuation $v$ \\ \hline
    $\vipos$ & Refers to the valuation that is the same as $v$ except $v_i=\true$ \\ \hline
    \vatomvi{} & All valuations for which the first $i-1$ elements are the same as $v$ \\ \hline
  \refnode{} & The set of all valuations which satisfy \node{}'s predicate subexpression  \\ \hline
  $\node[v]$ & Evaluating valuation $v$ against the subexpression represented by \node{} \\ \hline
  $\rank(\node, i)$ & The step index of \node{}'s last \patom{} descendant before $i$  \\ \hline
  $\frank(\node)$ & The step index of \node{}'s final \patom{} descendant (i.e., $\rank(\node, n+1)$)  \\ \hline
  $\chlc(\node, i)$ & The set of complete children of \node{} on step $i$ \\ \hline
  $\chlp(\node, i)$ & The set of positively determinable (but not complete) children of \node{} on step $i$ \\ \hline
  $\chln(\node, i)$ & The set of negatively determinable (but not complete) children of \node{} on step $i$ \\ \hline
  \ccmp{} & A complete child \\ \hline
  \cpos{} & A positively determinable child \\ \hline
  \cneg{} & A negatively determinable child \\ \hline
  \mystate[\node].\cmp[v] & Contains valuations which satisfy \node{}'s subexpression \\ \hline
  \mystate[\node].\detpos[v] & Contains a subset of valuations which satisfy \node{}'s subexpression \\ \hline
  \mystate[\node].\detneg[v] & Contains a subset of valuations which do not satisfy \node{}'s subexpression \\ \hline
  \end{tabular}
  \caption{Notation for common symbols and expressions.}
  \label{tab:notation}
\end{table}

\subsubsection{Definitions}

\defval*

\defpredapp*

\defsetop*

\defstep*

\defuniv*

\defvalid*

\defcost*

\defsetform*

\defsetformspace*

\defderive*

\defextuniv*

\defsolseq*

\defcmp*

\defdetpos*

\defdetneg*

\defvalgroup*

\defrank*

\deflineage*

\defeval*

\defstat*

\subsubsection{Properties}

\propcosta*

\propcostb*

\propcostc*

\propbasic*

\propcmponce*

\propdetneg*

\propdetpos*

\begin{restatable}{property}{propsubset}
  Let node \node{} be a node in the predicate tree.
  For any step index $i$:
  \[
    \bestd(\node, i) \subseteq \bestd(\parent(\node), i)
  \]
  \label{prop:subset}
\end{restatable}

\subsubsection{Theorems}

\thmneedpred*

\thmonepred*

\thmoptimal*

\thmevalnode*

\thmbestdcorrect*

\subsubsection{Lemmas}

\lemcriti*

\lemnosplitpi*

\lematomstatic*

\lemdepth*

\lemsubset*

\lemorsetred*

\lemdetpos*

\lemdetpostwo*

\lemcaveat*

\lemdfs*

\begin{restatable}{lemma}{lemgrowtwo}
  Given a predicate expression and an ordering of its \patom[s] $[\predi{1},...,\predi{n}]$, let \node{} be a non-root node in the associated predicate tree.
  If \node{}'s parent is an OR node, \node{} is positively determinable on step $i+1$, and \autoref{lem:subset} holds true up to step $i$, then for all $j \le i$:
  \[
    \mystate_{i+1}[\node].\detpos[v] \supseteq \mystate_j[\node].\detpos[v] \cap \bestd(\parent(\parent(\node)), \rank(\node,i+1))
  \]
  If \node{}'s parent is an AND node, \node{} is negatively determinable on step $i+1$, and \autoref{lem:subset} holds true up to step $i$, then for all $j \le i$:
  \[
    \mystate_{i+1}[\node].\detneg[v] \supseteq \mystate_j[\node].\detneg[v] \cap \bestd(\parent(\parent(\node)), \rank(\node,i+1))
  \]
  \label{lem:growtwo}
\end{restatable}

\begin{restatable}{lemma}{lemgrowtwoone}
  Let $[\predi{1},...,\predi{n}]$ be any ordering of \patom[s] for the given predicate expression, and let \node{} be a non-root node in the associated predicate tree.
  If \node{}'s parent is an OR node, \node{} is positively determinable on step $i+1$, and \autoref{lem:subset} holds true up to step $i$, then:
  \[
    \mystate_{i+1}[\node].\detpos[v] \supseteq \mystate_i[\node].\detpos[v] \cap \bestd(\parent(\parent(\node)), \rank(\node,i+1))
  \]
  If \node{}'s parent is an AND node, \node{} is negatively determinable on step $i+1$, and \autoref{lem:subset} holds true up to step $i$, then:
  \[
    \mystate_{i+1}[\node].\detneg[v] \supseteq \mystate_i[\node].\detneg[v] \cap \bestd(\parent(\parent(\node)), \rank(\node,i+1))
  \]
  \label{lem:growtwoone}
\end{restatable}

\begin{restatable}{lemma}{lemgrow}
  Let $[\predi{1},...,\predi{n}]$ be any ordering of \patom[s] for the given predicate expression, and let \node{} be a node in the associated predicate tree.
  If \node{} is positively determinable on step $i+1$, and \autoref{lem:subset} holds true up to step $i$, then for all $j \le i$:
  \[
    \mystate_{i+1}[\node].\detpos[v] \supseteq \mystate_j[\node].\detpos[v] \cap \bestd(\parent(\node), \rank(\node,i+1))
  \]
  If \node{} is negatively determinable on step $i+1$, and \autoref{lem:subset} holds true up to step $i$, then for all $j \le i$:
  \[
    \mystate_{i+1}[\node].\detneg[v] \supseteq \mystate_j[\node].\detneg[v] \cap \bestd(\parent(\node), \rank(\node,i+1))
  \]
  \label{lem:grow}
\end{restatable}

\begin{restatable}{lemma}{lemgrowone}
  Let $[\predi{1},...,\predi{n}]$ be any ordering of \patom[s] for the given predicate expression, and let \node{} be a node in the associated predicate tree.
  If \node{} is positively determinable on step $i+1$, and \autoref{lem:subset} holds true up to step $i$, then:
  \[
    \mystate_{i+1}[\node].\detpos[v] \supseteq \mystate_i[\node].\detpos[v] \cap \bestd(\parent(\node), \rank(\node,i+1))
  \]
  If \node{} is negatively determinable on step $i+1$, and \autoref{lem:subset} holds true up to step $i$, then:
  \[
    \mystate_{i+1}[\node].\detneg[v] \supseteq \mystate_i[\node].\detneg[v] \cap \bestd(\parent(\node), \rank(\node,i+1))
  \]
  \label{lem:grow-one}
\end{restatable}

\begin{restatable}{lemma}{lemdetpostwoc}
    Let $[\predi{1},...,\predi{n}]$ be any ordering of \patom[s] for the given predicate expression, and let \node{} be a non-leaf node in the associated predicate tree.
    Furthermore, Let \cany{} and \canyp{} be two different children of \node{}.
    If \node{} is an OR node, \autoref{lem:subset} holds true up to step $\jmax$, and \autoref{lem:grow-one} holds true up to \node{}'s children, then for all steps $i$ and $j$ such that $\max(i,j) \le \jmax$:
  \[
    \mystate_i[\cany].\detpos[v] \cap \mystate_j[\canyp].\detpos[v] = \varnothing
  \]
    If \node{} is an AND node, \autoref{lem:subset} holds true up to step $\jmax$, and \autoref{lem:grow-one} holds true up to \node{}'s children, then for all steps $i$ and $j$ such that $\max(i,j) \le \jmax$:
  \[
    \mystate_i[\cany].\detneg[v] \cap \mystate_j[\canyp].\detneg[v] = \varnothing
  \]
  \label{lem:detpos2-cond}
\end{restatable}

\begin{restatable}{lemma}{lemdiv}
  For any set of valuation sets \vsets{}, \patom[] \pred{}, and any $\vset \in \vsets$, let \vseto{} be some valuation set derived from $\vsets \cup \{\pred(\vset)\}$ without using \pred{} as an operator.
  There exists a valuation set $\vseto'$ derived from \vsets{} without using \pred{} as an operator, such that:
  \[
    \vseto \setminus \vset = \vseto' \setminus \vset
  \]
  \label{lem:div}
\end{restatable}

\begin{restatable}{lemma}{lemincok}
  Given a predicate expression and an ordering of its \patom[s] $[\predi{1},...,\predi{n}]$,
  let valuation set $\vseti{i}$ be the $i$th operand generated by \bestd{} and \update{}.
  In addition, let \anc{} be an ancestor in \predi{i}'s lineage.
  For all $v \in \vseti{i}$, there exists a $u \in \vatom[m](v,i)$ such that all incomplete children of \anc{} that are not ancestors of \predi{i} are statically true/false with respect to $\{\uipos,\uineg\}$ if \anc{} is an AND/OR node respectively.
  \label{lem:inc-ok}
\end{restatable}

\begin{restatable}{lemma}{lemincdynamic}
  Given a predicate expression and an ordering of its \patom[s] $[\predi{1},...,\predi{n}]$,
  let valuation set $\vseti{i}$ be the $i$th operand generated by \bestd{} and \update{}.
  In addition, let \anc{} be an ancestor in \predi{i}'s lineage.
  For all $v \in \vseti{i}$, an incomplete child of \anc{} which is not an ancestor of \predi{i} cannot be statically false/true with respect to $\vatom[m](v,i)$ if \anc{} is an AND/OR node respectively.
  \label{lem:inc-dynamic}
\end{restatable}

\begin{restatable}{lemma}{lemdeteffect}
  Given a predicate expression and an ordering of its \patom[s] $[\predi{1},...,\predi{n}]$,
  let valuation set $\vseti{i}$ be the $i$th operand generated by \bestd{} and \update{}.
  In addition, let \node{} be a node in the associated predicate tree.
  For all $v \in \vseti{i}$, if \node{} is not positively determinable on step $i$, then \node{} cannot be statically true with respect to $\vatomvi$.
  Similarly, for all $v \in \vseti{i}$, if \node{} is not negatively determinable on step $i$, then \node{} cannot be statically false with respect to \vatomvi.
  \label{lem:det-effect}
\end{restatable}

\begin{restatable}{lemma}{lemdetdef}
  Given a predicate expression and an ordering of its \patom[s] $[\predi{1},...,\predi{n}]$, let \node{} be a node in the associated predicate tree.
  If \node{} is positively determinable on step $i$, then $\mystate_i[\node].\detpos[v]$ is the set of valuations $v \in \bestd(\node, \rank(\node, i))$ such that every $u \in \vatomvi$ evaluated against \node{} is true:
  \[
    \mystate_i[\node].\detpos[v] = \{v \in \bestd(\parent(\node), \rank(\node,i)) \mid \forall u \in \vatomvi, \node[u] = \true\}
  \]
  Similarly, if \node{} is negatively determinable on step $i$, then $\mystate_i[\node].\detneg[v]$ is the set of valuations $v \in \bestd(\node, \rank(\node, i))$ such that every $u \in \vatomvi$ evaluated against \node{} is false:
  \[
    \mystate_i[\node].\detneg[v] = \{v \in \bestd(\parent(\node), \rank(\node,i)) \mid \forall u \in \vatomvi, \node[u] = \false\}
  \]
  \label{lem:det-def}
\end{restatable}


\clearpage

\subsection{Proofs}

Before we begin, a special note on Lemmas~\ref{lem:subset}, \ref{lem:grow-one}, and \ref{lem:detpos2-cond}.
The co-dependent nature of \bestd{} and \update{} require that these lemmas use each other in their proofs.
However, both Lemmas~\ref{lem:subset} and~\ref{lem:grow-one} are inductive proofs, and
by depending on the inductive hypotheses, we can avoid a cyclical dependency.
To take an example, in the proof for \autoref{lem:subset}, it is assumed the lemma holds up to step $k$.
Then, for the $(k+1)$th step, \autoref{lem:grow-one} is used\footnote{%
\autoref{lem:subset} actually uses \autoref{lem:growtwo}, but this uses \autoref{lem:growtwoone}, which uses \autoref{lem:grow}, which uses \autoref{lem:grow-one}.
However, the assumption is carried all the way through.
} but the conditions to use \autoref{lem:grow-one} only requires that \autoref{lem:subset} hold up to step $k$.
Thus, a cyclical dependency is avoided.
A similar situation occurs with \autoref{lem:detpos2-cond}.

\subsubsection{Proof of Lemma~\ref{lem:crit-i}}
\lemcriti*
\begin{proof}
  We prove
  the lemma by construction with the following algorithm:
  \begin{enumerate}
    \item
      Start with an empty $n$-length valuation $v =(\_,\_,...,\_)$.
    \item
      Starting from \predi{i}'s direct parent, for each ancestor \anc{} in \pred{i}'s lineage:
      \begin{itemize}
        \item For each \patom[] descendant $\predi{j}$ of \anc{} that is not \predi{i}:
          \begin{itemize}
            \item If $\vj$ is unset, set $\vj = \true$ if \anc{} is an AND node and $\vj = \false$ if
              \anc{} is an OR
          node.
          \end{itemize}
      \end{itemize}
    \item
      Return the pair of valuations $(\vipos, \vineg)$.
  \end{enumerate}

  To see why \critipos{} satisfies the overall predicate expression but \critineg{} does
  not, let \anc{} refer to an ancestor node in \predi{i}'s lineage, and let $\ancp{}$ be \anc{}'s child which is either also in \predi{i}'s lineage or is \predi{i} itself.
  Furthermore, let all \patom[s]
  that are descendants of \anc{} but not $\ancp{}$ be called ``other''
  \patom[s].
  If \anc{} is an AND node, setting the elements of $v$ corresponding to ``other'' \patom[] descendants to true ensures that every child other than $\ancp{}$ will evaluate to true.
  As a result, \anc{}'s overall evaluation on $v$ is wholly dependent on whether $\ancp{}$ evaluates to true or false.
  In the case that \anc{} is an OR node, setting the elements of $v$ corresponding to the
  ``other'' \patom[s] to false also ensures that every child other $\ancp{}$ will evaluate to false.
  As a result, \anc{}'s overall evaluation on $v$ once again is wholly dependent on whether $\ancp{}$ evaluates to true or false.
  The above algorithm applies this logic recursively to construct \critipos{}
  and \critineg{}, so
  the sole outcome of \predi{i}
  determines the outcome of the entire predicate expression.
\end{proof}

\subsubsection{Proof of Lemma~\ref{lem:nosplit-pi}}
\lemnosplitpi*
\begin{proof}
  We give a proof by case analysis. Let $X$ be a valuation set which has either
  both \vipos{} and \vineg{} or neither \vipos{} nor \vineg{}. If we apply a \patom[] \predi{j} which is not \predi{i} to $X$,  the result $\predi{j}(X)$ will have both
  $(\vipos,\vineg)$ if $\vj =\true$ and neither valuations if $\vj =\false$.
  Let $Y$ be another valuation set which also has either
  both or neither of \vipos{} and \vineg{}. We show for every set operation
  between the two valuation sets, the result will also have either have both or
  neither valuations. Table~\ref{tab:case-both-vi} shows the results.
  The table elements reflect whether both or neither valuations are present in the results of set
  operations between $X$ and $Y$.
  Thus, if every valuation set in \vsets{} contains either both or neither of \vipos{} and \vineg{}, then all derived sets which do not use \predi{i} as an operator must also contain either both or neither of \vipos{} and \vineg{}.
\end{proof}

\begin{table}[ht]
  \centering
  \begin{tabular}{|c|c|c|c|c|}
    \hline
    \multirow{2}{*}{\bf Set Op} & \multicolumn{2}{c|}{\bf Both in $X$} &
    \multicolumn{2}{c|}{\bf Neither in $X$} \\ \cline{2-5}
                                   & {\bf Both in $Y$} & {\bf Neither in $Y$} &
    {\bf Both in $Y$} & {\bf Neither in $Y$} \\ \hline
    \hline
    $X \cup Y$ & Both & Both & Both & Neither \\ \hline
    $X \cap Y$ & Both & Neither & Neither & Neither \\ \hline
    $X \setminus Y$ & Neither & Both & Neither & Neither \\ \hline
    $Y \setminus X$ & Neither & Neither & Both & Neither \\ \hline
  \end{tabular}
  \caption{Outcomes of set operations between valuation sets that have either both
  or neither \vipos{} and \vineg{}.}
  \label{tab:case-both-vi}
\end{table}

\subsubsection{Proof of Lemma~\ref{lem:atom-static}}
\lematomstatic*
\begin{proof}
  We first introduce the concept of \emph{static} and \emph{dynamic} nodes.
  \defstat*
  More intuitively, if every valuation within a valuation set would evaluate to true when evaluated against \node{}, then \node{} is considered statically true with respect to that valuation set.
  Similarly, if every valuation within a valuation set would evaluate to false, then \node{} is considered statically false with respect to that valuation set.
  If different valuations in a valuation set would result in different values when evaluated against \node{}, then \node{} is considered dynamic with respect to that valuation set.

  Based on this definition, we present the following lemma about incomplete nodes.
  \lemincok*

  We are now ready to prove Lemma~\ref{lem:atom-static}.
  Assume to the contrary that there exists a valuation $v \in \vseti{i}$ such that for all $u \in \vatom[m](v,i)$, the root node is static with respect to $\{\uipos, \uineg\}$.
  By definition, \predi{i} must be dynamic with respect to $\{\uipos, \uineg\}$.
  Therefore, somewhere along \predi{i}'s lineage, the nodes change from being dynamic to static with respect to $\{\uipos, \uineg\}$.
  Let \ancp{} be the ancestor in \predi{i}'s lineage closest to the root such that \ancp{} is dynamic with respect to $\{\uipos, \uineg\}$ while \ancp{}'s parent \anc{} is static with respect to $\{\uipos,\uineg\}$.
  We show that \anc{} cannot be static, leading to a contradiction.

  If \anc{} is an AND node, based on Algorithm~\ref{alg:bestd}, we see that $\bestd(\anc, i)$ returns only valuations which satisfy every one of \anc{}'s complete children on step $i$.
  Based on Property~\ref{prop:subset} (shown below), \vseti{i} must be a subset of $\bestd(\anc, i)$, so all complete children of \anc{} must be statically true with respect to \vseti{i}.
  Corollary~\ref{cor:vatom} states that $\vatom[m](v,i) \subseteq \vseti{i}$ and $\{\uipos, \uineg\} \subseteq \vatom[m](v,i)$, so all of \anc{}'s complete children must also be statically true with respect to $\{\uipos,\uineg\}$ as well.
  Based on Lemma~\ref{lem:inc-ok}, for all $v \in \vseti{i}$, there must exist a $u \in \vatom[m](v,i)$ such that all incomplete children of \anc{} except \ancp{} are statically true with respect to $\{\uipos, \uineg\}$.
  In other words, for this $u$, all children except \ancp{} are statically true with respect to $\{\uipos, \uineg\}$.
  Since we only have positive \patom[] expressions, \uipos{} must evaluate to true against \ancp{}
  and therefore evaluates to true against \anc{} itself as well.
  On the other hand, \uineg{} must evaluate to false against \ancp{} and evaluates to false against \anc{} as well.
  Therefore, \anc{} is dynamic with respect to $\{\uipos,\uineg\}$, and a contradiction is reached.

  The reasoning is basically the same if \anc{} is an OR node.
  Based on Algorithm~\ref{alg:bestd}, we see that $\bestd(\anc, i)$ returns only valuations which do not satisfy any one of \anc{}'s complete children on step $i$,
  meaning all of \anc{}'s complete children are statically false with respect to $\bestd(\anc,i)$.
  Based on Property~\ref{prop:subset}, \vseti{i} must be a subset of $\bestd(\anc, i)$, and Corollary~\ref{cor:vatom} states that $\vatom[m](v,i) \subseteq \vseti{i}$.
  Since, $\{\uipos, \uineg\} \subseteq \vatom[m](v,i)$, all of \anc{}'s complete children must also be statically false with respect to $\{\uipos,\uineg\}$.
  Based on Lemma~\ref{lem:inc-ok}, for all $v \in \vseti{i}$, there must exist a $u \in \vatom[m](v,i)$ such that all incomplete children of \anc{} except \ancp{} are statically false with respect to $\{\uipos, \uineg\}$.
  In other words, for this $u$, all children except \ancp{} are statically false with respect to $\{\uipos, \uineg\}$.
  Since we only have positive \patom[] expressions, \uipos{} must evaluate to true against \ancp{} and also against \anc{}.
  On the other hand, \uineg{} evaluates to false against \ancp{} and must also evaluate to false against \anc{}.
  Therefore, \anc{} is dynamic with respect to $\{\uipos,\uineg\}$ and a contradiction is reached.
\end{proof}

\propsubset*
This property is easy to see since $\bestd(\node,i)$ only ever removes sets from $\bestd(\parent(\node),i)$ (by intersection or sub subtraction) and never adds any sets.

\subsubsection{Proof of Lemma~\ref{lem:depth2}}
\lemdepth*
\begin{proof}
  We prove this by case analysis on the type of \node{}.
  \begin{enumerate}
    \item If \node{} is a leaf node, the conditions for being complete are the same as the conditions for being positively and negatively determinable, so the lemma is true.
    \item If \node{} is the child of an AND node, it must be an OR node (leaf nodes go to the first case).
      As an OR node, if \node{} is negatively determinable, each child of \node{} must also be negatively determinable by definition.
      However, the entire predicate tree has a maximum depth of 2, so all of \node{}'s children must be leaf nodes.
      All leaf nodes which are negatively determinable are also complete, so \node{} must also be complete.
    \item
      If \node{} is the child of an OR node and positively determinable, the reasoning is the same as when \node{} is the child of AND node and negatively determinable.
      As an AND node, if \node{} is positively determinable, each child of \node{} must also be positively determinable by definition.
      However, the entire predicate tree has a maximum depth of 2, so all of \node{}'s children must be leaf nodes.
      All leaf nodes which are positively determinable are also complete, so \node{} must also be complete.
  \end{enumerate}
\end{proof}

\subsubsection{Proof of Lemma~\ref{lem:subset}}
\lemsubset*
\begin{proof}
  We prove this by performing induction over the maximum value that $j$ can be (i.e., $\jmax \ge j$).
  The base case occurs when $\jmax = 2$.
  Since both \predi{i} an \predi{j} are descendants of \node{}, there can be no positively or negatively determinable children along \node{}'s ancestors.
  Thus, $\bestd(\parent(\node), 1) = \bestd(\parent(\node), 2) = \{T,F\}^n$, and the lemma is satisfied.

  As the inductive hypothesis, assume the lemma holds for $\jmax = k$.
  For the inductive step, $\jmax = k+1$.
  Let us assume that $\predi{j} = \predi{k+1}$ is a descendant of \node{}. Otherwise, the lemma is trivially true based on the inductive hypothesis.
  Furthermore, assume that $i = \rank(\node,j)$.
  Due to the monotonicity of $\subseteq$, if the lemma holds for $i = \rank(\node,j)$, then the lemma holds for all $i < j$.

  We prove the case for $\jmax = k+1$ by induction over the distance of \node{} from the root node (henceforth referred to as the ``inner'' induction).
  For the base case of the inner induction, \node{} is the root node.
  In this case, $\bestd(\parent(\node), i) = \bestd(\parent(\node), j) = \{T,F\}^n$ for all $i$ and $j$, so the lemma is satisfied.
  Next, we assume as the inner inductive hypothesis that the lemma holds for when \node{} is $k$ distance away from the root node.
  For the inner inductive step, we assume \node{} is $k+1$ distance away from the root node.

  {\bf If \node{}'s parent is an AND node,}
  based on Algorithm~\ref{alg:bestd}, we can expand $\bestd(\parent(\node), j)$ as:
  \begin{multline}
    \bestd(\parent(\node), j) = \bestd(\parent(\parent(\node)), j) \\
    \cap \left(\left(\bigcap_{\ccmp \in \chlc(\parent(\node), j)} \mystate_j[\ccmp].\cmp[v] \right) \setminus \left( \bigcup_{\cneg \in \chln(\parent(\node), j) \setminus \{ \node \}} \mystate_j[\cneg].\detneg[v] \right)\right)
    \label{eq:lemsubset:2}
  \end{multline}
  For any sets $A$, $B$, and $C$ it is trivially true that $A \supseteq A \cap (B \setminus C)$.
  Thus:
  \begin{multline*}
    \bestd(\parent(\node), i) \supseteq \bestd(\parent(\node), i) \\
    \cap \left(\bigcap_{\ccmp \in \chlc(\parent(\node), j)} \mystate_j[\ccmp].\cmp[v] \setminus \mystate_j[\ccmp].\detneg[v] \right) \\
  \setminus \left( \bigcup_{\cneg \in \chln(\parent(\node), j) \setminus \{ \node \}} \mystate_j[\cneg].\detneg[v] \right)
  \end{multline*}
  We show that the right-hand side of this equation is a superset of $\bestd(\parent(\node), j)$, completing the proof for the AND case.
  We start by expanding $\bestd(\parent(\node), i)$ according to Algorithm~\ref{alg:bestd}:
  \begin{multline*}
    \bestd(\parent(\node), i) \supseteq \bestd(\parent(\parent(\node)), i) \\
    \cap \left(\left(\bigcap_{\ccmp \in \chlc(\parent(\node), i)} \mystate_i[\ccmp].\cmp[v] \right) \setminus \left( \bigcup_{\cneg \in \chln(\parent(\node), i) \setminus \{ \node \}} \mystate_i[\cneg].\detneg[v] \right)\right) \\
    \cap \left(\bigcap_{\ccmp \in \chlc(\parent(\node), j)} \mystate_j[\ccmp].\cmp[v] \setminus \mystate_j[\ccmp].\detneg[v] \right) \\
  \setminus \left( \bigcup_{\cneg \in \chln(\parent(\node), j) \setminus \{ \node \}} \mystate_j[\cneg].\detneg[v] \right)
  \end{multline*}
  We can use the identity $(A \setminus B) \cap (C \setminus D) = (A \cap C) \setminus (B \cup D)$ for all sets $A$, $B$, $C$, and $D$ (\autoref{apx:set-alg}) to rearrange:
  \begin{multline*}
    \bestd(\parent(\node), i) \supseteq \bestd(\parent(\parent(\node)), i) \\
    \cap \left(\bigcap_{\ccmp \in \chlc(\parent(\node), i)} \mystate_i[\ccmp].\cmp[v] \bigcap_{\ccmpp \in \chlc(\parent(\node), j)} \mystate_j[\ccmpp].\cmp[v] \right)\\
    \setminus \left(\rule{0pt}{2em}\right.\left(  \bigcup_{\cneg \in \chln(\parent(\node), i) \setminus \{\node\}} \mystate_i[\cneg].\detneg[v]  \right)\\
    \cup \left(\bigcup_{\cnegp \in (\chln(\parent(\node), j) \cup \chlc(\parent(\node), j) \setminus \{\node\})} \mystate_j[\cnegp].\detneg[v] \right)\left.\rule{0pt}{2em}\right)
  \end{multline*}
  Since $j > i$, the set of complete children at step $j$ must be a superset of the complete children at step $i$ ($\chlc(\parent(\node), j) \supseteq \chlc(\parent(\node), i)$).
  Furthermore, based on the Property~\ref{prop:cmp-once}, the \cmp[v] values of these complete children do not change, so the second line can be simplified:
  \begin{multline}
    \bestd(\parent(\node), i) \supseteq \bestd(\parent(\parent(\node)), i) \cap \left(\bigcap_{\ccmp \in \chlc(\parent(\node), j)} \mystate_j[\ccmp].\cmp[v] \right)\\
    \setminus \left(\rule{0pt}{2em}\right.\left(  \bigcup_{\cneg \in \chln(\parent(\node), i) \setminus \{\node\}} \mystate_i[\cneg].\detneg[v]  \right)\\
    \cup \left(\bigcup_{\cnegp \in (\chln(\parent(\node), j) \cup \chlc(\parent(\node), j) \setminus \{\node\})} \mystate_j[\cnegp].\detneg[v] \right)\left.\rule{0pt}{2em}\right)
      \label{eq:lemsubset:1}
  \end{multline}
  Next, the set of negatively determinable children at step $j$ must be also superset of the negatively determinable children at step $i$.
  Some of these children may have graduated to also becoming complete, but the union in the third line
   is also over complete children. 
   For each of these negatively determinable children, we apply \autoref{lem:growtwo}\footnote{%
    To apply \autoref{lem:growtwo} to \autoref{eq:lemsubset:5}, \autoref{lem:subset} must hold true up until $(j-1)$th step.
    The inductive hypothesis applies in this case, so there is no cyclical dependency.
  } (shown below):
  \begin{equation}
    \mystate_j[\cneg].\detneg[v] \supseteq \mystate_i[\cneg].\detneg[v] \cap \bestd(\parent(\parent(\node)), \rank(\cneg, j))
    \label{eq:lemsubset:5}
  \end{equation}
  By the inner inductive hypothesis, $\bestd(\parent(\parent(\node)), i) \subseteq \bestd(\parent(\parent(\node)), \rank(\cneg, j))$.
  Using the identity that $A \setminus B = A \setminus (B \cap A)$ for any sets $A$ and $B$ (\autoref{apx:set-alg}), we move a copy of the $\bestd(\parent(\parent(\node)), i)$ term in \autoref{eq:lemsubset:1} inward and intersect with $\mystate_i[\cneg].\detneg[v]$ to reduce to:
  \begin{multline*}
    \bestd(\parent(\node), i) \supseteq \bestd(\parent(\parent(\node)), i) \\
    \cap \left(\bigcap_{\ccmp \in \chlc(\parent(\node), j)} \mystate_j[\ccmp].\cmp[v] \right)\\
    \setminus \left(  \bigcup_{\cneg \in (\chln(\parent(\node), j) \cup \chlc(\parent(\node), j) \setminus \{\node\})} \mystate_j[\cneg].\detneg[v] \right)
      \label{eq:lemsubset:1}
  \end{multline*}
  Using Property~\ref{prop:detneg}'s mutual exclusivity once again, we can get rid of the \detneg[v] values of complete children:
  \begin{multline*}
    \bestd(\parent(\node), i) \supseteq \bestd(\parent(\parent(\node)), i) \\
    \cap \left(\bigcap_{\ccmp \in \chlc(\parent(\node), j)} \mystate_j[\ccmp].\cmp[v] \right)\\
    \setminus \left(  \bigcup_{\cneg \in \chln(\parent(\node), j) \setminus \{\node\}} \mystate_j[\cneg].\detneg[v] \right)
  \end{multline*}
  At this point, we can take the intersection of $\bestd(\parent(\parent(\node)), j)$ to both sides and apply the inner inductive hypothesis which states that:
  \[
    \bestd(\parent(\parent(\node)), j) \subseteq \bestd(\parent(\parent(\node)), i)
  \]
  We end up with:
  \begin{multline*}
    \bestd(\parent(\node), i) \cap \bestd(\parent(\parent(\node)), j) \supseteq \bestd(\parent(\parent(\node)), j) \\
    \cap \left(\bigcap_{\ccmp \in \chlc(\parent(\node), j)} \mystate_j[\ccmp].\cmp[v] \right)\\
    \setminus \left(  \bigcup_{\cneg \in \chln(\parent(\node), j) \setminus \{\node\}} \mystate_j[\cneg].\detneg[v] \right)
  \end{multline*}
  Note that the right-hand side of this equation is equivalent to \autoref{eq:lemsubset:2}.
  Thus:
  \[
    \bestd(\parent(\node), i) \supseteq \bestd(\parent(\node), j)
  \]

  \lemgrowtwo*

  {\bf If \node{}'s parent is an OR node,}
  based on Algorithm~\ref{alg:bestd}, we can expand $\bestd(\parent(\node), j)$ as:
  \begin{multline*}
    \bestd(\parent(\node), j) = \bestd(\parent(\parent(\node)), j) \\
    \setminus \left(\left(\bigcup_{\ccmp \in \chlc(\parent(\node), j)} \mystate_j[\ccmp].\cmp[v] \right) \cup \left( \bigcup_{\cpos \in \chlp(\parent(\node), j) \setminus \{ \node \}} \mystate_j[\cpos].\detpos[v] \right)\right)
  \end{multline*}
  \autoref{prop:detpos} allows us to convert the \cmp[v] values of complete children into \detpos[v], so this can be rewritten as:
  \begin{multline}
    \bestd(\parent(\node), j) = \bestd(\parent(\parent(\node)), j) \\
    \setminus \left(\bigcup_{\cany \in (\chlc(\parent(\node), j) \cup \chlp(\parent(\node), j) \setminus \{\node\})} \mystate_j[\cany].\detpos[v] \right)
    \label{eq:lemsubset:3}
  \end{multline}
  For any sets $A$ and $B$ it is trivially true that $A \supseteq A \setminus B$.
  Thus:
  \begin{multline*}
    \bestd(\parent(\node), i) \supseteq \bestd(\parent(\node), i) \\
    \setminus \left( \bigcup_{\cany \in (\chlc(\parent(\node), j) \cup \chlp(\parent(\node), j) \setminus \{\node\})} \mystate_j[\cany].\detpos[v]  \right)
  \end{multline*}
  We once again show that the right-hand side of this equation is a superset of $\bestd(\parent(\node), j)$, completing the proof for the OR case.
  We start by expanding $\bestd(\parent(\node), i)$ according to Algorithm~\ref{alg:bestd}:
  \begin{multline*}
    \bestd(\parent(\node), i) \supseteq \bestd(\parent(\parent(\node)), i) \\
    \setminus \left(\rule{0pt}{2em}\right. \left( \bigcup_{\cany \in (\chlc(\parent(\node), j) \cup \chlp(\parent(\node), j) \setminus \{\node\})} \mystate_j[\cany].\detpos[v]  \right) \\
  \cup \left( \bigcup_{\ccmp \in \chlc(\parent(\node), i)} \mystate_i[\ccmp].\cmp[v] \bigcup_{\cpos \in \chlp(\parent(\node), i) \setminus \{\node\}} \mystate_i[\cpos].\detpos[v] \right) \left.\rule{0pt}{2em}\right)
  \end{multline*}
  We use \autoref{prop:detpos} again to convert all \cmp[v] values to \detneg[v]:
  \begin{multline}
    \bestd(\parent(\node), i) \supseteq \bestd(\parent(\parent(\node)), i) \\
    \setminus \left(\rule{0pt}{2em}\right. \left( \bigcup_{\cany \in (\chlc(\parent(\node), j) \cup \chlp(\parent(\node), j) \setminus \{\node\})} \mystate_j[\cany].\detpos[v]  \right) \\
    \cup \left( \bigcup_{\canyp \in (\chlc(\parent(\node), i) \cup \chlp(\parent(\node), i) \setminus \{\node \})} \mystate_i[\canyp].\detpos[v] \right) \left.\rule{0pt}{2em}\right)
    \label{eq:lemsubset:4}
  \end{multline}
  Since $j > i$, the set of complete and positively determinable children at step $j$ must be a superset of the complete and positively determinable children at step $i$.
  Furthermore, \autoref{lem:growtwo} states if \cany{} is a positively determinable child of an OR node:
  \[
    \mystate_j[\cany].\detpos[v] \supseteq  \mystate_i[\cany].\detpos[v] \cap \bestd(\parent(\parent(\node)), \rank(\cany, j))
  \]
  Using the inner inductive hypothesis once again, \autoref{eq:lemsubset:4} simplifies to:
  \begin{multline*}
    \bestd(\parent(\node), i) \supseteq \bestd(\parent(\parent(\node)), i) \\
    \setminus \left( \bigcup_{\cany \in (\chlc(\parent(\node), j) \cup \chlp(\parent(\node), j) \setminus \{\node\})} \mystate_j[\cany].\detpos[v]  \right)
  \end{multline*}
  At this point, we can take the intersection of $\bestd(\parent(\parent(\node)), j)$ to both sides and apply the inner inductive hypothesis which states that:
  \[
    \bestd(\parent(\parent(\node)), j) \subseteq \bestd(\parent(\parent(\node)), i)
  \]
  We end up with:
  \begin{multline*}
    \bestd(\parent(\node), i) \cap \bestd(\parent(\parent(\node)), j)\supseteq \bestd(\parent(\parent(\node)), j) \\
    \setminus \left( \bigcup_{\cany \in (\chlc(\parent(\node), j) \cup \chlp(\parent(\node), j) \setminus \{\node\})} \mystate_j[\cany].\detpos[v]  \right)
  \end{multline*}
  Note that the right-hand side of this equation is equivalent to \autoref{eq:lemsubset:3}.
  Thus:
  \[
    \bestd(\parent(\node), i) \supseteq \bestd(\parent(\node), j)
  \]
\end{proof}

\subsubsection{Proof of Lemma~\ref{lem:orsetred}}
\lemorsetred*
\begin{proof}
    Here, $A$ is a common term on the outside, so it is sufficient to prove $(B
    \setminus C) \cup (D \setminus (E \cup F)) = (B \setminus C) \cup (D
    \setminus E)$.
    \begin{align*}
      (B \setminus C) \cup (D \setminus (E \cup F)) &= ((B \setminus C) \cup D) \setminus ((E \cup F) \setminus (B \setminus C)) \\
                      &= ((B \setminus C) \cup D) \setminus ((E \setminus (B \setminus C)) \cup ( F \setminus (B \setminus C))) \\
                      &= ((B \setminus C) \cup D) \setminus ((E \setminus (B \setminus C)) \cup ((F \cap C) \cup (F \setminus B))) \\
                      &= ((B \setminus C) \cup D) \setminus ((E \setminus (B \setminus C)) \cup (\varnothing \cup \varnothing)) \\
                      &= ((B \setminus C) \cup D) \setminus (E \setminus (B \setminus C)) \\
                      &= (B \setminus C) \cup (D \setminus E)
    \end{align*}
    The first, second, and third steps use the following identities respectively (Appendix~\ref{apx:set-alg}).
    For any sets $A$, $B$, and $C$:
    \begin{align*}
      A \cup (B \setminus C) &= (A \cup B) \setminus (C \setminus A) & (A \cup B) \setminus C &= (A\setminus C) \cup (B \setminus C) \\
      A \setminus (B \setminus C) &= (A \setminus B) \cup (A \cap C)
    \end{align*}
    The fourth step uses the assumptions $B \supseteq F$ and $F \cap C = \varnothing$.
    The last step uses the same identity as the first step in reverse.
\end{proof}

\subsubsection{Proof of Lemma~\ref{lem:detpos}}
\lemdetpos*
\begin{proof}
  We prove this by strong induction on the height of the tree referred to by \node{}.
  As the base case, let \node{} be a leaf node which refers to \patom[] \predi{i}.
  The variable $\mystate[\node].\detpos[v]$ is initially set to the empty set, and until \predi{i} is positively determinable, the lemma is trivially true.
  Once \node{} is positively determinable, $\mystate[\node].\detpos[v]$ is updated with $\predi{i}(\vseti{i})$.
  By definition $\refnode = \predi{i}(\{T,F\}^n)$, and the largest \vseti{i} can be is $\{T,F\}^n$.
  As a result, the lemma holds even when \node{} is positively determinable.

  As the inductive hypothesis, assume that the lemma holds for all predicate tree nodes of height $k$ or less.
  If \node{} has a height of $k+1$, \node{} must either be an AND or OR node.
  Once again if \node{} is not positively determinable, then the lemma holds trivially.
  Thus, we assume \node{} is positively determinable on step $i$.

  {\bf If \node{} is an AND node,} based on the definition of \refnode{}:
  \[
    \refnode = \bigcap_{\cany \in \chl(\node)} \refnodei{\cany}
  \]
  By the inductive hypothesis, for any child $\cany{}$ of \node{} and any step index $i$, $\refnodei{\cany} \supseteq \mystate_i[\cany].\detpos[v]$, so:
  \begin{align*}
    \refnode &\supseteq \bigcap_{\cany \in \chl(\node)} \mystate_i[\cany].\detpos[v] \\
             &\supseteq \bigcap_{\cany \in \chl(\node)} \mystate_i[\cany].\detpos[v] \cap \bestd(\parent(\node),\rank(\node, i))
  \end{align*}
  Recall that $\rank(\node, i)$ is the largest index among \node's \patom[] descendants that is smaller than $i$; in other words, $\rank(\node,i)$ is the last time \node{} was updated before step $i$.
  By definition if \node{} is an AND node and positively determinable, all of it children must also be positively determinable.
  Then, by Algorithm~\ref{alg:update}:
  \[
    \mystate_i[\node].\detpos[v] = \bigcap_{\cany \in \chl(\node)} \mystate_i[\cany].\detpos[v] \cap \bestd(\parent(\node), \rank(\node, i))
  \]
  Thus, $\refnode \supseteq \mystate_i[\node].\detpos[v]$ for any step $i$.

  {\bf If \node{} is an OR node,} based on the definition of \refnode{}:
  \[
    \refnode = \bigcup_{\cany \in \chl(\node)} \refnodei{\cany}
  \]
  By the inductive hypothesis, for any child $\cany{}$ of \node{} and any step index $i$, $\refnodei{\cany} \supseteq \mystate_i[\cany].\detpos[v]$, so:
  \begin{align*}
    \refnode &\supseteq \bigcup_{\cany \in \chl(\node)} \mystate_i[\cany].\detpos[v] \\
             &\supseteq \bigcup_{\cpos \in \chlp(\node, i)} \mystate_i[\cpos].\detpos[v] \\
             &\supseteq \bigcup_{\cpos \in \chlp(\node, i)} \mystate_i[\cpos].\detpos[v] \cap \bestd(\parent(\node), \rank(\node, i))
  \end{align*}
  The second step is valid because the set of positively determinable children will be at most the set of all children.
  By Algorithm~\ref{alg:update}:
  \[
    \mystate_i[\node].\detpos[v] = \bigcup_{\cpos \in \chlp(\node, i)} \mystate[\cpos].\detpos[v] \cap \bestd(\parent(\node), \rank(\node, i))
  \]
  Thus, $\refnode \supseteq \mystate_i[\node].\detpos[v]$ for any step $i$.
\end{proof}

\subsubsection{Proof of Lemma~\ref{lem:detpos2}}
\lemdetpostwo*
\begin{proof}
  This is just the non-conditional form of \autoref{lem:detpos2-cond}.
  The proof for  \autoref{lem:detpos2-cond} depends on \autoref{lem:subset} and \autoref{lem:grow-one}.
  Since we have already shown \autoref{lem:subset} to be true for any node in the predicate tree, \autoref{lem:grow-one} must apply to predicate tree nodes of all height, and  \autoref{lem:detpos2-cond} must apply to all step indices as well, giving us this non-conditional form.
\end{proof}

\subsubsection{Proof of Lemma~\ref{lem:caveat}}
\lemcaveat*
\begin{proof}
  We prove Lemma~\ref{lem:caveat} by showing that if $\vseti{j} \cap \vseti{i} = \varnothing$,
  all valuation sets \vseti{j} which can be derived from $\univi{i-1} \cup \{\pred(\vseti{i})\}$ without using \pred{} as an operator can also be constructed from $\univi{i-1}$ without using \pred{} as an operator.
  We do this by applying the identity $A = (A \setminus B) \cup (A \cap B)$ for any sets $A$ and $B$ (Appendix~\ref{apx:set-alg}):
  \begin{align*}
    \vseti{j} & = (\vseti{j} \setminus \vseti{i}) \cup (\vseti{j} \cap \vseti{i}) \\
              & = (\vseti{j} \setminus \vseti{i}) \cup \varnothing \\
              & = \vseti{j} \setminus \vseti{i}
  \end{align*}
  The second step comes from the precondition.
  Lemma~\ref{lem:div} (shown below) states that there must exist some $\vseti{j}'$ derived from \univi{i-1} without using \pred{} as an operator such that $\vseti{j} \setminus \vseti{i} = \vseti{j}' \setminus \vseti{i}$.
  Since $\vseti{j} = \vseti{j}' \setminus \vseti{i}$, \vseti{j} can also be derived from \univi{i-1} without using \pred{} as an operator.
\end{proof}

\lemdiv*

\subsubsection{Proof of Lemma~\ref{lem:dfs}}
\lemdfs*
\begin{proof}
  We prove this by case analysis on the type of \node{}.
  \begin{enumerate}
    \item If \node{} is a leaf node, the conditions for being complete are the same as the conditions for being positively and negatively determinable, so the lemma is true.
    \item If \node{} is the child of an AND node, it must be an OR node (leaf nodes go to the first case).
      As an OR node, if \node{} is negatively determinable, each child of \node{} must also be negatively determinable by definition.
      A node can only be negatively determinable if at least one \patom[] descendant of it has already been applied, so all of \node{}'s children must have at least one applied \patom[] descendant each.
      However, we have a DFS ordering, so if \node{}'s child has one applied \patom[] descendant, all of its \patom[] descendants must be applied, marking the child as complete.
      Since all of \node{}'s children are complete, \node{} is also complete.
    \item If \node{} is the child of an OR node and positively determinable, the reasoning is the same as when \node{} is the child of AND node and negatively determinable.
      As an AND node, if \node{} is positively determinable, each child of \node{} must also be positively determinable by definition.
      A node can only be positively determinable if at least one \patom[] descendant of it has already been applied, so all of \node{}'s children must have at least one applied \patom[] descendant.
      However, we have a DFS ordering, so if \node{}'s child has one applied \patom[] descendant, all of its \patom[] descendants must be applied, marking the child as complete.
      Since all of \node{}'s children are complete, \node{} is also complete.
  \end{enumerate}
\end{proof}

\subsubsection{Proof of Lemma~\ref{lem:growtwo}}
\lemgrowtwo*
\begin{proof}
  We break the proof up into two parts.
  First, we claim that:
  \lemgrowtwoone*
  \noindent
  Note that the difference between \autoref{lem:growtwoone} and \autoref{lem:growtwo} is that \autoref{lem:growtwoone} only applies between consecutive steps $i$ and $i+1$.
  Once \autoref{lem:growtwoone} has been shown to be true, we can show by strong induction on $j$ that the lemma is true for all $j \le i$.
  Assume as the base case $j = 1$.
  This case is trivially satisfied because $\mystate_1[\node] = \varnothing$ for any \node{} based on initialization.
  Next, assume as the inductive hypothesis, the lemma holds for when $j=k$.
  For the inductive step, $j = k+1$.
  For all $\rank(\node,i+1) < j \le i$, $\mystate[\node]$ is not updated.
  Hence $\mystate_j[\node] = \mystate_i[\node]$ for these $j$, and by \autoref{lem:growtwoone}, if \node{}'s parent is an OR node:
  \begin{equation}
    \mystate_{i+1}[\node].\detpos[v] \supseteq \mystate_j[\node].\detpos[v] \cap \bestd(\parent(\parent(\node)), \rank(\node, i+1))
    \label{eq:lemgrowtwo:1}
  \end{equation}
  The inductive hypothesis states that for all $j \le \rank(\node,i+1)$:
  \[
    \mystate_{\rank(\node,i+1)+1}[\node].\detpos[v] \supseteq \mystate_j[\node].\detpos[v] \cap \bestd(\parent(\parent(\node)), \rank(\node, \rank(\node, i+1)))
  \]
  Intersecting both sides with $\bestd(\parent(\parent(\node)), \rank(\node, i+1))$ and applying \autoref{lem:subset} yields:
  \begin{multline*}
    \mystate_{\rank(\node,i+1)+1}[\node].\detpos[v] \cap \bestd(\parent(\parent(\node)), \rank(\node, i+1)) \\
    \supseteq \mystate_j[\node].\detpos[v] \cap \bestd(\parent(\parent(\node)),  \rank(\node, i+1))
  \end{multline*}
  The left-hand side of this equation is one of the cases of the right-hand side of \autoref{eq:lemgrowtwo:1}. Thus, for all $j \le i$:
  \[
    \mystate_{i+1}[\node].\detpos[v] i+1)) \supseteq \mystate_j[\node].\detpos[v] \cap \bestd(\parent(\parent(\node)),  \rank(\node, i+1))
  \]

  The exact same reasoning applies to \detneg[v] if \node{}'s parent is an AND node.
\end{proof}

\subsubsection{Proof of Lemma~\ref{lem:growtwoone}}
\lemgrowtwoone*
\begin{proof}
  Let us assume that \node{} is positively or negatively determinable on step $i$ if \node{}'s parent is an OR or AND node respectively.
  Otherwise, $\mystate_i[\node].\detpos[v] = \mystate_i[\node].\detneg[v] = \varnothing$ from initialization, and the lemma is satisfied trivially.

  Next, we introduce \autoref{lem:grow}.
  \lemgrow*

  Note that the relationship in \autoref{lem:grow} refers to just the parent of \node{}, whereas \autoref{lem:growtwoone} refers to the grandparent of \node{}.

  {\bf If \node{}'s parent is an OR node,} based on \autoref{lem:grow}:
  \begin{equation*}
    \mystate_{i+1}[\node].\detpos[v] \supseteq \mystate_i[\node].\detpos[v] \cap \bestd(\parent(\node), \rank(\node, i+1))
  \end{equation*}
  Expanding out \bestd{} based on Algorithm~\ref{alg:bestd} gives us:
  \begin{multline*}
    \mystate_{i+1}[\node].\detpos[v] \supseteq \mystate_i[\node].\detpos[v] \cap \bestd(\parent(\parent(\node)), \rank(\node, i+1)) \\
    \setminus  \Bigggl \left(\bigcup_{\ccmp \in \chlc(\parent(\node), \rank(\node,i+1))} \mystate_{\rank(\node,i+1)}[\ccmp].\cmp[v] \right)  \\
    \cup  \left(\bigcup_{\cpos \in \chlp(\parent(\node), \rank(\node,i+1)) \setminus \{\node\}} \mystate_{\rank(\node,i+1)}[\cpos].\detpos[v] \right) \Bigggr
  \end{multline*}
  Property~\ref{prop:detpos} states that $\mystate_i[\ccmp].\cmp[v] = \mystate_i[\ccmp].\detpos[v]$ for all complete children \ccmp{}, so:
  \begin{multline*}
    \mystate_{i+1}[\node].\detpos[v] \supseteq \mystate_i[\node].\detpos[v] \cap \bestd(\parent(\parent(\node)), \rank(\node, i+1)) \\
    \setminus  \left(\bigcup_{\cpos \in (\chlc(\parent(\node), \rank(\node,i+1)) \cup \chlp(\parent(\node), \rank(\node,i+1)) \setminus \{\node\})} \mystate_{\rank(\node,i+1)}[\cpos].\detpos[v] \right)
  \end{multline*}
  \autoref{lem:detpos2-cond} (a conditional version of \autoref{lem:detpos2}; shown below) states for two different nodes \cany{} and \canyp{}, $\mystate[\cany].\detpos[v]$ and $\mystate[\canyp].\detpos[v]$ are mutually exclusive.
  Thus, all set subtraction terms can be removed, and:
  \[
    \mystate_{i+1}[\node].\detpos[v] \supseteq \mystate_i[\node].\detpos[v] \cap \bestd(\parent(\parent(\node)), \rank(\node, i+1))
  \]

  \lemdetpostwoc*

  {\bf If \node{}'s parent is an AND node,} based on \autoref{lem:grow}:
  \begin{equation}
    \mystate_{i+1}[\node].\detneg[v] \supseteq \mystate_i[\node].\detneg[v] \cap \bestd(\parent(\node), \rank(\node, i+1))
    \label{eq:corbestd:4}
  \end{equation}
  Expanding out \bestd{} based on Algorithm~\ref{alg:bestd} gives us:
  \begin{multline*}
    \mystate_{i+1}[\node].\detneg[v] \supseteq \mystate_i[\node].\detneg[v] \cap \bestd(\parent(\parent(\node)), \rank(\node, i+1)) \\
    \cap   \left(\bigcap_{\ccmp \in \chlc(\parent(\node), \rank(\node,i+1))} \mystate_{\rank(\node,i+1)}[\ccmp].\cmp[v] \right)  \\
    \setminus  \left(\bigcup_{\cneg \in \chln(\parent(\node), \rank(\node,i+1)) \setminus \{\node\}} \mystate_{\rank(\node,i+1)}[\cneg].\detneg[v] \right)
  \end{multline*}
  \autoref{lem:detpos2-cond} states that the \detneg[v] values of two different node must also be mutually exclusive.
  Thus, all set subtraction terms can be removed once again:
  \begin{multline}
    \mystate_{i+1}[\node].\detneg[v] \supseteq \mystate_i[\node].\detneg[v] \cap \bestd(\parent(\parent(\node)), \rank(\node, i+1)) \\
    \cap   \left(\bigcap_{\ccmp \in \chlc(\parent(\node), \rank(\node,i+1))} \mystate_{\rank(\node,i+1)}[\ccmp].\cmp[v] \right)
    \label{eq:growtwoone:1}
  \end{multline}
  At this point, we establish the following relationships thanks to Algorithm~\ref{alg:update}:
  \begin{align*}
    \mystate_i[\node].\detneg[v] &\subseteq \bestd(\node, \rank(\node, i)) \\
    \mystate_{\rank(\node,i+1)}[\ccmp].\cmp[v] &\subseteq \bestd(\node, \rank(\ccmp, \rank(\node,i+1)))
  \end{align*}
  If $\rank(\node, i) < \rank(\ccmp, \rank(\node,i+1))$ (meaning a \patom[] descendant of \ccmp{} was applied more recently than one of \node{}), as part of calculating \bestd{}, the \detneg[v] values of all negatively determinable children are subtracted away.
  As such, \node{} must have been one of these children:
  \[
    \mystate_{\rank(\node,i+1)}[\ccmp].\cmp[v] \subseteq \bestd(\node, \rank(\ccmp, \rank(\node,i+1))) \setminus \mystate_{\rank(\ccmp, \rank(\node, i+1))}[\node].\detneg[v]
  \]
  Since $\rank(\node, i) < \rank(\ccmp, \rank(\node, i+1))$, $\mystate_{\rank(\ccmp,\rank(\node, i+1))}[\node] = \mystate_i[\node]$, thus $\mystate_i[\ccmp].\cmp[v]$ must be mutually exclusive with $\mystate_i[\node].\detneg[v]$.

  On the other hand, if  $\rank(\node, i) > \rank(\ccmp, \rank(\node, i+1))$ (meaning a \patom[] descendant of \node{} was applied more recently than one of \ccmp{}), as part of calculating \bestd{}, an intersection with the \cmp[v] values of all complete children is performed.
  Since $\rank(\node, i) > \rank(\ccmp, \rank(\node,i+1))$, \ccmp{} must have be one of these children:
  \[
    \mystate_i[\node].\detneg[v] \subseteq \bestd(\node, \rank(\node, i)) \cap \mystate_{\rank(\node, i)}[\ccmp].\cmp[v]
  \]
  Since $\rank(\node, i) > \rank(\ccmp, \rank(\node, i+1))$, $\mystate_{\rank(\node,i)}[\ccmp] = \mystate_i[\ccmp]$, thus $\mystate_i[\node].\detneg[v] \subseteq \mystate_i[\ccmp].\cmp[v]$.

  If the first case is true for any complete child \node{} in \autoref{eq:growtwoone:1}, then the right-hand side simplifies to the empty set, and the lemma is trivially true.
  If all complete children \ccmp{} of belong to the second case, then $\mystate_i[\node].\detneg[v]$ is a subset of that intersection, and  \autoref{eq:growtwoone:1} simplifies to:
  \[
    \mystate_{i+1}[\node].\detneg[v] \supseteq \mystate_i[\node].\detneg[v] \cap \bestd(\parent(\parent(\node)), \rank(\node, i))
  \]
\end{proof}

\subsubsection{Proof of Lemma~\ref{lem:grow}}
\lemgrow*
\begin{proof}
  Similar to the proof for \autoref{lem:growtwo}, this proof is divided into two parts.
  First, we present:
  \lemgrowone*
  \noindent
  Similar to before, the difference between \autoref{lem:grow-one} and \autoref{lem:grow} is that \autoref{lem:grow-one} only applies between consecutive steps $i$ and $i+1$.
  Once \autoref{lem:grow-one} has shown to be true, we can show by strong induction on $j$ that the lemma is true for all $j \le i$.
  Assume as the base case $j = 1$.
  This case is trivially satisfied since $\mystate_1[\node] = \varnothing$ for any \node{} based on initialization.
  Next, assume as the inductive hypothesis, the lemma holds for when $j=k$.
  For the inductive step, $j = k+1$.
  For all $\rank(\node,i+1) < j \le i$, $\mystate[\node]$ is not updated.
  Hence $\mystate_j[\node] = \mystate_i[\node]$ for these $j$, and by \autoref{lem:grow-one}:
  \begin{equation}
    \mystate_{i+1}[\node].\detpos[v] \supseteq \mystate_j[\node].\detpos[v] \cap \bestd(\parent(\node), \rank(\node, i+1))
    \label{eq:lemgrow:50}
  \end{equation}
  The inductive hypothesis states that for all $j \le \rank(\node,i+1)$:
  \[
    \mystate_{\rank(\node,i+1)+1}[\node].\detpos[v] \supseteq \mystate_j[\node].\detpos[v] \cap \bestd(\parent(\node), \rank(\node, \rank(\node, i+1)))
  \]
  Intersecting both sides with $\bestd(\parent(\node), \rank(\node, i+1))$ and applying \autoref{lem:subset} yields:
  \begin{multline*}
    \mystate_{\rank(\node,i+1)+1}[\node].\detpos[v] \cap \bestd(\parent(\node), \rank(\node, i+1)) \\
    \supseteq \mystate_j[\node].\detpos[v] \cap \bestd(\parent(\node),  \rank(\node, i+1))
  \end{multline*}
  The left-hand side of this equation is one of the cases of the right-hand side of \autoref{eq:lemgrow:50}. Thus, for all $j \le i$:
  \[
    \mystate_{i+1}[\node].\detpos[v] \supseteq \mystate_j[\node].\detpos[v] \cap \bestd(\parent(\node),  \rank(\node, i+1))
  \]

  The exact same reasoning applies to $\mystate[\node].\detneg[v]$.
\end{proof}

\subsubsection{Proof of Lemma~\ref{lem:grow-one}}
\lemgrowone*
\begin{proof}
  We prove this by strong induction on the height of the tree referred to by \node{}.
  As the base case, let \node{} be a leaf node, which refers to \predi{j}.
  A leaf node's \detpos[v] value is only ever updated once (at the time of its completion).
  If $j < i$, then \predi{j} has already been applied.
  Thus, $\mystate_{i+1}[\node] = \mystate_j[\node]$, and the lemma is satisfied.
  If $j = i$, then $\mystate_j[\node].\detpos[v] = \varnothing$ because it still holds its initialized state, and the lemma is trivially satisfied.

  Assume as the inductive hypothesis that the lemma holds for any \node{} with a height of $k$.
  Next, for the inductive step, let \node{} refer to a node with a height of $k+1$.
  Let us also assume that \predi{i} is a descendant of \node{}.
  Otherwise, \node{}'s state will not be updated and $\mystate_{i+1}[\node] = \mystate_i[\node]$, and the lemma is trivially true.
  Furthermore, we assume that \node{} is positively and negatively determinable on step $i$.
  Otherwise, $\mystate_i[\node].\detpos[v] = \mystate_i[\node].\detneg[v] = \varnothing$ based on initialization, and the lemma is trivially satisfied once again.

  {\bf If \node{} is an AND node and positively determinable,} then all of its children must be positively determinable by definition.
  Based on Algorithm~\ref{alg:update}:
  \begin{align}
    \mystate_{i+1}[\node].\detpos[v] &= \bigcap_{\cany \in \chl(\node)} \mystate_{i+1}[\cany].\detpos[v] \cap \bestd(\parent(\node), i) \label{eq:lemgrow:10} \\
    \mystate_i[\node].\detpos[v] &= \bigcap_{\cany \in \chl(\node)} \mystate_{i}[\cany].\detpos[v] \cap \bestd(\parent(\node), \rank(\node,i)) \label{eq:lemgrow:11}
  \end{align}
  As stated in the precondition, we assume that \autoref{lem:subset} holds up to step $i$.
  Thus, $\bestd(\parent(\node), i) \subseteq \bestd(\parent(\node), \rank(\node,i))$, and:
  \begin{equation}
    \mystate_i[\node].\detpos[v] \cap \bestd(\parent(\node), i)= \bestd(\parent(\node), i) \cap \left(\bigcap_{\cany \in \chl(\node)} \mystate_{i}[\cany].\detpos[v] \right)
    \label{eq:lemgrow:14}
  \end{equation}
  All children of \node{} except \predi{i}'s ancestor do not have their \mystate{} values updated between steps $i$ and $i+1$.
  For these children $\mystate_i = \mystate_{i+1}$.
  Let \cspc{} be \node{}'s child that is \predi{i}'s ancestor and let:
  \begin{equation*}
    X = \bigcap_{\cany \in \chl(\node) \setminus\{\cspc\}} \mystate_i[\cany].\detpos[v]
  \end{equation*}
  We can rewrite Equations~\ref{eq:lemgrow:10} and~\ref{eq:lemgrow:14} as:
  \begin{align}
    \mystate_{i+1}[\node].\detpos[v] &= \bestd(\parent(\node), i) \cap X \cap \mystate_{i+1}[\cspc].\detpos[v] \label{eq:lemgrow:15}\\
    \mystate_i[\node].\detpos[v] \cap \bestd(\parent(\node), i)&= \bestd(\parent(\node), i) \cap X \cap \mystate_i[\cspc].\detpos[v]  \label{eq:lemgrow:16}
  \end{align}
  The inductive hypothesis states that for child \cspc{}:
  \[
    \mystate_{i+1}[\cspc].\detpos[v] \supseteq \mystate_i[\cspc].\detpos[v] \cap \bestd(\node, i)
  \]
  Expanding out \bestd{} based on Algorithm~\ref{alg:bestd}, we get:
  \begin{multline*}
    \mystate_{i+1}[\cspc].\detpos[v] \supseteq \mystate_i[\cspc].\detpos[v] \cap \bestd(\parent(\node), i) \\
    \cap \left( \left(\bigcap_{\ccmp \in \chlc(\node, i)} \mystate_i[\ccmp].\cmp[v] \right) \setminus \left( \bigcup_{\cneg \in \chln(\node, i) \setminus \{\cspc\}} \mystate_i[\cneg].\detneg[v] \right) \right)
  \end{multline*}
  If we intersect both sides with $X \cap \bestd(\parent(\node), i)$, the left-hand side is the same as Equation~\ref{eq:lemgrow:15}.
  Thus:
  \begin{multline*}
    \mystate_{i+1}[\node].\detpos[v] \supseteq \mystate_i[\cspc].\detpos[v] \cap \bestd(\parent(\node), i) \cap X \\
    \cap \left( \left(\bigcap_{\ccmp \in \chlc(\node, i)} \mystate_i[\ccmp].\cmp[v] \right) \setminus \left( \bigcup_{\cneg \in \chln(\node, i) \setminus \{\cspc\}} \mystate_i[\cneg].\detneg[v] \right) \right)
  \end{multline*}
  Property~\ref{prop:detpos} states that $\mystate[\ccmp].\cmp[v] = \mystate[\ccmp].\detpos[v]$ for all complete children, so these terms can be absorbed into $X$.
  Expanding out $X$ leaves us with:
  \begin{multline*}
    \mystate_{i+1}[\node].\detpos[v] \supseteq \mystate_i[\cspc].\detpos[v] \cap \bestd(\parent(\node), i) \\
    \cap \left( \left(\bigcap_{\cany \in \chl(\node, i) \setminus \{\cspc\}} \mystate_i[\cany].\detpos[v] \right) \setminus \left( \bigcup_{\cneg \in \chln(\node, i) \setminus \{\cspc\}} \mystate_i[\cneg].\detneg[v] \right) \right)
  \end{multline*}
  Using the identity $(A \cap B) \setminus (C \cup D) = (A \setminus C) \cap (B \setminus D)$ for all sets $A$, $B$, $C$, and $D$ (\autoref{apx:set-alg}), we can rearrange this to:
  \begin{multline*}
    \mystate_{i+1}[\node].\detpos[v] \supseteq \mystate_i[\cspc].\detpos[v] \cap \bestd(\parent(\node), i) \\
    \cap  \left(\bigcap_{\cany \in \chl(\node, i) \setminus \{\cspc\}} \mystate_i[\cany].\detpos[v] \setminus \mystate_i[\cany].\detneg[v] \right)
  \end{multline*}
  Not all children may be negatively determinable, but the $\detneg[v]$ value of those children will be $\varnothing$ anyway based on initialization.
  Based on Property~\ref{prop:detneg}, the \detpos[v] and \detneg[v] values for any node must mutually exclusive for any time step, so we can remove the set subtraction terms:
  \begin{equation*}
    \mystate_{i+1}[\node].\detpos[v] \supseteq \mystate_i[\cspc].\detpos[v] \cap \bestd(\parent(\node), i) \cap  \left(\bigcap_{\cany \in \chl(\node, i) \setminus \{\cspc\}} \mystate_i[\cany].\detpos[v] \right)
  \end{equation*}
  The right-hand side of this equation is equivalent to Equation~\ref{eq:lemgrow:16}, so:
  \[
    \mystate_{i+1}[\node].\detpos[v] \supseteq \mystate_i[\node].\detpos[v] \cap \bestd(\parent(\node), i)
  \]

  {\bf If \node{} is an OR node and positively determinable,} then based on Algorithm~\ref{alg:update}:
  \begin{align}
    \mystate_{i+1}[\node].\detpos[v] &= \bigcup_{\cpos \in \chlp(\node, i+1)} \mystate_{i+1}[\cpos].\detpos[v] \cap \bestd(\parent(\node), i) \label{eq:lemgrow:20}\\
    \mystate_i[\node].\detpos[v] &= \bigcup_{\cpos \in \chlp(\node, i)} \mystate_i[\cpos].\detpos[v] \cap \bestd(\parent(\node), \rank(\node, i)) \label{eq:lemgrow:21}
  \end{align}
  Similar to before, based on Lemma~\ref{lem:subset}, Equation~\ref{eq:lemgrow:21} can be intersected with $\bestd(\parent(\node), i)$ to get:
  \begin{equation}
    \mystate_i[\node].\detpos[v] \cap \bestd(\parent(\node), i)= \bigcup_{\cpos \in \chlp(\node, i)} \mystate_i[\cpos].\detpos[v] \cap \bestd(\parent(\node), i) \label{eq:lemgrow:22}
  \end{equation}
  All children of \node{} except \predi{i}'s ancestor do not have their \mystate{} values updated between steps $i$ and $i+1$.
  For these children $\mystate_i = \mystate_{i+1}$.
  Let \cspc{} be \node{}'s child that is \predi{i}'s ancestor and let:
  \begin{equation*}
    X = \bigcup_{\cpos \in \chlp(\node, i) \setminus\{\cspc\}} \mystate_i[\cpos].\detpos[v]
  \end{equation*}
  We can rewrite Equations~\ref{eq:lemgrow:20} and~\ref{eq:lemgrow:22} as:
  \begin{align}
    \mystate_{i+1}[\node].\detpos[v] &= \bestd(\parent(\node), i) \cap (X \cup \mystate_{i+1}[\cspc].\detpos[v] \label{eq:lemgrow:23})\\
    \mystate_i[\node].\detpos[v] \cap \bestd(\parent(\node), i)&= \bestd(\parent(\node), i) \cap (X \cup \mystate_i[\cspc].\detpos[v])  \label{eq:lemgrow:24}
  \end{align}
  The inductive hypothesis states that for child \cspc{}:
  \[
    \mystate_{i+1}[\cspc].\detpos[v] \supseteq \mystate_i[\cspc].\detpos[v] \cap \bestd(\node, i)
  \]
  Expanding out \bestd{} based on Algorithm~\ref{alg:bestd}, we get:
  \begin{multline*}
    \mystate_{i+1}[\cspc].\detpos[v] \supseteq \mystate_i[\cspc].\detpos[v] \cap \bestd(\parent(\node), i) \\
    \setminus \left( \bigcup_{\ccmp \in \chlc(\node, i)} \mystate_i[\ccmp].\cmp[v] \bigcup_{\cpos \in \chlp(\node, i) \setminus \{\cspc\}} \mystate_i[\cpos].\detpos[v] \right)
  \end{multline*}
  Once again, based on Property~\ref{prop:detpos}, for complete children \ccmp{}, $\mystate[\ccmp].\cmp[v] = \mystate[\ccmp].\detpos[v]$, so:
  \begin{multline*}
    \mystate_{i+1}[\cspc].\detpos[v] \supseteq \mystate_i[\cspc].\detpos[v] \cap \bestd(\parent(\node), i) \\
    \setminus \left( \bigcup_{\cany \in (\chlc(\node, i) \cup \chlp(\node, i) \setminus \{\cspc\})} \mystate_i[\cany].\detpos[v] \right)
  \end{multline*}
  Lemma~\ref{lem:detpos2-cond} states that for two different children $\cany$ and $\canyp$, $\mystate_i[\cany].\detpos[v]$ and $\mystate_i[\canyp].\detpos[v]$ are mutually exclusive\footnote{%
    To apply \autoref{lem:detpos2-cond} to \node{}'s children here, \autoref{lem:grow-one} must hold true up up to \node{}'s children.
    The inductive hypothesis applies in this case, so there is no cyclical dependency.
  }, so this simplifies to:
  \begin{equation*}
    \mystate_{i+1}[\cspc].\detpos[v] \supseteq \mystate_i[\cspc].\detpos[v] \cap \bestd(\parent(\node), i)
  \end{equation*}
  We can take the union with respect to $\bestd(\parent(\node), i) \cap X$ on both sides.
  This results in the left-hand side being equivalent to \autoref{eq:lemgrow:23}, and the right-hand side being equivalent to \autoref{eq:lemgrow:24}. Thus:
  \[
    \mystate_{i+1}[\node].\detpos[v] \supseteq \mystate_i[\node].\detpos[v] \cap \bestd(\parent(\node), i)
  \]

  {\bf If \node{} is an AND node and negatively determinable,} then based on Algorithm~\ref{alg:update}:
  \begin{align}
    \mystate_{i+1}[\node].\detneg[v] &= \bigcup_{\cneg \in \chln(\node, i+1)} \mystate_{i+1}[\cneg].\detneg[v] \cap \bestd(\parent(\node), i) \label{eq:lemgrow:30}\\
    \mystate_i[\node].\detneg[v] &= \bigcup_{\cneg \in \chln(\node, i)} \mystate_i[\cneg].\detneg[v] \cap \bestd(\parent(\node), \rank(\node, i)) \label{eq:lemgrow:31}
  \end{align}
  Once again, based on Lemma~\ref{lem:subset}, Equation~\ref{eq:lemgrow:31} can be intersected with $\bestd(\parent(\node), i)$ to get:
  \begin{equation}
    \mystate_i[\node].\detneg[v] \cap \bestd(\parent(\node), i)= \bigcup_{\cneg \in \chln(\node, i)} \mystate_i[\cneg].\detneg[v] \cap \bestd(\parent(\node), i) \label{eq:lemgrow:32}
  \end{equation}
%
  All children of \node{} except \predi{i}'s ancestor do not have their \mystate{} values updated between steps $i$ and $i+1$.
  For these children $\mystate_i = \mystate_{i+1}$.
  Let \cspc{} be \node{}'s child that is \predi{i}'s ancestor and let:
  \begin{equation*}
    X = \bigcup_{\cneg \in \chln(\node, i) \setminus\{\cspc\}} \mystate_i[\cneg].\detneg[v]
  \end{equation*}
  We can rewrite Equations~\ref{eq:lemgrow:30} and~\ref{eq:lemgrow:32} as:
  \begin{align}
    \mystate_{i+1}[\node].\detneg[v] &= \bestd(\parent(\node), i) \cap (X \cup \mystate_{i+1}[\cspc].\detneg[v] \label{eq:lemgrow:33})\\
    \mystate_i[\node].\detneg[v] \cap \bestd(\parent(\node), i)&= \bestd(\parent(\node), i) \cap (X \cup \mystate_i[\cspc].\detneg[v])  \label{eq:lemgrow:34}
  \end{align}
  The inductive hypothesis states that for child \cspc{}:
  \[
    \mystate_{i+1}[\cspc].\detneg[v] \supseteq \mystate_i[\cspc].\detneg[v] \cap \bestd(\node, i)
  \]
  Expanding out \bestd{} based on Algorithm~\ref{alg:bestd}, we get:
  \begin{multline*}
    \mystate_{i+1}[\cspc].\detneg[v] \supseteq \mystate_i[\cspc].\detneg[v] \cap \bestd(\parent(\node), i) \\
    \cap \left( \left(\bigcap_{\ccmp \in \chlc(\node, i)} \mystate_i[\ccmp].\cmp[v] \right) \setminus \left( \bigcup_{\cneg \in \chln(\node, i) \setminus \{\cspc\}} \mystate_i[\cneg].\detneg[v] \right) \right)
  \end{multline*}
  \autoref{lem:detpos2-cond} tells us that the \detneg[v] values of two different children are mutually exclusive, so we can remove the set subtraction terms:
  \begin{equation}
    \mystate_{i+1}[\cspc].\detneg[v] \supseteq \mystate_i[\cspc].\detneg[v] \cap \bestd(\parent(\node), i) \cap  \left(\bigcap_{\ccmp \in \chlc(\node, i)} \mystate_i[\ccmp].\cmp[v] \right)
    \label{eq:lemgrow:35}
  \end{equation}
  At this point, we establish the following relationships thanks to Algorithm~\ref{alg:update}:
  For \cspc{} and any complete child \ccmp{} of \node{} on step $i$:
  \begin{align*}
    \mystate_i[\cspc].\detneg[v] &\subseteq \bestd(\node, \rank(\cspc, i)) \\
    \mystate_i[\ccmp].\cmp[v] &\subseteq \bestd(\node, \rank(\ccmp, i))
  \end{align*}
  If $\rank(\cspc, i) < \rank(\ccmp, i)$ (meaning a \patom[] descendant of \ccmp{} was applied more recently than one of \cspc{}), as part of calculating \bestd{}, the \detneg[v] values of all negatively determinable children are subtracted away.
  Since $\rank(\cspc, i) < \rank(\ccmp, i)$, \cspc{} must be one of these children:
  \[
    \mystate_i[\ccmp].\cmp[v] \subseteq \bestd(\node, \rank(\ccmp, i)) \setminus \mystate_{\rank(\ccmp, i)}[\cspc].\detneg[v]
  \]
  Since $\rank(\cspc, i) < \rank(\ccmp, i)$, $\mystate_{\rank(\ccmp,i)}[\cspc] = \mystate_i[\cspc]$, thus $\mystate_i[\ccmp].\cmp[v]$ must be mutually exclusive with $\mystate_i[\cspc].\detneg[v]$.

  On the other hand, if $\rank(\cspc, i) > \rank(\ccmp, i)$ (meaning a \patom[] descendant of \cspc{} was applied more recently than one of \ccmp{}), as part of calculating \bestd{}, an intersection with the \cmp[v] values of all complete children is performed.
  Since $\rank(\cspc, i) > \rank(\ccmp, i)$, \ccmp{} must have be one of these children:
  \[
    \mystate_i[\cspc].\detneg[v] \subseteq \bestd(\node, \rank(\cspc, i)) \cap \mystate_{\rank(\cspc, i)}[\ccmp].\cmp[v]
  \]
  Since $\rank(\cspc, i) > \rank(\ccmp, i)$, $\mystate_{\rank(\cspc,i)}[\ccmp] = \mystate_i[\ccmp]$, thus $\mystate_i[\cspc].\detneg[v] \subseteq \mystate_i[\ccmp].\cmp[v]$.

  If the first case is true for any complete child \ccmp{} in \autoref{eq:lemgrow:35}, then the right-hand side simplifies to the empty set, and the lemma is trivially true.
  If all complete children \ccmp{} of belong to the second case, then $\mystate_i[\child].\detneg[v]$ must be a subset of the intersection, and Equation~\ref{eq:lemgrow:35} simplifies to:
  \[
    \mystate_{i+1}[\child].\detneg[v] \supseteq \mystate_i[\child].\detneg[v] \cap \bestd(\parent(\node), i)
  \]
  We can take the union with respect to $\bestd(\parent(\node), i) \cap X$ on both sides.
  This results in the left-hand side being equivalent to \autoref{eq:lemgrow:33}, and the right-hand side being equivalent to \autoref{eq:lemgrow:34}. Thus:
  \[
    \mystate_{i+1}[\node].\detneg[v] \supseteq \mystate_i[\node].\detneg[v] \cap \bestd(\parent(\node), i)
  \]

  {\bf If \node{} is an OR node and negatively determinable,} then all of its children must be negatively determinable by definition.
  Based on Algorithm~\ref{alg:update}:
  \begin{align}
    \mystate_{i+1}[\node].\detneg[v] &= \bigcap_{\cany \in \chl(\node)} \mystate_{i+1}[\cany].\detneg[v] \cap \bestd(\parent(\node), i) \label{eq:lemgrow:40} \\
    \mystate_i[\node].\detneg[v] &= \bigcap_{\cany \in \chl(\node)} \mystate_{i}[\cany].\detneg[v] \cap \bestd(\parent(\node), \rank(\node,i)) \label{eq:lemgrow:41}
  \end{align}
  Since we assumed that \autoref{lem:subset} holds up to step $i$, $\bestd(\parent(\node), i) \subseteq \bestd(\parent(\node), \rank(\node,i))$, and:
  \begin{equation}
    \mystate_i[\node].\detneg[v] \cap \bestd(\parent(\node), i)= \bestd(\parent(\node), i) \cap \left(\bigcap_{\cany \in \chl(\node)} \mystate_{i}[\cany].\detneg[v] \right)
    \label{eq:lemgrow:42}
  \end{equation}
  All children of \node{} except \predi{i}'s ancestor do not have their \mystate{} values updated between steps $i$ and $i+1$.
  For these children $\mystate_i = \mystate_{i+1}$.
  Let \cspc{} be \node{}'s child that is \predi{i}'s ancestor and let:
  \begin{equation*}
    X = \bigcap_{\cany \in \chl(\node) \setminus\{\cspc\}} \mystate_i[\cany].\detneg[v]
  \end{equation*}
  We can rewrite Equations~\ref{eq:lemgrow:40} and~\ref{eq:lemgrow:42} as:
  \begin{align}
    \mystate_{i+1}[\node].\detneg[v] &= \bestd(\parent(\node), i) \cap X \cap \mystate_{i+1}[\cspc].\detneg[v] \label{eq:lemgrow:43}\\
    \mystate_i[\node].\detneg[v] \cap \bestd(\parent(\node), i)&= \bestd(\parent(\node), i) \cap X \cap \mystate_i[\cspc].\detneg[v]  \label{eq:lemgrow:44}
  \end{align}
  The inductive hypothesis states that for child \cspc{}:
  \[
    \mystate_{i+1}[\cspc].\detneg[v] \supseteq \mystate_i[\cspc].\detneg[v] \cap \bestd(\node, i)
  \]
  Expanding out \bestd{} based on Algorithm~\ref{alg:bestd}, we get:
  \begin{multline*}
    \mystate_{i+1}[\cspc].\detneg[v] \supseteq \mystate_i[\cspc].\detneg[v] \cap \bestd(\parent(\node), i) \\
    \setminus \left( \bigcup_{\ccmp \in \chlc(\node, i)} \mystate_i[\ccmp].\cmp[v] \bigcup_{\cpos \in \chlp(\node, i) \setminus \{\cspc\}} \mystate_i[\cpos].\detpos[v] \right)
  \end{multline*}
  Once again, based on Property~\ref{prop:detpos}, for complete children \ccmp{}, $\mystate[\ccmp].\cmp[v] = \mystate[\ccmp].\detpos[v]$, so:
  \begin{multline*}
    \mystate_{i+1}[\cspc].\detneg[v] \supseteq \mystate_i[\cspc].\detneg[v] \cap \bestd(\parent(\node), i) \\
    \setminus \left( \bigcup_{\cany \in (\chlc(\node, i) \cup \chlp(\node, i) \setminus \{\cspc\})} \mystate_i[\cany].\detpos[v] \right)
  \end{multline*}
  If we intersect both sides with $X \cap \bestd(\parent(\node), i)$, the left-hand side is the same as Equation~\ref{eq:lemgrow:43}.
  Thus:
  \begin{multline*}
    \mystate_{i+1}[\node].\detneg[v] \supseteq \mystate_i[\cspc].\detneg[v] \cap \bestd(\parent(\node), i) \cap X \\
    \setminus \left( \bigcup_{\cany \in (\chlc(\node, i) \cup \chlp(\node, i) \setminus \{\cspc\})} \mystate_i[\cany].\detpos[v] \right)
  \end{multline*}
  Using the identity $(A \cap B) \setminus (C \cup D) = (A \setminus C) \cap (B \setminus D)$ for all sets $A$, $B$, $C$, and $D$ (\autoref{apx:set-alg}) and expanding $X$, we can rearrange this to:
  \begin{multline*}
    \mystate_{i+1}[\node].\detneg[v] \supseteq \mystate_i[\cspc].\detneg[v] \cap \bestd(\parent(\node), i) \\
    \cap \left( \bigcap_{\cany \in \chl(\node) \setminus \{\cspc\}}  \mystate_i[\cany].\detneg[v] \setminus \mystate_i[\cany].\detpos[v] \right)
  \end{multline*}
  Not all children may be positively determinable, but the $\detpos[v]$ value of those children will be $\varnothing$ anyway based on initialization.
  Based on Property~\ref{prop:detneg}, the \detneg[v] and \detpos[v] values for any node must mutually exclusive for any time step, so we can remove the set subtraction terms:
  \begin{multline*}
    \mystate_{i+1}[\node].\detneg[v] \supseteq \mystate_i[\cspc].\detneg[v] \cap \bestd(\parent(\node), i) \cap \left( \bigcap_{\cany \in \chl(\node) \setminus \{\cspc\}}  \mystate_i[\cany].\detneg[v] \right)
  \end{multline*}
  The right-hand side of this equation is equivalent to Equation~\ref{eq:lemgrow:44}, so:
  \[
    \mystate_{i+1}[\node].\detneg[v] \supseteq \mystate_i[\node].\detneg[v] \cap \bestd(\parent(\node), i)
  \]
\end{proof}

\subsubsection{Proof of Lemma~\ref{lem:detpos2-cond}}
\lemdetpostwoc*
\begin{proof}


  We prove this by induction over the value of $\jmax$.
  Let us first assume \node{} is an OR node.
  As the base case, $i = j = 1$.
  Since all \detpos[v] values are initialized to the empty set: $\mystate_1[\cany].\detpos[v] = \mystate_1[\canyp].\detpos[v] = \varnothing$, and the lemma is trivially satisfied.

  As the inductive hypothesis, assume that for $\jmax = k$, the lemma holds.
  For the inductive step $\jmax = k+1$, there are three cases:
  \begin{enumerate}
    \item \patom[c] \predi{k} is not a descendant of either \cany{} or \canyp{}.
      In this case, the \mystate{} does not change for either \cany{} and \canyp{} between steps $k$ and $k+1$, and:
      \begin{align*}
        \mystate_{k+1}[\cany].\detpos[v] = \mystate_k[\cany].\detpos[v] && \mystate_{k+1}[\canyp].\detpos[v] = \mystate_k[\canyp].\detpos[v]
      \end{align*}
      Thus, We can apply the inductive hypothesis directly.

    \item \patom[c] \predi{k} is a descendant of \cany{}.
      We assume that $\mystate_{k+1}[\cany].\detpos[v] \ne \mystate_k[\cany].\detpos[v]$.
      Otherwise, this situation reduces to the first case.
      If \mystate{} is updated, based on Algorithm~\ref{alg:update}, we know that: $\mystate_{k+1}[\cany].\detpos[v] \subseteq \bestd(\node, k)$.
      However, as a part of calculating $\bestd(\node, k)$, $\mystate_k[\canyp].\detpos[v]$ is explicitly removed from result, so:
      \[
        \mystate_{k+1}[\cany].\detpos[v] \cap \mystate_k[\canyp].\detpos[v] = \varnothing
      \]
      In addition, since \predi{k} is not a descendant of \canyp{}, $\mystate[\canyp].\detpos[v]$ remains unchanged between steps $k$ and $k+1$.
      In fact, $\mystate[\canyp]$ remains unchanged for all steps since the last \patom[] descendant of \patom[], so for all $\rank(\canyp,k) < j \le (k+1)$:
      \[
        \mystate_{k+1}[\cany].\detpos[v] \cap \mystate_j[\canyp].\detpos[v] = \varnothing
      \]
%
      Based on Lemma~\ref{lem:grow-one}, we know that $\mystate_{\rank(\canyp,k)+1}[\canyp].\detpos[v] \supseteq \mystate_{\rank(\canyp,k)}[\canyp].\detpos[v] \cap \bestd(\node, \rank(\canyp,k))$.
      Furthermore, Lemma~\ref{lem:subset} tells us that $\bestd(\node, k) \subseteq \bestd(\node, \rank(\canyp,k))$, so:
      \begin{align*}
        \mystate_{\rank(\canyp,k)+1}[\canyp].\detpos[v] &\supseteq \mystate_{\rank(\canyp,k)}[\canyp].\detpos[v] \cap \bestd(\node, \rank(\canyp,k)) \\
      \mystate_{\rank(\canyp,k)+1}[\canyp].\detpos[v] &\supseteq \mystate_{\rank(\canyp,k)}[\canyp].\detpos[v] \cap \bestd(\node, k) \\
      \mystate_{k+1}[\cany].\detpos[v] \cap \mystate_{\rank(\canyp,k)+1}[\canyp].\detpos[v] &\supseteq \mystate_{\rank(\canyp,k)}[\canyp].\detpos[v] \cap \bestd(\node, k) \cap \mystate_{k+1}[\cany].\detpos[v] \\
     \varnothing  &\supseteq \mystate_{\rank(\canyp,k)}[\canyp].\detpos[v] \cap \bestd(\node, k) \cap \mystate_{k+1}[\cany].\detpos[v] \\
     \varnothing  &\supseteq \mystate_{\rank(\canyp,k)}[\canyp].\detpos[v] \cap \mystate_{k+1}[\cany].\detpos[v]
      \end{align*}
      The final step comes from the fact that $\mystate_{k+1}[\cany].\detpos[v] \subseteq \bestd(\node, k)$.
      This reasoning can be applied recursively for all $j \le \rank(\node, k)$, such that for all $1 \le j \le k+1$:
      \[
        \mystate_{k+1}[\cany].\detpos[v] \cap \mystate_j[\canyp].\detpos[v] = \varnothing
      \]
      For cases in which $i \le k$ and $j = k+1$, $\mystate[\canyp].\detpos[v]$ remains unchanged between steps $k$ and $k+1$, so the inductive hypothesis can be directly applied.

    \item
      \patom[c] \predi{k} is a descendant of \canyp{}.
      This is the same as the second case, except \cany{} and \canyp{} are flipped.
      The same reasoning applies.
  \end{enumerate}

  The exact same reasoning applies to $\mystate[\cany].\detneg[v]$ and $\mystate[\canyp].\detneg[v]$ if \node{} is an AND node.
\end{proof}

\subsubsection{Proof of Lemma~\ref{lem:div}}
\lemdiv*
\begin{proof}
  We prove this by structural induction over all sets \vseto{} which can be derived from $\vsets{} \cup \{\pred(\vset)\}$.
  If $\vseto \in \vsets$, the lemma is trivially satisfied with $\vseto' = \vseto$.
  Thus, \vseto{} must be a valuation set that can be derived from $\vsets \cup \{\pred(\vset)\}$ but not \vsets{}.
  The base case is $\vseto = \pred(\vset)$. By definition, $\pred(\vset) \subseteq \vset$, so: $\vseto \setminus \vset = \varnothing$, and the lemma is satisfied with $\vseto' = \vset$.

  Next, assume as the inductive hypothesis that \vsetp{} and \vsetq{} are some valuation sets derived from $\vsets \cup \{\pred(\vset)\}$ and \vsetpp{} and \vsetqp{} are the corresponding valuation set derived from $\vsets$ such that $\vsetp \setminus \vset = \vsetpp \setminus \vset$ and $\vsetq \setminus \vset = \vsetqp \setminus \vset$.
  We can construct a new valuation set \vseto{} derived from $\vsets \cup \{\pred(\vset)\}$ by performing an operation on \vsetp{} (and \vsetq{}):
  \begin{enumerate}
    \item $\vseto = \pred'(\vsetp)$ for some \patom[] $\pred' \ne \pred$:
      \begin{align*}
        \vseto \setminus \vset &= \pred'(\vsetp) \setminus \vset \\
                               &= \pred'(\vsetp \setminus \vset) \\
                               &= \pred'(\vsetpp \setminus \vset) \\
                               &= \pred'(\vsetpp) \setminus \vset
      \end{align*}
      Thus, $\vsetop = \pred'(\vsetpp)$.
      The second and fourth steps come from Property~\ref{prop:basic}, and the third step is the application of the inductive hypothesis.

    \item $\vseto = \vsetp \cap \vsetq$:
      \begin{align*}
        \vseto \setminus \vset &= (\vsetp \cap \vsetq) \setminus \vset \\
                               &= (\vsetp \setminus \vset) \cap (\vsetq \setminus \vset) \\
                               &= (\vsetpp \setminus \vset) \cap (\vsetqp \setminus \vset) \\
                               &= (\vsetpp \cap \vsetqp) \setminus \vset
      \end{align*}
      Thus, $\vsetop = \vsetpp \cap \vsetqp$.
      The second and fourth steps are from the identity $(A \cap B) \setminus C = (A \setminus C) \cap (B \setminus C)$ for any sets $A$, $B$, and $C$ (Appendix~\ref{apx:set-alg}), and the third step is the application of the inductive hypothesis.

    \item $\vseto = \vsetp \cup \vsetq$:
      \begin{align*}
        \vseto \setminus \vset &= (\vsetp \cup \vsetq) \setminus \vset \\
                               &= (\vsetp \setminus \vset) \cup (\vsetq \setminus \vset) \\
                               &= (\vsetpp \setminus \vset) \cup (\vsetqp \setminus \vset) \\
                               &= (\vsetpp \cup \vsetqp) \setminus \vset
      \end{align*}
      Thus, $\vsetop = \vsetpp \cup \vsetqp$.
      The second and fourth steps are from the identity $(A \cup B) \setminus C = (A \setminus C) \cup (B \setminus C)$ for any sets $A$, $B$, and $C$ (Appendix~\ref{apx:set-alg}), and the third step is the application of the inductive hypothesis.

    \item $\vseto = \vsetp \setminus \vsetq$:
      \begin{align*}
        \vseto \setminus \vset &= (\vsetp \setminus \vsetq) \setminus \vset \\
                               &= (\vsetp \setminus \vset) \setminus \vsetq \\
                               &= (\vsetpp \setminus \vset)  \setminus \vsetq \\
                               &= \vsetpp \setminus (\vsetq \cup \vset) \\
                               &= \vsetpp \setminus ((\vsetq \setminus \vset) \cup \vset) \\
                               &= \vsetpp \setminus ((\vsetqp \setminus \vset) \cup \vset) \\
                               &= \vsetpp \setminus (\vsetqp \cup \vset) \\
                               &= (\vsetpp \setminus \vsetqp) \setminus \vset
      \end{align*}
      Thus, $\vsetop = \vsetpp \setminus \vsetqp$.
      The following identities are used (Appendix~\ref{apx:set-alg}). For any sets $A$, $B$, and $C$:
      \begin{align*}
        (A \setminus B) \setminus C = (A \setminus C) \setminus B && (A \setminus B) \setminus C = A \setminus (B \cup C) && A \cup B = (A \setminus B) \cup B
      \end{align*}
      The third and sixth steps are the applications of the inductive hypothesis.
  \end{enumerate}
\end{proof}

\subsubsection{Proof of Lemma~\ref{lem:inc-ok}}
\lemincok*
\begin{proof}
  In our setup, all \patom[] expressions are positive, so the given predicate expression and all subexpressions are monotone Boolean functions~\cite{stephenCountingInequivalentMonotone2014}.
  Monotone Boolean functions are functions $f: \{0,1\}^n \rightarrow \{0,1\}$ which have the property that if $x \le y$, then $f(x) \le f(y)$,
  in which $x \le y$ is defined as $x_i \le y_i$ for all $i$ in $1 \le i \le n$.
  Here 1 is used to represent true values and 0 is used to represent false values.

  Let ancestor \anc{} be an AND node.
  We prove by construction that for all $v \in \vseti{i}$, there exists a $u \in \vatom[m](v,i)$ that evaluates to true against all incomplete children of \anc{}.
  Using the same notation as monotone Boolean functions, the construction of valuation $u$ is as follows:
  \begin{enumerate}
    \item For all $j$ in  $(1 \le j \le i-1)$, set $u_j = v_j$.
    \item For all $j$ in $(i \le j \le n)$, set $u_j = 1$.
  \end{enumerate}
  Lemma~\ref{lem:inc-dynamic} (shown below) states that an incomplete child of \anc{} that is not an ancestor of \predi{i} cannot be statically false with respect to \vatomvi{}.
  Thus, for each incomplete child of \anc{}, there must exist at least one $u' \in \vatom[m](v,i)$ that evaluates to true against that child.
  In other words, if we let \node{} be the incomplete child of \anc{}, then there exists a $u' \in \vatomvi$ such that $\node[u'] = 1$.
  When compared with the above constructed $u$, $u' \le u$ because $u' \in \vatomvi$.
  Thus, based on the property of monotone Boolean functions, if $\node[u'] = 1$, then $\node[u] = 1$.
  This reasoning holds for all incomplete children of \anc{}, so $u$ must evaluate to true against all incomplete children of \anc{}.

  \lemincdynamic*

  Similarly, if \anc{} is an OR node, we can construct a $u \in \vatom[m](v,i)$ that evaluates to false against all incomplete children of \anc{} with:
  \begin{enumerate}
    \item For all $j$ in  $(1 \le j \le i-1)$, set $u_j = v_j$.
    \item For all $j$ in $(i \le j \le n)$, set $u_j = 0$.
  \end{enumerate}
  Once again, Lemma~\ref{lem:inc-dynamic} states that for each incomplete child of \anc{} that is not an ancestor of \predi{i} cannot be statically true with respect to \vatomvi{}.
  Thus, if \node{} is an incomplete child of \anc{}, then there exists a $u' \in \vatomvi$ such that $\node[u'] = 0$.
  When compared with the above constructed $u$, $u \le u'$ because $u' \in \vatomvi$.
  Thus, based on the property of monotone Boolean functions, if $\node[u'] = 0$, then $\node[u] = 0$.
  This reasoning holds for all incomplete children of \anc{}, so $u$ must evaluate to false against all incomplete children of \anc{}.
\end{proof}

\subsubsection{Proof of Lemma~\ref{lem:inc-dynamic}}
\lemincdynamic*
\begin{proof}
  First we introduce Lemmas~\ref{lem:det-effect} and~\ref{lem:det-def}.
  \lemdeteffect*
  \lemdetdef*

We are now ready to prove Lemma~\ref{lem:inc-dynamic}.
Let \anc{} be an AND node, and let \node{} be an incomplete child of \anc{}.
If \node{} is not negatively determinable on step $i$, Lemma~\ref{lem:det-effect} states that \node{} cannot be statically false with respect to \vatomvi, trivially satisfying the lemma.
Thus, let us assume \node{} is negatively determinable.
According to Algorithm~\ref{alg:bestd}, part of calculating $\bestd(\anc, i)$ involves set subtracting $\mystate_i[\node].\detneg[v]$ from $\bestd(\parent(\anc),i)$ if \node{} is negatively determinable.
More specifically, $\bestd(\anc,i) \subseteq \bestd(\parent(\anc),i) \setminus \mystate_i[\node].\detneg[v]$.
Property~\ref{prop:subset} states that $\vseti{i} \subseteq \bestd(\anc,i)$, so $\vseti{i} \cap \mystate_i[\node].\detneg[v] = \varnothing$.
We can substitute the value of $\mystate_i[\node].\detneg[v]$ based on Lemma~\ref{lem:det-def} to get:
\begin{align*}
  \vseti{i} \cap \{v \in \bestd(\parent(\node), \rank(\node,i)) \mid \forall u \in \vatomvi, \node[u] = \false\} & = \varnothing
\end{align*}
Since, $\vseti{i} \subseteq \bestd(\parent(\node), i)$, and $\bestd(\parent(\node), i) \subseteq \bestd(\parent(\node), \rank(\node,i))$ based on Lemma~\ref{lem:subset}:
\begin{align*}
  \{v \in \vseti{i} \mid \forall u \in \vatomvi, \node[u] = \false\} & = \varnothing
\end{align*}
In other words, \node{} is not statically false with respect to \vatomvi, for any $v \in \vseti{i}$.

The same reasoning applies if \anc{} is an OR node.
If \node{} is not positively determinable on step $i$, Lemma~\ref{lem:det-effect} states that \node{} cannot be statically true with respect to \vatomvi, trivially satisfying the lemma.
Thus, let us assume \node{} is positively determinable.
According to Algorithm~\ref{alg:bestd}, part of calculating $\bestd(\anc, i)$ involves set subtracting $\mystate_i[\node].\detpos[v]$ from $\bestd(\parent(\anc),i)$ if \node{} is positively determinable.
More specifically, $\bestd(\anc,i) \subseteq \bestd(\parent(\anc),i) \setminus \mystate_i[\node].\detpos[v]$.
Property~\ref{prop:subset} states that $\vseti{i} \subseteq \bestd(\anc,i)$, so $\vseti{i} \cap \mystate_i[\node].\detpos[v] = \varnothing$.
We can substitute the value of $\mystate_i[\node].\detpos[v]$ based on Lemma~\ref{lem:det-def} to get:
\begin{align*}
  \vseti{i} \cap \{v \in \bestd(\parent(\node), \rank(\node,i)) \mid \forall u \in \vatomvi, \node[u] = \true\} & = \varnothing
\end{align*}
Since, $\vseti{i} \subseteq \bestd(\parent(\node), i)$, and $\bestd(\parent(\node), i) \subseteq \bestd(\parent(\node), \rank(\node,i))$ based on Lemma~\ref{lem:subset}:
\begin{align*}
  \{v \in \vseti{i} \mid \forall u \in \vatomvi, \node[u] = \true\} & = \varnothing
\end{align*}
In other words, \node{} is not statically true with respect to \vatomvi, for any $v \in \vseti{i}$.
\end{proof}

\subsubsection{Proof of Lemma~\ref{lem:det-effect}}
\lemdeteffect*
\begin{proof}
  We prove this by strong induction on the height of the tree referred to by \node{}.
  As the base case, let \node{} be a leaf node which refers to \patom[] \predi{j}.
  If $j \ge i$, then \node{} cannot be statically true or false with respect to \vatomvi{} for any $v$.
  If $j < i$, then \node{} must be complete by step $i$, so it must also be positively and negatively determinable by step $i$.
  Thus, the base case is trivially resolved.

  As the inductive hypothesis, assume that the lemma holds for any predicate tree nodes of height $k$ or less.
  Next, as the inductive step, let \node{} have a height of $k+1$.
  We prove the contrapositive of the lemma for \node{}:
  For all $v \in \vseti{i}$, if \node{} is statically true with respect to \vatomvi{}, then \node{} is positively determinable on step $i$.
  If \node{} is statically false with respect to \vatomvi{}, then \node{} is negatively determinable on step $i$.

  If \node{} is an AND node and statically true with respect to \vatomvi{}, then each $u \in \vatomvi$ must resolve to true against every child of \node{}.
  Thus, all of \node{}'s children must be statically true with respect to \vatomvi{}, and by the inductive hypothesis, positively determinable on step $i$.
  Therefore, by definition, \node{} itself must also be positively determinable.

  In the case that \node{} is an AND node and statically false with respect to \vatomvi{}, we claim that there exists a child of \node{} which is statically false with respect to \vatomvi{}.
  By the inductive hypothesis, if a child of \node{} is statically false, it must be negatively determinable.
  By definition, if one of \node{}'s child is negatively determinable, \node{} must be negatively determinable, completing the proof for this case.
  Assume to the contrary that all children of \node{} are not statically false with respect to \vatomvi{}.
  There are two cases for each child of \node{}:
  \begin{enumerate*}[(1)]
    \item The child is statically true with respect to \vatomvi{}.
    \item The child is dynamic with respect to \vatomvi{}.
  \end{enumerate*}
  The first case is valid, but not all children can fall into this category because that would imply that \node{} is statically true with respect to \vatomvi{}, leading to a contradiction.
  Therefore, there must be at at least one child that is dynamic with respect to \vatomvi{}.
  Now consider once again that our predicate expression is a monotone Boolean function~\cite{stephenCountingInequivalentMonotone2014}.
  Using the same notation as before, consider the valuation $u$ which is constructed according to:
  \begin{enumerate}
    \item For all $j$ in  $(1 \le j \le i-1)$, set $u_j = v_j$.
    \item For all $j$ in $(i \le j \le n)$, set $u_j = 1$.
  \end{enumerate}
  Since \node{} is statically false with respect to \vatomvi{}, there must exist a child node \child{} of \node{} such that $\child[u] = 0$.
  However, for all $u' \in \vatomvi$, $u' \le u$, so based on the property of monotone Boolean functions, $\child[u'] = 0$.
  This would mean that \child{} is statically false with respect to \vatomvi{}, leading to another contradiction.

  If \node{} is an OR node, the same reasoning applies but in reverse.
  In the case that \node{} is an OR node and statically true with respect to \vatomvi{}, we claim that there exists a child of \node{} which is statically true with respect to \vatomvi{}.
  By the inductive hypothesis, if a child of \node{} is statically true, it must be positively determinable.
  By definition, if one of \node{}'s child is positively determinable, \node{} must be positively determinable, completing the proof for this case.
  Assume to the contrary that all children of \node{} are not statically true with respect to \vatomvi{}.
  There are two cases for each child of \node{}:
  \begin{enumerate*}[(1)]
    \item The child is statically false with respect to \vatomvi{}.
    \item The child is dynamic with respect to \vatomvi{}.
  \end{enumerate*}
  The first case is valid, but not all children can fall into this category because that would imply that \node{} is statically false with respect to \vatomvi{}, leading to a contradiction.
  Therefore, there must be at at least one child that is dynamic with respect to \vatomvi{}.
  Now consider once again that our predicate expression is a monotone Boolean function.
  Using the same notation as before, consider the valuation $u$ which is constructed according to:
  \begin{enumerate}
    \item For all $j$ in  $(1 \le j \le i-1)$, set $u_j = v_j$.
    \item For all $j$ in $(i \le j \le n)$, set $u_j = 0$.
  \end{enumerate}
  Since \node{} is statically true with respect to \vatomvi{}, there must exist a child node \child{} of \node{} such that $\child[u] = 1$.
  However, for all $u' \in \vatomvi$, $u \le u'$, so based on the property of monotone Boolean functions, $\child[u'] = 1$.
  However, this would mean that \child{} is statically true with respect to \vatomvi{}, leading to another contradiction.

  If \node{} is an OR node and statically false with respect to \vatomvi{}, then each $u \in \vatomvi$ must resolve to false against every child of \node{}.
  Thus, all of \node{}'s children must be statically false with respect to \vatomvi{}, and by the inductive hypothesis, negatively determinable on step $i$.
  Therefore, by definition, \node{} itself must also be negatively determinable.
\end{proof}

\subsubsection{Proof of Lemma~\ref{lem:det-def}}
\lemdetdef*
\begin{proof}
  We prove this by strong induction on the height of the tree referred to by \node{}.
  As the base case, let \node{} be a leaf node which refers to \patom[] \predi{j}.
  If \node{} is positively determinable on step $i$, it must be that $j < i$ and $j = \rank(\node,i)$.
  When \detpos[v] and \detneg[v] were updated at the end of step $j$, they were updated with:
  \begin{align*}
    \mystate_i[\node].\detpos[v] &= \predi{j}(\bestd(\parent(\node),j)) \\
    \mystate_i[\node].\detneg[v] &= \bestd(\parent(\node),j) \setminus \predi{j}(\bestd(\parent(\node),j))
  \end{align*}
  These expressions expand to:
  \begin{align*}
    \mystate_i[\node].\detpos[v] &= \{v \in \bestd(\parent(\node),j) \mid \node[v] = \true\} \\
    \mystate_i[\node].\detneg[v] &= \{v \in \bestd(\parent(\node),j) \mid \node[v] = \false\}
  \end{align*}
  Since the value of $\node[v]$ for \patom[] \predi{j} is solely determined by $v_j$, if $\node[v] = \true$, then for all $u \in \vatomvi, \node[u] = \true$.
  Similarly, if $\node[v] = \false$, then for all $u \in \vatomvi, \node[u] = \false$.
  Thus, the above expressions reduce to:
  \begin{align*}
    \mystate_i[\node].\detpos[v] &= \{v \in \bestd(\parent(\node),j) \mid \forall u \in \vatomvi, \node[u] = \true\} \\
    \mystate_i[\node].\detneg[v] &= \{v \in \bestd(\parent(\node),j) \mid \forall u \in \vatomvi, \node[u] = \false\}
  \end{align*}
  and the base case is resolved.

  As the inductive hypothesis, assume that the lemma holds for all predicate tree nodes of height $k$ or less.
  For the inductive step, let \node{} be a predicate tree node of height $k+1$.

  {\bf If \node{} is an AND node and positively determinable on step $i$,} all of its children must be positively determinable.
  Based on Algorithm~\ref{alg:update}:
  \begin{equation}
    \mystate_i[\node].\detpos[v] = \bigcap_{\cany \in \chl(\node)} \mystate_i[\cany].\detpos[v] \cap \bestd(\parent(\node), \rank(\node,i))
    \label{eq:lemdetdef:1}
  \end{equation}
  The outer inductive hypothesis states that for child \cany{} of \node{}:
  \[
    \mystate_i[\cany].\detpos[v] = \{v \in \bestd(\node, \rank(\cany, i)) \mid \forall u \in \vatomvi, \cany[u] = \true\}
  \]
  Based on Algorithm~\ref{alg:bestd}, we can expand \bestd:
  \begin{multline}
    \mystate_i[\cany].\detpos[v] = \{v \in \bestd(\parent(\node), \rank(\cany, i)) \mid \forall u \in \vatomvi, \cany[u] = \true\} \\
    \cap \left( \left(\bigcap_{\ccmp \in \chlc(\node, \rank(\cany, i))} \mystate_{\rank(\cany,i)}[\ccmp].\cmp[v] \right) \setminus\left(\bigcup_{\cneg \in \chln(\node,\rank(\cany,i)) \setminus \{\cany\}} \mystate_{\rank(\cany,i)}[\cneg].\detneg[v]\right) \right)
    \label{eq:lemdetdef:20}
  \end{multline}
  Substituting into \autoref{eq:lemdetdef:1} gives us:
  \begin{multline*}
    \mystate_i[\node].\detpos[v] = \bigcap_{\cany \in \chl(\node)} \{v \in \bestd(\parent(\node), \rank(\cany, i)) \mid \forall u \in \vatomvi, \cany[u] = \true\} \\
    \cap \left( \left(\bigcap_{\ccmp \in \chlc(\node, \rank(\cany, i))} \mystate_{\rank(\cany,i)}[\ccmp].\cmp[v] \right) \setminus\left(\bigcup_{\cneg \in \chln(\node,\rank(\cany,i)) \setminus \{\cany\}} \mystate_{\rank(\cany,i)}[\cneg].\detneg[v]\right) \right) \\
    \cap \bestd(\parent(\node), \rank(\node, i))
  \end{multline*}
  After using \autoref{prop:detpos} to convert \cmp[v] values to \detpos[v] values, the expression $\mystate_{\rank(\cany,i)}[\ccmp].\cmp[v]$ in the second line can be substituted using \autoref{eq:lemdetdef:20}.
  By repeatedly expanding this term, applying \autoref{lem:subset} ($\rank(\cany,i)$ must be the latest step for any $\cany$), and applying the identities $A \cap (B \setminus C) = (A\cap B) \setminus C$ and $(A \setminus B) \setminus C = A \setminus (B \cup C)$ for all sets $A$, $B$, and $C$ (\autoref{apx:set-alg}), we are left with:
  \begin{multline*}
    \mystate_i[\node].\detpos[v] = \bigcap_{\cany \in \chl(\node)} \{v \in \bestd(\parent(\node), \rank(\cany, i)) \mid \forall u \in \vatomvi, \cany[u] = \true\} \\
    \cap \bestd(\parent(\node), \rank(\node, i)) \setminus \left( \bigcup_{j_{\cany}} \mystate_{j_\cany}[\cany].\detneg[v] \right)
  \end{multline*}
  for some set of step indices $j_{\cany}$ for each child $\cany$.
  By definition, $\mystate_{j_{\cany}}[\cany].\detneg[v]$ does not contain any valuations $v$ for which $\cany[v] = \true$, whereas the expression $\{v \in \bestd(\parent(\node), \rank(\cany, i)) \mid \forall u \in \vatomvi, \cany[u] = \true\}$ only contains valuations $v$ for which $\cany[v] = \true$ (because $v \in \vatomvi$).
  Thus, these sets are mutually exclusive, and we can get rid of the set subtraction terms:
  \begin{multline*}
    \mystate_i[\node].\detpos[v] = \bigcap_{\cany \in \chl(\node)} \{v \in \bestd(\parent(\node), \rank(\cany, i)) \mid \forall u \in \vatomvi, \cany[u] = \true\} \\
    \cap \bestd(\parent(\node), \rank(\node, i))
  \end{multline*}
  Applying \autoref{lem:subset} yields:
  \[
    \mystate_i[\node].\detpos[v] = \bigcap_{\cany \in \chl(\node)} \{v \in \bestd(\parent(\node), \rank(\node, i)) \mid \forall u \in \vatomvi, \cany[u] = \true\}
  \]
  The expression $\{v \in \bestd(\parent(\node), j) \mid \forall u \in \vatomvi, \cany[u] = \true\}$ returns only valuations $v$ for which child $\cany$ is statically true with respect to \vatomvi{}.
  The intersection of these valuations is the valuations $v$ for which every child of \node{} is statically true with respect to \vatomvi{}.
  For an AND node, having every child be statically true with respect to a valuation set is the requirement to be statically true to the valuation set.
  Thus:
  \[
    \mystate_i[\node].\detpos[v] =  \{v \in \bestd(\parent(\node), \rank(\node, i)) \mid \forall u \in \vatomvi, \node[u] = \true\}
  \]
  {\bf If \node{} is an AND node and negatively determinable on step $i$,} we prove this case by induction on step index $i$ (henceforth referred to as the ``inner'' induction).
  For \node{} to be negatively determinable, one of its children must be negatively determinable.
  Let \cnegp{} be the first negatively determinable child of \node{}, and
  let \cnegp{} be first negatively determinable on step $j+1$.
  The base case for the inner induction is $i = j+1$.
  Based on Algorithm~\ref{alg:update}:
  \[
    \mystate_{j+1}[\node].\detneg[v] = \bigcup_{\cneg \in \chln(\node, j+1)} \mystate_{j+1}[\cneg].\detneg[v] \cap \bestd(\parent(\node), j)
  \]
  For the base case, on step $j$, only one of \node{}'s children is negatively determinable:
  \[
    \mystate_{j+1}[\node].\detneg[v] = \mystate_{j+1}[\cnegp].\detneg[v] \cap \bestd(\parent(\node), j)
  \]
  After applying the inductive hypothesis of the outer induction, this expands to:
  \[
    \mystate_i[\node].\detneg[v] = \{v \in \bestd(\node,j) \mid \forall u \in \vatomvi, \cnegp[u] = \false \} \cap \bestd(\parent(\node), j)
  \]
  None of \node{}'s children are negatively determinable nor complete on step $j$, so based on Algorithm~\ref{alg:bestd}, $\bestd(\node,j) = \bestd(\parent(\node),j)$, and the above equation reduces to:
  \[
    \mystate_i[\node].\detneg[v] = \{v \in \bestd(\parent(\node),j) \mid \forall u \in \vatomvi, \cnegp[u] = \false \}
  \]
  Since \node{} is an AND node, if one of \node{}'s children is statically false with respect to \vatomvi{}, then \node{} itself must also be statically false with respect to \vatomvi{}:
  \[
    \mystate_i[\node].\detneg[v] = \{v \in \bestd(\parent(\node),j) \mid \forall u \in \vatomvi, \node[u] = \false \}
  \]
  Thus, the base case is resolved.

  Next, as the inner inductive hypothesis, assume that the lemma holds for any step $i =k$.
  For the inductive step, let us observe step $i=k+1$.
  If \predi{k} is not a descendant of \node{}, $\mystate_i = \mystate_k$, and the inner inductive hypothesis applies.
  Thus, let us assume \predi{k} is a descendant of \node{} and updates $\mystate[\node].\detneg[v]$.
  Based on Algorithm~\ref{alg:update}:
  \begin{align*}
    \mystate_{k+1}[\node].\detneg[v] &= \bigcup_{\cneg \in \chln(\node,k+1)} \mystate_{k+1}[\cneg].\detneg[v] \cap \bestd(\parent(\node), k)) \\
    \mystate_{k}[\node].\detneg[v] &= \bigcup_{\cneg \in \chln(\node,k)} \mystate_{k}[\cneg].\detneg[v] \cap \bestd(\parent(\node), \rank(\node, k)))
  \end{align*}
  There is only one step difference between $k$ and $k+1$, so at most one of \node{}'s children can be updated with respect to \predi{k}.
  Let \cspc{} be this ancestor of \predi{k}.
  For all other negatively determinable children $\cneg \ne \cspc$, $\mystate_{k+1}[\cneg] = \mystate_k[\cneg]$.
  Lemma~\ref{lem:subset} states that $\bestd(\parent(\node), k) \subseteq \bestd(\parent(\node), \rank(\node,k))$, and \autoref{lem:growtwo} states that $\mystate_{k+1}[\cspc].\detneg[v] \supseteq \mystate_k[\cspc].\detneg[v] \cap \bestd(\parent(\node), k)$.
  Thus, we can write $\mystate_{k+1}[\node].\detneg[v]$ in terms of $\mystate_{k}[\node].\detneg[v]$:
  \begin{equation}
    \mystate_{k+1}[\node].\detneg[v] = \left(\mystate_k[\node].\detneg[v] \cap \bestd(\parent(\node), k)\right) \cup \mystate_{k+1}[\cspc].\detneg[v]
    \label{eq:lemdetdef:8}
  \end{equation}
  The outer inductive hypothesis states that:
  \[
    \mystate_{k+1}[\cspc].\detneg[v] = \{v \in \bestd(\node, k) \mid \forall u \in \vatomvi, \cspc[u] = \false \}
  \]
  By Property~\ref{prop:subset}, $\bestd(\node, k) \subseteq \bestd(\parent(\node), k)$, so:
  \[
    \mystate_{k+1}[\cspc].\detneg[v] \subseteq \{v \in \bestd(\parent(\node), k) \mid \forall u \in \vatomvi, \cspc[u] = \false \}
  \]
  Based on the inner inductive hypothesis and Lemma~\ref{lem:subset}:
  \[
    \mystate_{k}[\node].\detneg[v] \cap \bestd(\parent(\node), k)= \{v \in \bestd(\parent(\node), k) \mid \forall u \in \vatomvi, \node[u] = \false \}
  \]
  In both cases the domain ($\bestd(\parent(\node), k)$) is the same, but the condition $\cspc[u] = \false$ implies $\node[u] = \false$:
  \[
    \mystate_{k+1}[\cspc].\detneg[v] \subseteq \mystate_{k}[\node].\detneg[v] \cap \bestd(\parent(\node), k)
  \]
  Thus, Equation~\ref{eq:lemdetdef:8} simplifies to:
  \begin{align*}
    \mystate_{k+1}[\node].\detneg[v] &= \mystate_k[\node].\detneg[v] \cap \bestd(\parent(\node), k) \\
                                   & =\{v \in \bestd(\parent(\node), k) \mid \forall u \in \vatomvi, \node[u] = \false \}
  \end{align*}


  {\bf If \node{} is an OR node and positively determinable on step $i$,} the reasoning is similar as when \node{} is an AND node and negatively determinable on step $i$.
  We prove this case by induction on step index $i$ (henceforth referred to as the ``inner'' induction).
  Let \cposp{} be the first positively determinable child of \node{}, and
  let \cposp{} be first positively determinable on step $j+1$.
  The reasoning for the base case of the inner induction is the exact same as the reasoning for the negatively determinable AND node's base case.
  The only difference is that each positively determinable child is statically true with respect to \vatomvi{} instead of statically false.

  Next, as the inner inductive hypothesis, assume that the lemma holds for any step $i =k$.
  For the inductive step, let us observe step $i=k+1$.
  If \predi{k} is not a descendant of \node{}, $\mystate_i = \mystate_k$, and the inner inductive hypothesis applies.
  Thus, let us assume \predi{k} is a descendant of \node{} and updates $\mystate[\node].\detpos[v]$.
  Based on Algorithm~\ref{alg:update}:
  \begin{align*}
    \mystate_{k+1}[\node].\detpos[v] &= \bigcup_{\cpos \in \chln(\node,k+1)} \mystate_{k+1}[\cpos].\detpos[v] \cap \bestd(\parent(\node), k)) \\
    \mystate_{k}[\node].\detpos[v] &= \bigcup_{\cpos \in \chln(\node,k)} \mystate_{k}[\cpos].\detpos[v] \cap \bestd(\parent(\node), \rank(\node, k)))
  \end{align*}
  There is only one step difference between $k$ and $k+1$, so at most one of \node{}'s children can be updated with respect to \predi{k}.
  Let \cspc{} be this ancestor of \predi{k}.
  For all other positively determinable children $\cpos \ne \cspc$, $\mystate_{k+1}[\cpos] = \mystate_k[\cpos]$.
  Lemma~\ref{lem:subset} states that $\bestd(\parent(\node), k) \subseteq \bestd(\parent(\node), \rank(\node,k))$, and \autoref{lem:growtwo} states that $\mystate_{k+1}[\cspc].\detpos[v] \supseteq \mystate_k[\cspc].\detpos[v] \cap \bestd(\parent(\node), k)$.
  Therefore, we can write $\mystate_{k+1}[\node].\detpos[v]$ in terms of $\mystate_{k}[\node].\detpos[v]$:
  \begin{equation}
    \mystate_{k+1}[\node].\detpos[v] = \left(\mystate_k[\node].\detpos[v] \cap \bestd(\parent(\node), k)
\right) \cup \mystate_{k+1}[\cspc].\detpos[v]
    \label{eq:lemdetdef:9}
  \end{equation}
  The outer inductive hypothesis states that:
  \[
    \mystate_{k+1}[\cspc].\detpos[v] = \{v \in \bestd(\node, k) \mid \forall u \in \vatomvi, \cspc[u] = \true \}
  \]
  By Property~\ref{prop:subset}, $\bestd(\node, k) \subseteq \bestd(\parent(\node), k)$, so:
  \[
    \mystate_{k+1}[\cspc].\detpos[v] \subseteq \{v \in \bestd(\parent(\node), k) \mid \forall u \in \vatomvi, \cspc[u] = \true \}
  \]
  Based on the inner inductive hypothesis and Lemma~\ref{lem:subset}:
  \[
    \mystate_{k}[\node].\detpos[v] \cap \bestd(\parent(\node), k)= \{v \in \bestd(\parent(\node), k) \mid \forall u \in \vatomvi, \node[u] = \true \}
  \]
  In both cases the domain ($\bestd(\parent(\node), k)$) is the same, but the condition $\cspc[u] = \true$ implies $\node[u] = \true$:
  \[
    \mystate_{k+1}[\cspc].\detpos[v] \subseteq \mystate_{k}[\node].\detpos[v] \cap \bestd(\parent(\node), k)
  \]
  Thus, Equation~\ref{eq:lemdetdef:9} simplifies to:
  \begin{align*}
    \mystate_{k+1}[\node].\detpos[v] &= \mystate_k[\node].\detpos[v] \cap \bestd(\parent(\node), k) \\
                                   & =\{v \in \bestd(\parent(\node), k) \mid \forall u \in \vatomvi, \node[u] = \true \}
  \end{align*}

  {\bf If \node{} is an OR node and negatively determinable on step $i$,} all of its children must also be negatively determinable.
  Based on Algorithm~\ref{alg:update}:
  \begin{equation}
    \mystate_i[\node].\detneg[v] = \bigcap_{\cany \in \chl(\node)} \mystate_i[\cany].\detneg[v] \cap \bestd(\parent(\node), \rank(\node,i))
    \label{eq:lemdetdef:10}
  \end{equation}
  The outer inductive hypothesis states that for child \cany{} of \node{}:
  \[
    \mystate_i[\cany].\detneg[v] = \{v \in \bestd(\node, \rank(\cany, i)) \mid \forall u \in \vatomvi, \cany[u] = \false\}
  \]
  Based on Algorithm~\ref{alg:bestd}, we can expand \bestd:
  \begin{multline*}
    \mystate_i[\cany].\detneg[v] = \{v \in \bestd(\parent(\node), \rank(\cany, i)) \mid \forall u \in \vatomvi, \cany[u] = \false\} \\
    \setminus \left( \bigcup_{\ccmp \in \chlc(\node, \rank(\cany, i))} \mystate_{\rank(\cany,i)}[\ccmp].\cmp[v]  \bigcup_{\cpos \in \chlp(\node,\rank(\cany,i)) \setminus \{\cany\}} \mystate_{\rank(\cany,i)}[\cpos].\detpos[v] \right)
  \end{multline*}
  Using \autoref{prop:detpos}, all \cmp[v] values can be converted into \detpos[v]:
  \begin{multline*}
    \mystate_i[\cany].\detneg[v] = \{v \in \bestd(\parent(\node), \rank(\cany, i)) \mid \forall u \in \vatomvi, \cany[u] = \false\} \\
    \setminus \left( \bigcup_{\cpos \in (\chlc(\node, \rank(\cany, i)) \cup \chlp(\node, \rank(\cany,i)) \setminus \{\cany \})} \mystate_{\rank(\cany,i)}[\cpos].\detpos[v] \right)
  \end{multline*}
  Substituting into \autoref{eq:lemdetdef:10} gives us:
  \begin{multline*}
    \mystate_i[\node].\detneg[v] = \bigcap_{\cany \in \chl(\node)} \Bigggl \{v \in \bestd(\parent(\node), \rank(\cany, i)) \mid \forall u \in \vatomvi, \cany[u] = \false\} \\
    \cap \bestd(\parent(\node), \rank(\node, i)) \Bigggr \\
    \setminus \left( \bigcup_{\cpos \in (\chlc(\node, \rank(\cany, i)) \cup \chlp(\node, \rank(\cany,i)) \setminus \{\cany \})} \mystate_{\rank(\cany,i)}[\cpos].\detpos[v] \right)
  \end{multline*}
  After applying \autoref{lem:subset} ($\rank(\cany,i) \le \rank(\node, i)$), this simplifies to:
  \begin{multline*}
    \mystate_i[\node].\detneg[v] = \bigcap_{\cany \in \chl(\node)} \{v \in \bestd(\parent(\node), \rank(\node, i)) \mid \forall u \in \vatomvi, \cany[u] = \false\} \\
    \setminus \left( \bigcup_{\cpos \in (\chlc(\node, \rank(\cany, i)) \cup \chlp(\node, \rank(\cany,i)) \setminus \{\cany \})} \mystate_{\rank(\cany,i)}[\cpos].\detpos[v] \right)
  \end{multline*}
  By definition, $\mystate_{j_{\cany}}[\cany].\detpos[v]$ does not contain any valuations $v$ for which $\cany[v] = \false$, whereas the expression $\{v \in \bestd(\parent(\node), \rank(\node, i)) \mid \forall u \in \vatomvi, \cany[u] = \false\}$ only contains valuations $v$ for which $\cany[v] = \false$ (because $v \in \vatomvi$).
  Thus, these sets are mutually exclusive, and we can get rid of the set subtraction terms:
  \[
    \mystate_i[\node].\detneg[v] = \bigcap_{\cany \in \chl(\node)} \{v \in \bestd(\parent(\node), \rank(\node, i)) \mid \forall u \in \vatomvi, \cany[u] = \false\}
  \]
  The expression $\{v \in \bestd(\parent(\node), j) \mid \forall u \in \vatomvi, \cany[u] = \false\}$ returns only valuations $v$ for which child $\cany$ is statically false with respect to \vatomvi{}.
  The intersection of these valuations is the valuations $v$ for which every child of \node{} is statically false with respect to \vatomvi{}.
  For an OR node, having every child be statically false with respect to a valuation set is the requirement to be statically false to the valuation set.
  Thus:
  \[
    \mystate_i[\node].\detneg[v] =  \{v \in \bestd(\parent(\node), \rank(\node, i)) \mid \forall u \in \vatomvi, \node[u] = \false\}
  \]

\end{proof}

\clearpage

\section{Set Identities}
\label{apx:set-alg}
Here, we show the proofs for the various set identities we use
throughout the paper. Identities which are not proven here come from related
works~\cite{stollSetTheoryLogic1979}~\cite{alma990000570490106761}~\cite{monk1973introduction}.

\begin{enumerate}[leftmargin=*,itemsep=3pt]
  \item
    $A \cap (B \setminus C) = (A \cap B) \setminus C$.

  \item
    $A \cap (B \setminus C) = (A \cap B) \setminus (A \cap C)$.

  \item
    $(A \cap B) \setminus C = (A \setminus C) \cap (B \setminus C)$.

  \item
    $A \cup B = (A \setminus B) \cup B$.

  \item
    $(A \setminus B) \setminus C  = A \setminus (B \cup C)$.

  \item
    $A \setminus B = A \setminus (B \cap A)$.

    \item
      $(A \cup B) \setminus C) =  (A \setminus  C) \cup (B \setminus C)$.

    \item
      $A \setminus (B \setminus C) = (A \setminus B) \cup (A \cap C)$.

  \item
    $A \cap (B \setminus (C \cup D)) = (A \setminus C) \cap (B \setminus D)$.
    \begin{align*}
      A \cap (B \setminus (C \cup D))
      &= A \cap ((B\setminus C) \cap (B \setminus D)) \\
      &= B \cap (A \setminus C) \cap (B \setminus D) \\
      &= (A \setminus C) \cap (B \setminus D)
    \end{align*}
  \item
    $(A \cap B) \setminus (C \cup D) =  (A \setminus C) \cap (B \setminus D)$.
    \begin{align*}
      (A \cap B) \setminus (C \cup D) &= A \cap (B \setminus (C \cup D)) \\
                                      &= (A \setminus C) \cap (B \setminus D)
    \end{align*}

  \item
    $A \cup (B \cap (C \setminus A)) = A \cup (B \cap C)$.
    \begin{align*}
      A \cup (B \cap (C \setminus A)) & = A \cup (B \cap (A \cup (C \setminus A))) \\
                                      & = A \cup (B \cap (A \cup C)) \\
                                      & = A \cup (B \cap C)
    \end{align*}

    \item
    $A \cup (B \setminus C) =  (A \cup B) \setminus (C \setminus A)$.
    \begin{align*}
      (A \cup B) \setminus (C \setminus A) & = ((A \cup B) \setminus C) \cup ((A \cup B) \cap A) \\
                                           & = ((A \cup B) \setminus C) \cup A \\
                                           & = A \cup (B \setminus C)
    \end{align*}

  \item
    $A = (A \setminus B) \cup (A \cap B)$.
    \begin{align*}
      (A \setminus B) \cup (A \cap B)
      &= (A \cup (A \cap B)) \setminus (B  \setminus ( A \cap B)) \\
      &= (A) \setminus (\varnothing) \\
      &= A
    \end{align*}

\end{enumerate}

}

\end{document}